\documentclass[traditabstract]{aa}  

\usepackage{booktabs}
\usepackage{graphicx}
\usepackage{txfonts}
\usepackage{multirow}
\usepackage{tabularx}
\usepackage{hyperref}
\hypersetup{colorlinks,citecolor=blue}
\usepackage{amstext}

\begin{document}

   \title{Exploring Galaxy Properties of eCALIFA with Contrastive Learning}

   \titlerunning{Exploring Galaxy Properties of eCALIFA with CL}

    \author{G.~Mart\'inez-Solaeche \inst{\ref{1}}
    \and R.~Garc\'ia-Benito \inst{\ref{1}} \and
    R. M.~Gonz\'alez Delgado \inst{\ref{1}} \and
    Luis~D\'iaz-Garc\'ia \inst{\ref{1}} \and
    S.F.~S\'anchez \inst{\ref{2}} \and
    A.M.~Conrado \inst{\ref{1}} \and
    J. E. Rodr\'iguez-Mart\'in \inst{\ref{1}}
    }
    
    \institute{Instituto de Astrof\'isica de Andaluc\'ia (CSIC), PO Box 3004, 18080 Granada, Spain (\email{gimarso@iaa.es}) \label{1} \and
    Instituto de Astronom\'ia, Universidad Nacional Aut\'onoma de M\'exico, A. P. 70-264, C.P. 04510 M\'exico, D.F., M\'exico \label{2}}

   \date{}

    \abstract{Contrastive learning (CL) has emerged as a potent tool for building meaningful latent representations of galaxy properties across a broad spectrum of wavelengths, ranging from optical and infrared to radio frequencies. These latent representations facilitate a variety of downstream tasks, including galaxy classification, similarity searches in extensive datasets, and parameter estimation, which is why they are often referred to as foundation models for galaxies. In this study, we employ CL on the latest extended data release from the Calar Alto Legacy Integral Field Area (CALIFA) survey, which encompasses a total of 895 galaxies with enhanced spatial resolution that reaches the limits imposed by natural seeing (FWHM$_{\text{PSF}}\sim 1.5$). We demonstrate that CL can be effectively applied to Integral Field Unit (IFU) surveys, even with relatively small training sets, to construct meaningful embedding where galaxies are well-separated based on their physical properties. We discover that the strongest correlations in the embedding space are observed with the equivalent width of H$\alpha$, galaxy morphology, stellar metallicity, luminosity-weighted age, stellar surface mass density, the [NII]/H$\alpha$ ratio, and stellar mass, in descending order of correlation strength. Additionally, we illustrate the feasibility of unsupervised separation of galaxy populations along the star formation main sequence, successfully identifying the blue cloud and the red sequence in a two-cluster scenario, and the green valley population in a three-cluster scenario. Our findings indicate that galaxy luminosity profiles have minimal impact on the construction of the embedding space, suggesting that morphology and spectral features play a more significant role in distinguishing between galaxy populations. Moreover, we explore the use of CL for detecting variations in galaxy population distributions across different large-scale structure, including voids, clusters, filaments and walls. Nonetheless, we acknowledge the limitations of the CL framework and our specific training set in detecting subtle differences in galaxy properties, such as the presence of an AGN or other minor scale variations that exceed the scope of primary parameters like stellar mass or morphology. Conclusively, we propose that CL can serve as an embedding function for the development of larger models capable of integrating data from multiple datasets, thereby advancing the construction of more comprehensive foundation models for galaxies.}

    \maketitle

%

\section{Introduction}
The new era of Big Data has revolutionised the field of astronomy´s approach to data analysis and interpretation. Traditional methodologies, while having served us well in the past, are being challenged and reformed by the sheer volume and complexity of data generated by contemporary and forthcoming astronomical surveys. This transformative shift is exemplified by projects such as the Dark Energy Spectroscopic Instrument \citep[DESI;][]{2013arXiv1308.0847L}, the Large Synoptic Survey Telescope \citep[LSST;][]{2019ApJ...873..111I}, the Square Kilometre Array \citep[SKA;][]{2009IEEEP..97.1482D}, or the Javalambre Physics of the Accelerating Universe Astrophysical Survey \citep[J-PAS;][]{2014arXiv1403.5237B}, which will implement a variety of observation techniques for a wide range of wavelengths.  
\par The concept of foundation models, which has recently gained significant traction across various scientific domains, represents a groundbreaking approach for harnessing the potential of massive datasets. Originating in the field of language processing with the advent of models like OpenAI's GPT \citep{2023arXiv230308774O} and Google's Gemini \citep{2023arXiv231211805G}, foundation models have fundamentally altered how we approach complex data analysis. These models are characterised by their ability to learn from a vast corpus of data and then apply this learned knowledge to a wide range of tasks, demonstrating remarkable versatility and adaptability.
\par In language processing, foundation models have achieved unprecedented success by synthesising vast amounts of textual data into coherent, contextually aware representations. This success has sparked interest in adapting the foundation model framework to other domains, including astronomy \citep{2022mla..confE..29W,2023RSOS...1021454S,2024MNRAS.527.1494L}. Unlike the sequential and textual nature of language data, astronomical data presents a unique challenge due to its diversity in formats, ranging from photometric and spectroscopic data to complex multidimensional observations like  Integral Field Unit Spectroscopy (IFS).
\par foundation models for astronomy would need to emulate the success of their language-processing counterparts by integrating and synthesising data from multiple astronomical surveys and sources. Thus, the ultimate goal would be to create a comprehensive, unified model that can analyze and interpret the complex phenomena observed in the universe. When it comes to galaxy evolution, these models would be capable of assimilating various types of data – such as the morphological, spectral, and kinematic properties of galaxies – and provide a holistic view of the processes governing galaxy formation and evolution. For example, while photometric data can reveal the distribution of stellar populations within a galaxy, spectroscopic data provides detailed information about the chemical composition and kinematics of galactic components. A foundation model for galaxies would leverage these varied datasets to deliver a more nuanced and comprehensive understanding of galaxies, transcending the limitations of analysing these datasets in isolation.
\par 
Central to the development of these foundation models is the need for sophisticated embedding mechanisms. Unlike natural language data, which is inherently sequential and textual, astronomical data spans a spectrum of formats and wavelengths, from photometric measurements to Integral Field Units (IFU) data. Efficiently embedding this multifaceted data is a precursor to building foundation models that can holistically analyze and interpret the vast expanse of astronomical information.
\par 
In this context, contrastive learning (CL) emerge as powerful tools. These methods have demonstrated remarkable efficacy in creating meaningful embedding in various fields, including image and language processing \citep{9226466}. By learning a similarity metric between input pairs, one can distil complex data into compressed, insightful representations, aiding in the identification of underlying patterns and correlations. CL in astronomy has been applied in various studies, demonstrating its potential for enhancing astronomical data analysis. Notable applications include \cite{2021ApJ...911L..33H} who applied this technique to Sloan Digital Sky Survey \citep[SDSS;][]{2000AJ....120.1579Y} \textit{ugriz} galaxy images. The authors show that the embedding representation can be used for a variety of \textit{downstream} tasks including galaxy morphology classification and photometric redshift estimation. \cite{2021ApJ...921..177S} applied CL to study stellar population and kinematic maps of galaxies obtained from the analysis of MaNGA data cubes \citep{2015ApJ...798....7B}. They emphasised CL's effectiveness in generating embedding that are robust against instrumental and observational variances. 
\par Additional applications are highlighted by the work of \cite{2021arXiv211013151S}, who developed a CL model trained on data from the DESI Legacy Imaging Surveys. Their primary aim was to create a tool that facilitates similarity searches within such extensive surveys, thereby accelerating the pace of crowd-sourcing campaigns. Similarly, \cite{2022MNRAS.517.1837G} employed CL for the unsupervised clustering of spiral galaxies observed by the Wide-Field Infrared Explorer (WISE) survey \citep{2010AJ....140.1868W}. Furthermore, in the domain of radio astronomy, CL has been utilised to conduct similarity searches and unsupervised classification of radio sources, as demonstrated by \cite{2024RASTI...3...19S}. These instances underscore the versatility and efficacy of CL in handling vast astronomical datasets, streamlining the identification of similar objects, and enabling the classification of astronmical sources in an unsupervised manner across different observational modalities.
\par Nevertheless, when it comes to IFS data, the implementation of CL for creating embedding presents unique challenges. This is because IFS represent some of the most sophisticated techniques in this field, surpassing traditional single spectroscopy or photometry. Their capability to simultaneously capture both 'images' and spectra of galaxies renders them exceptionally versatile to reveal new spatially resolved relationships governing star formation and chemical enrichment in galaxies \citep[see e.g.][]{2013ApJ...764L...1P,2014A&A...563A..49S,2016A&A...590A..44G,2021MNRAS.504.3478C}. However, this comes at the cost of producing data cubes that are significantly more complex in terms of dimensionality and memory requirements. 
\par Our work pioneers the application of CL to eCALIFA data cubes \citep{2023MNRAS.526.5555S}, which represents an extended and enhanced version of the CALIFA survey \citep{2012A&A...538A...8S,2015A&A...576A.135G}. These data cubes offer improved spatial resolution, nearly matching the limit imposed by natural seeing (FWHM $\sim 1.5''$), while preserving image quality and photometry. 
\par Our research task presents a substantial challenge: developing an effective embedding for the eCALIFA galaxies situated in a high-dimensional space ($[N_x,N_y,N_\lambda] = [159,151,1877]$). This task is particularly daunting given the relatively modest sample size of approximately 900 galaxies, especially when compared to the larger datasets typically utilised in deep learning applications within astronomy \citep{2019arXiv190407248B,huertas-company_lanusse_2023}.
\par This paper is structured as follows: Sect.~\ref{sec:data} introduces the eCALIFA data cubes utilised in this study. Subsequently, Sect.~\ref{sec:CL_framework} provides a comprehensive explanation of the CL learning model employed to train on eCALIFA galaxies. The main results are detailed in Sect.~\ref{sec:results}, followed by a discussion in Sect.~\ref{sec:discussion}. Finally, we summarise and conclude in Sect.~\ref{sec:conclusions}.

\section{Data}\label{sec:data}
\par 
In this paper we make use of the extended data release (eDR) of the CALIFA survey \citep{2023MNRAS.526.5555S}, which were acquired using the 3.5-m telescope at Calar Alto Observatory.  Utilizing the PMAS/PPak integral field spectrograph \citep{2005PASP..117..620R,2006PASP..118..129K} with the V500 grating setup, the spectra span a wavelength range of $3745-7500~\AA$. This setup achieves a resolution of approximately $R~\sim~850$ and an instrumental FWHM of $\sim 6.5~\AA$. The hexagonal field of view of the instrument is $74~\times~64$~arcsec$^2$, made possible by the PMAS/PPak fiber bundle which consists of 331 science fibers, each $2.7$ arcsec in diameter. 
\par To enhance the data quality, a dithering scheme was employed, ensuring overlapping between the apertures of adjacent fibers, which is crucial for accurate image reconstruction. The data reduction process saw remarkable advancements, in particular the implementation of a new image reconstruction algorithm and a narrower interpolation kernel. These modifications were aimed at improving spatial resolution. The data underwent a flux-conservative variation of Shepard's interpolation method and were re-sampled into a regular sampled datacube.
\par The culmination of these efforts resulted in a final data set with enhanced spatial resolution and image quality. The spatial resolution achieved is nearly at the limit of seeing conditions, with a FWHM PSF of approximately 1.25 arcsec. This level of data quality significantly augments the research capabilities in galaxy evolution, offering a more detailed and accurate representation of astronomical phenomena. 
The eCALIFA data release encompasses data cubes for a total of 1,116 galaxies, of which 895 are classified as good quality. The initial CALIFA sample was selected based on galaxy diameters, suited to the PPAK science bundle's field of view, and prioritised galaxies in the nearby universe while excluding dwarf galaxies by an apparent magnitude limit. This approach formed the CALIFA mother sample \citep{2014A&A...569A...1W}. Subsequently, the criteria were expanded to include underrepresented galaxy types, leading to the CALIFA extended sample. The eCALIFA sample covers a redshift range of $0.0005$ to $0.08$, with the majority (94~$\%$) of objects below $z < 0.03$, comprising a good representation of nearby universe galaxies. Interacting galaxies within the same field of view are presented as separate data cubes. Notably, VIIZw466a, ARP118, NGC5421NED02, NGC7436B, SQbF1, and Arp142 galaxies could be separated into two data cubes, and this was done. After visual inspection, galaxies such as PTF11owc, NGC0169\_1, and UGC00987 were excluded from the sample due to marginal spectral quality, resulting in a final sample of 898 galaxies. 
\newpage
\par
\begin{figure}[!ht]
    \centering
        \includegraphics[width=\hsize]{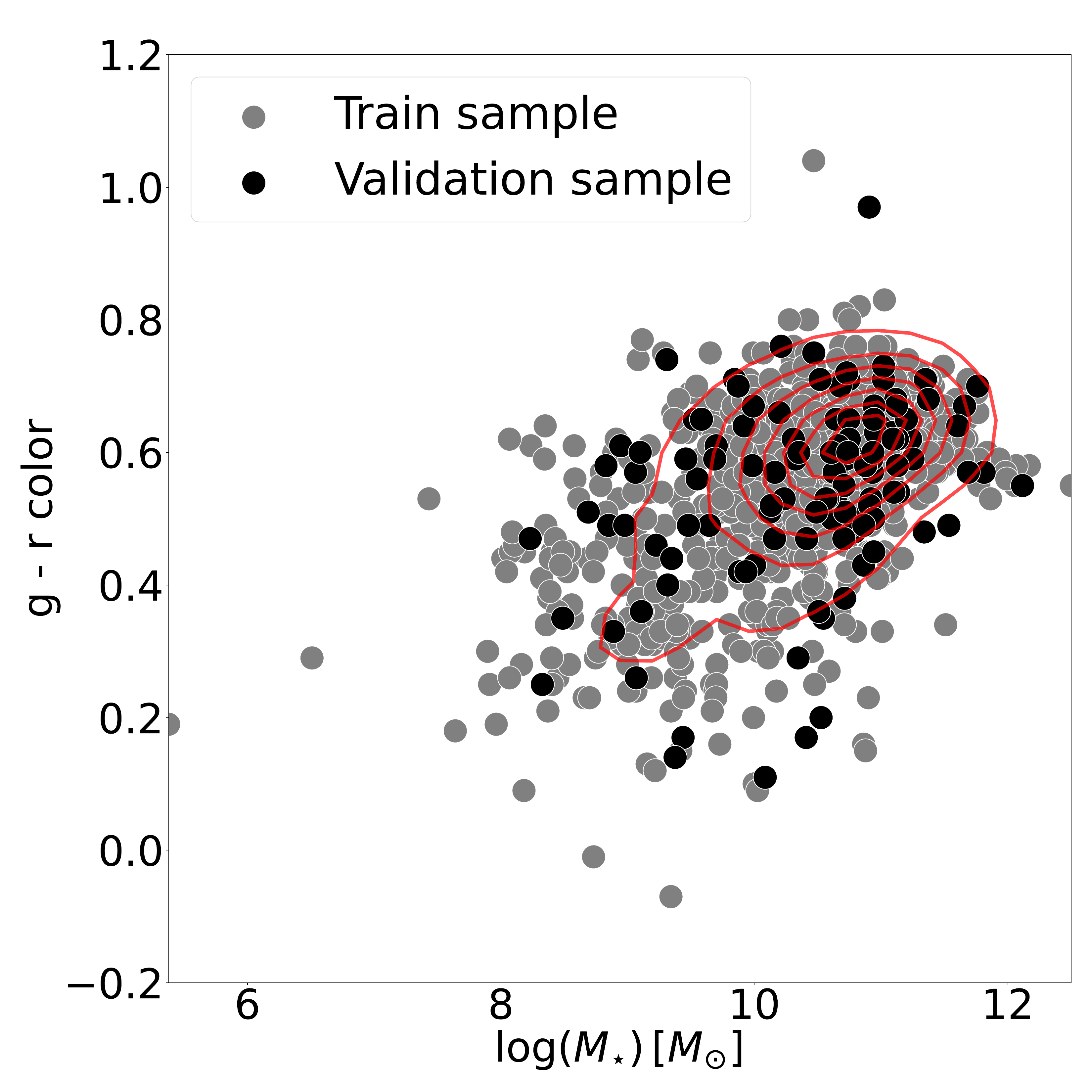}
        \vspace{1pt} 
        \includegraphics[width=\hsize]{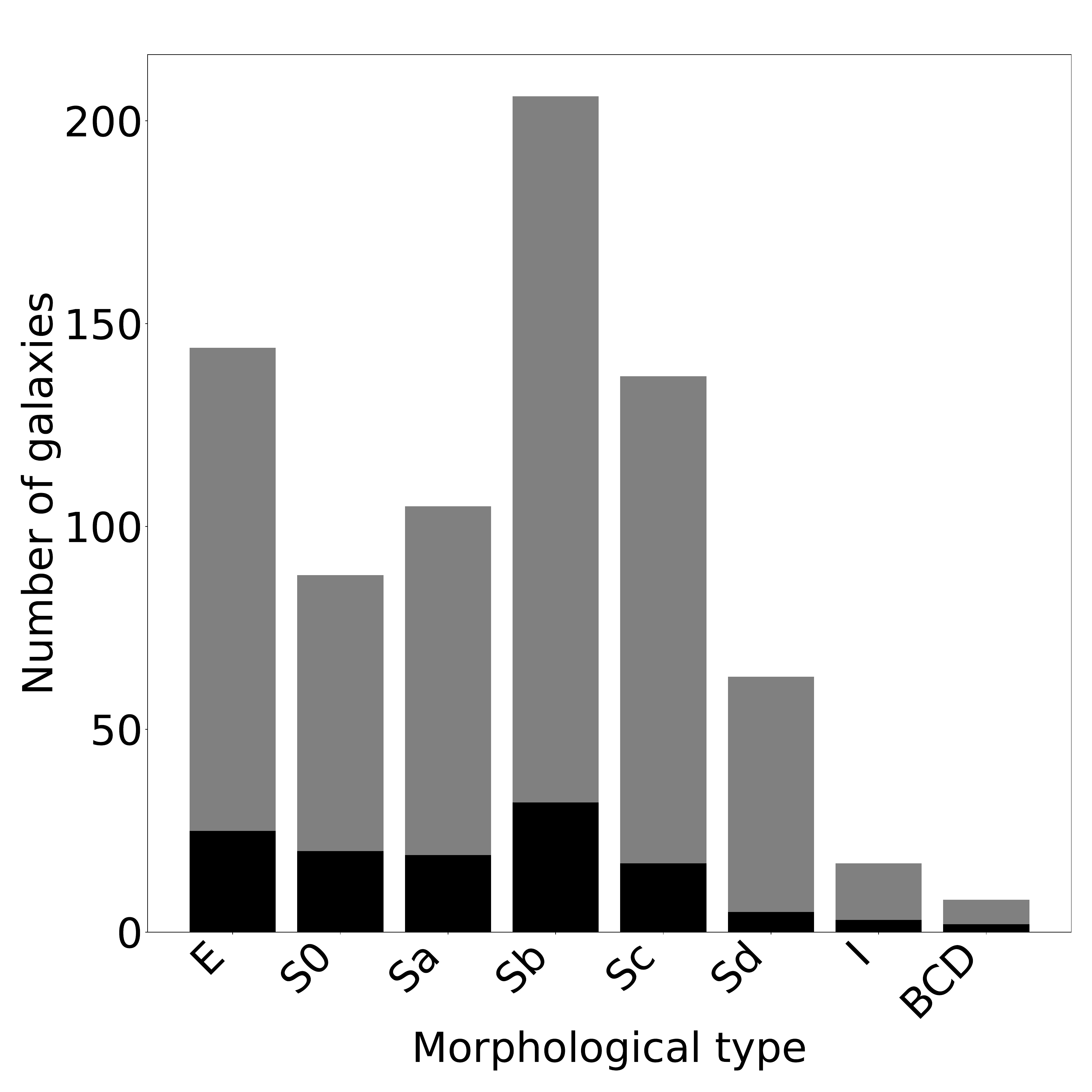}
        \caption{\small{Top: colour-mass diagram of galaxies in eCALIFA survey. We show the observed (g-r) colour in function of the stellar mass for both training and test samples. Density contours are drawn in red. Bottom: the bar plot highlights the count of galaxies across different morphological categories, such as E, S0, Sa, Sb, Sc, Sd, I, and BCD.}}
\label{fig:hist_sample}
\end{figure}

 
 We divided the eCALIFA sample into two distinct subsets: a training set comprising 772 galaxies and a validation set consisting of 126 galaxies. This division was carefully done to ensure a balanced representation of morphological types in both subsets. By maintaining similar proportions of each morphological type in both the training and validation sets, we also achieved a comparable colour-mass distribution across these subsets. This is summarised in Fig.~\ref{fig:hist_sample}.
\section{Contrastive learning framework}\label{sec:CL_framework}
\par Contrastive learning (CL) stands out for its ability to identify and maintain invariant features across a spectrum of transformations. Observational factors in astronomy often introduce non-physical discrepancies in the appearance and characteristics of astronomical objects. For instance, variations in observational conditions or instrumental calibrations can lead to distorted representations of a galaxy's properties. These discrepancies, if not accounted for, can skew our understanding and lead to inaccurate interpretations. CL, in this context, serves as a robust framework to mitigate such distortions, ensuring a more faithful representation of physical attributes.
\par The essence of CL in astronomy is encapsulated in its self-supervised nature. It thrives on a principle of minimising the distance between different views of the same astronomical object in a learned representation space, while simultaneously maximising the distance between representations of distinct objects. This is achieved through the use of contrastive losses, a technique that harnesses the power of randomised, semantic-preserving augmentations. These augmentations are carefully chosen to ensure they do not alter the inherent properties of the subject, such as the colours attributes of galaxies, which are crucial for tasks like determining the stellar populations properties.
\par At its core, CL involves a trainable function that maps a data set to the representation space. This mapping is optimised to ensure that the representations become invariant to different views of the same astronomical object. Positive pairs (different views of the same object) and negative pairs (combinations of different objects) form the basis of this learning process. Unlike other representation learning methods, CL does not rely on a predefined notion of distance in the input space. Instead, it defines positive pairs in a manner that is conducive to the specific pretext task the neural network is optimised for. This approach is particularly advantageous when dealing with complex noise or selection effects in the input data, enabling the identification of semantically similar data points that might not be easily discernible in the input space \citep[see][for a review of contrastive learning applied to astrophysics]{2023RASTI...2..441H}
\par In this section, we describe the augmentation process applied to eCALIFA data cubes. Subsequently, we elaborate on the methods employed for dimensionality reduction of these data cubes, a critical step to enhance the training efficiency of the neural network. Finally, we provide the details of the model architecture integral to our CL framework.
\subsection{Galaxy augmentation}\label{subsec:galaxy_augmentation}
\par \textit{Gaussian Noise.} To address the variability in signal-to-noise ratio (S/N) across galaxies, particularly those further away with potentially lower S/N due to uniform exposure times, we apply Gaussian noise to each spaxel. The noise level  is randomly varied for each realization, thereby generating galaxy spectra that exhibit a uniform S/N across the entire galaxy, with values ranging from 1 to 10. This method ensures that galaxies with lower S/N, a factor not indicative of their intrinsic properties, are not disproportionately represented or misinterpreted when compared to their higher S/N counterparts.
\newpage
\par \textit{Rotation and Flipping.} Each galaxy undergoes random flipping in either the 'x' or 'y' direction and a random rotation. This step is crucial as it introduces a level of agnosticism to the orientation of galaxies. In space, galaxies have no intrinsic 'up' or 'down', and our model should reflect this, ensuring that it learns to recognise galaxies irrespective of their orientation in the observational data.
\par \textit{Gaussian Blur.} We implemented a point spread function by applying a 2D Gaussian convolution with a 1$\sigma$ kernel to slightly blur the images. This process helps in softening structures that might be artefacts or excessively sharp, while still preserving essential morphological information. This step is crucial in ensuring that our model focuses on significant galactic features rather than being misled by potential anomalies in the data.
\par \textit{Simulated Stars.} To mitigate the potential bias introduced by foreground stars along the line of sight in eCALIFA galaxies, we employ a strategy of randomly generating simulated stars. These stars follow the radial and azimuthal distributions observed in eCALIFA galaxies, where stars are more frequently located at the periphery of the field of view. This approach ensures that the presence of foreground stars does not lead the network to erroneously identify them as intrinsic galactic features. Both real and simulated stars are masked by overlaying them with a simplified model of the sky background, which is produced by generating Gaussian noise spectra at the detection limit of  CALIFA observations.
\par \textit{Translation.} We randomly recenter the galaxy within a radius of 16 pixels. This translation ensures that the model remains invariant to the exact centering of the galaxy within the field of view, an important consideration given the variability in galaxy positioning due to observational factors.
\par \textit{Resize.} The galaxies are resized using factors ranging from a 2x2 to a 5x5 kernel. This resizing is essential because, despite the galaxies being in the nearby universe and having a narrow redshift range, their apparent size still varies as a function of redshift. This resizing ensures that our model can recognise galaxies across this range of apparent sizes.
\par
\begin{figure}[!ht]
    \centering
    \begin{tabular}{cc}
        \includegraphics[width=4.2cm,height=4.2cm]{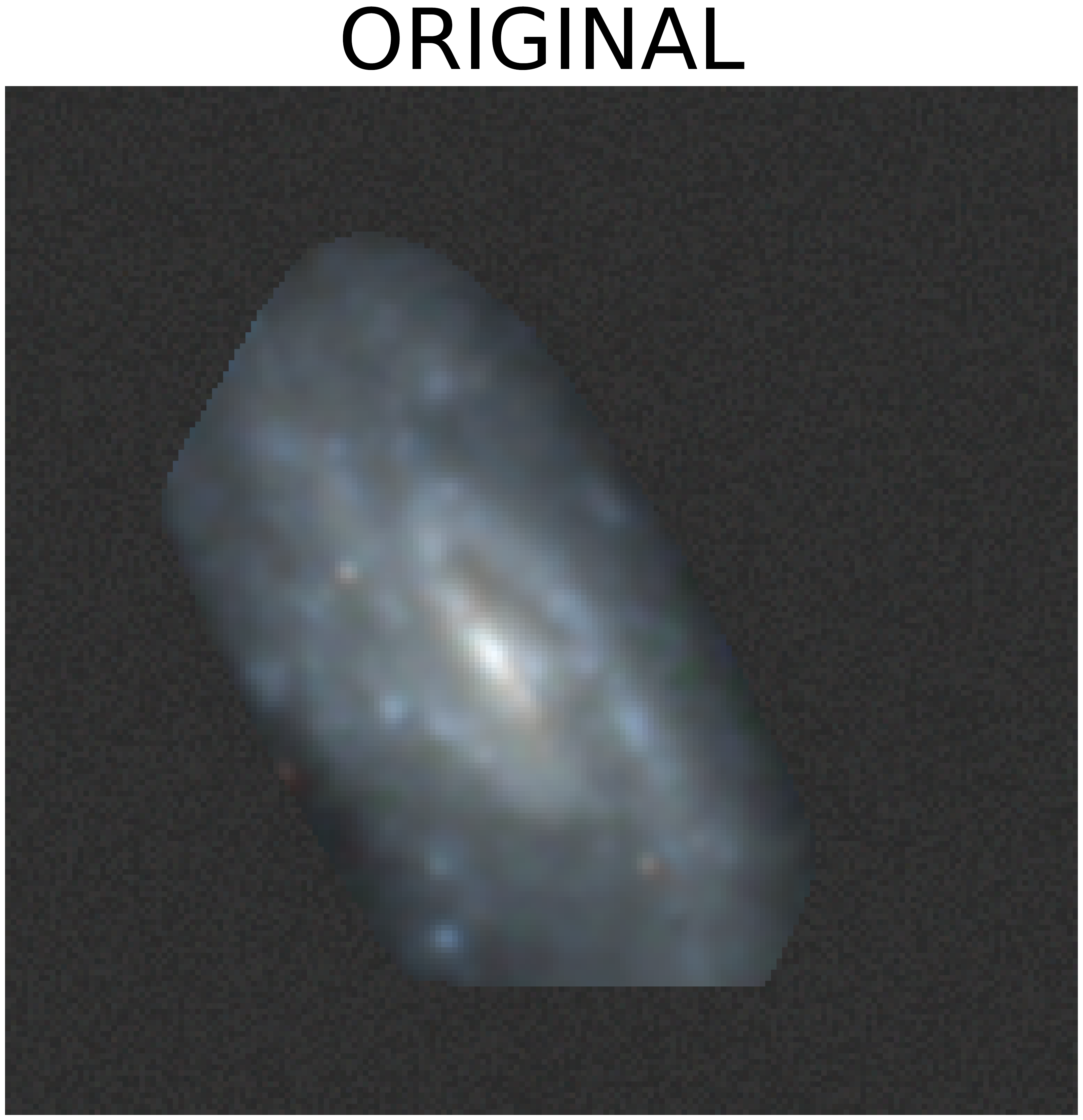} & 
        \includegraphics[width=4.2cm,height=4.2cm]{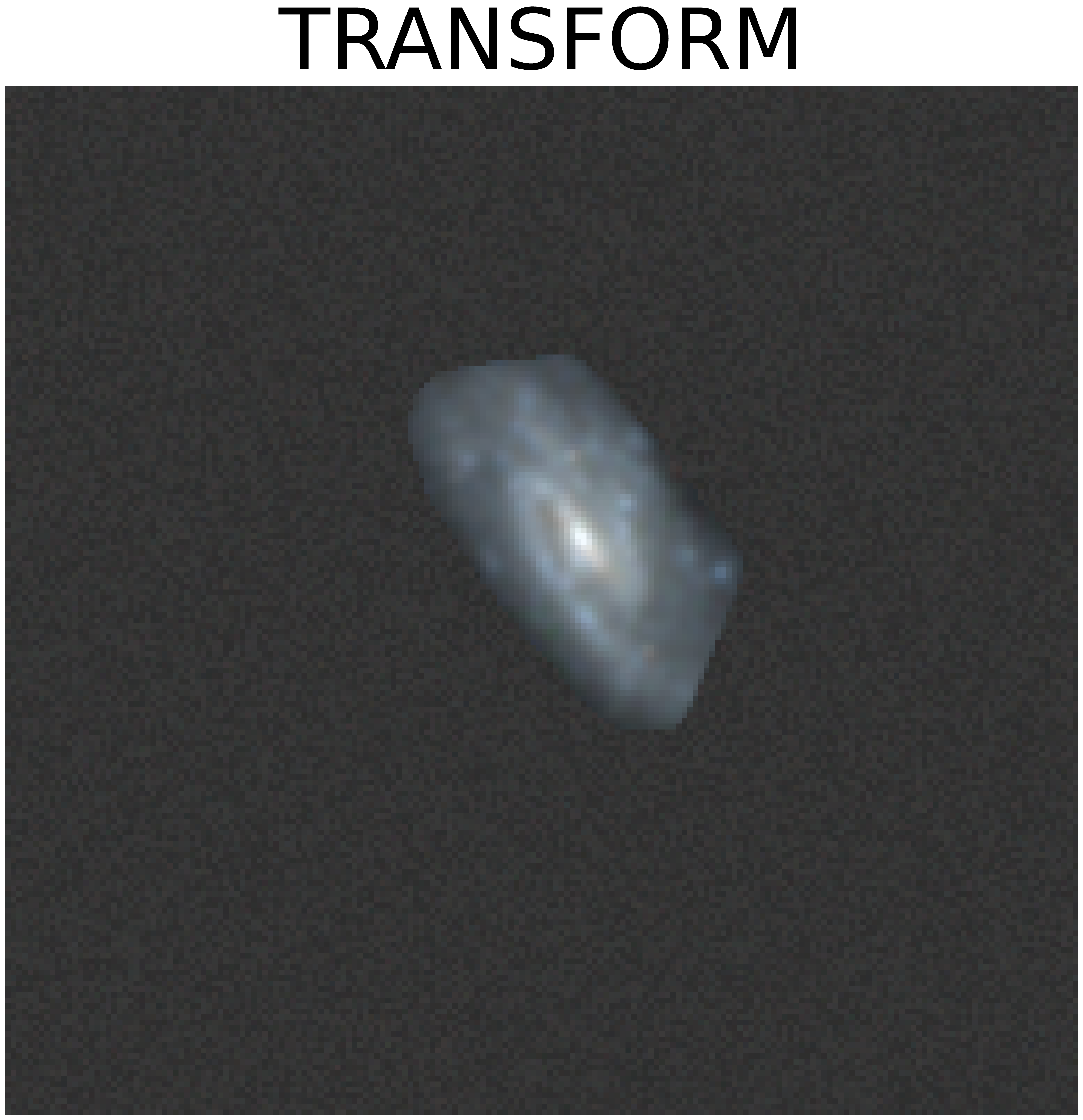} \\
        \includegraphics[width=4.2cm,height=4.2cm]{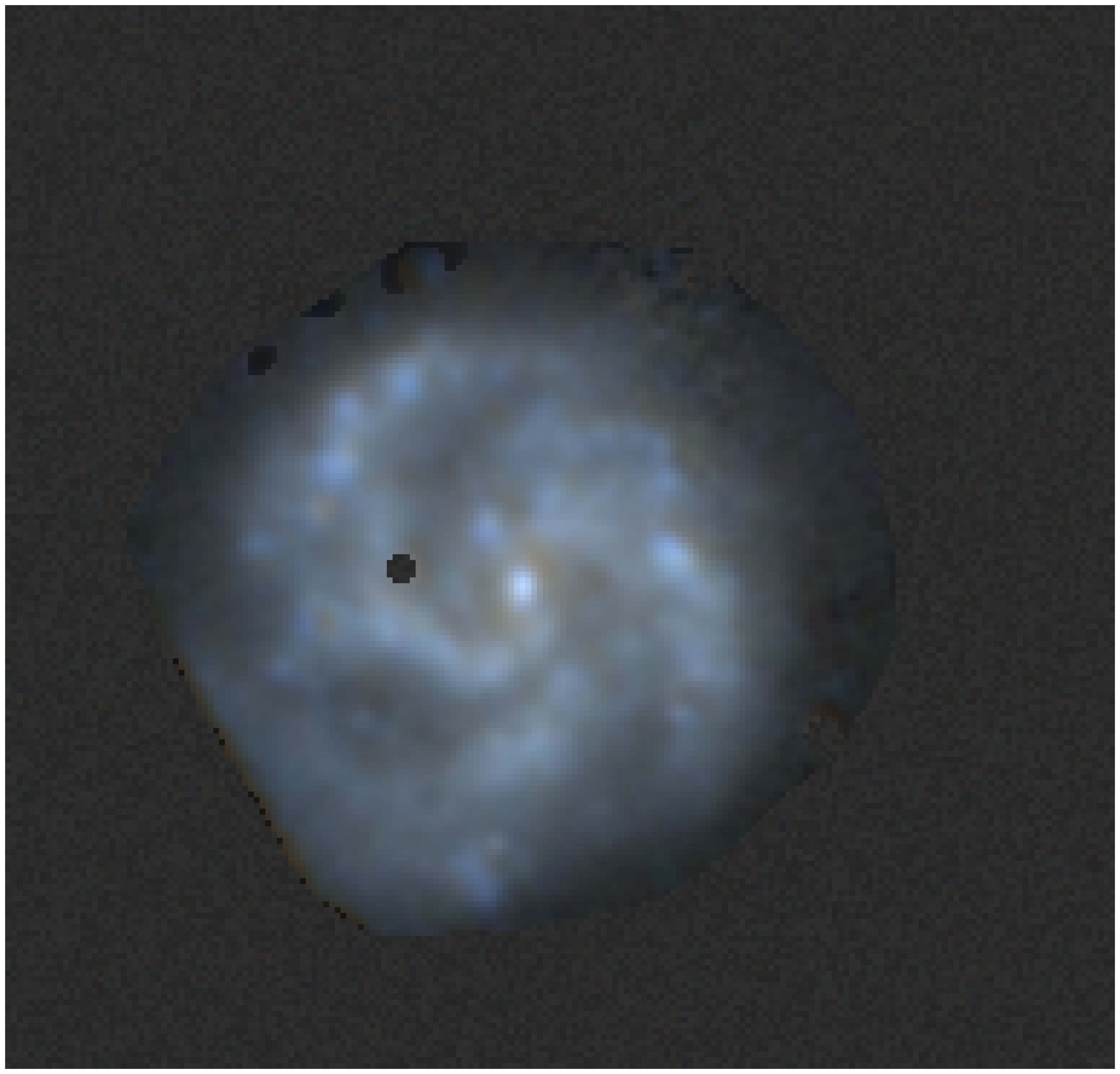} & 
        \includegraphics[width=4.2cm,height=4.2cm]{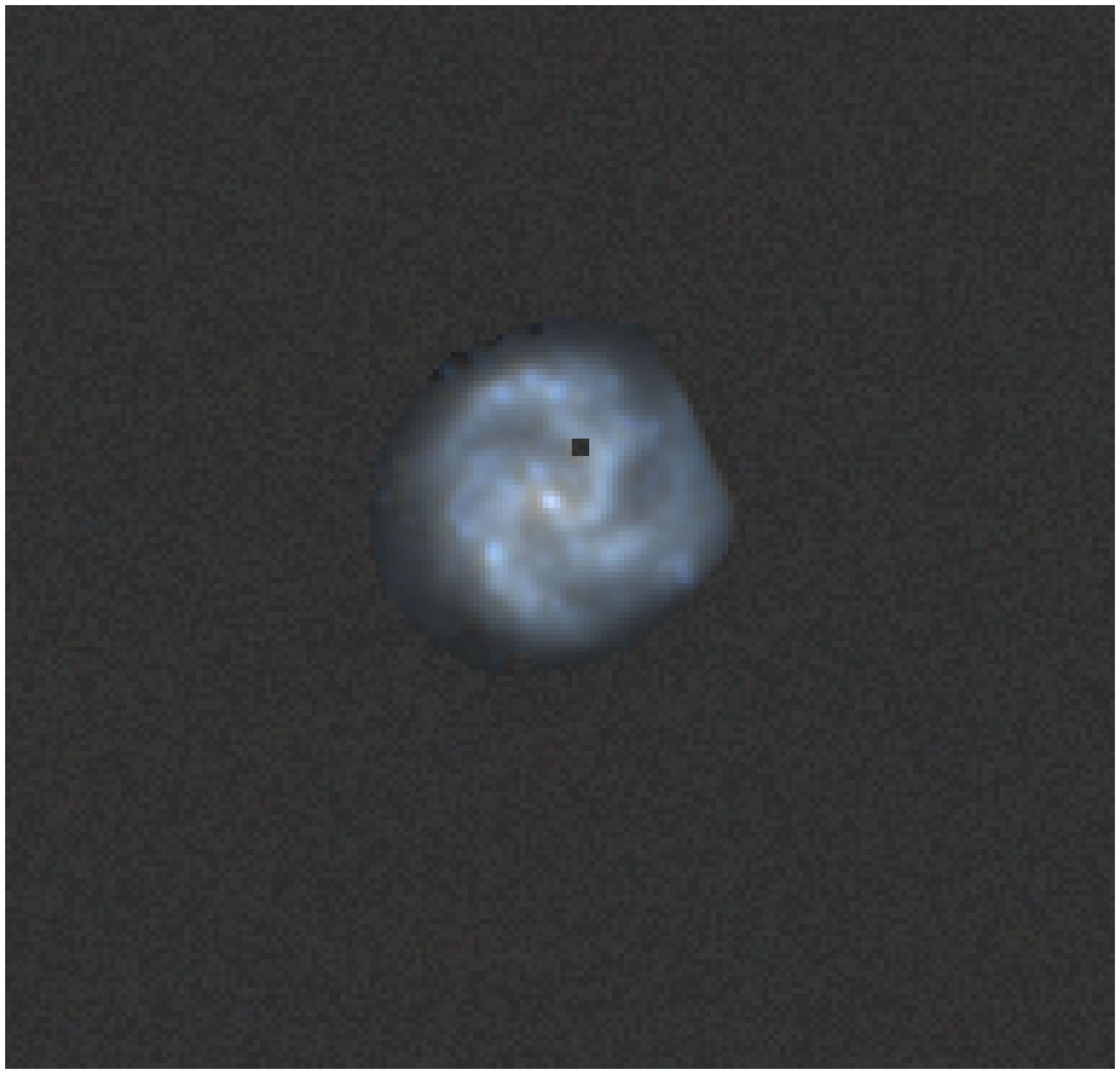} \\
        \includegraphics[width=4.2cm,height=4.2cm]{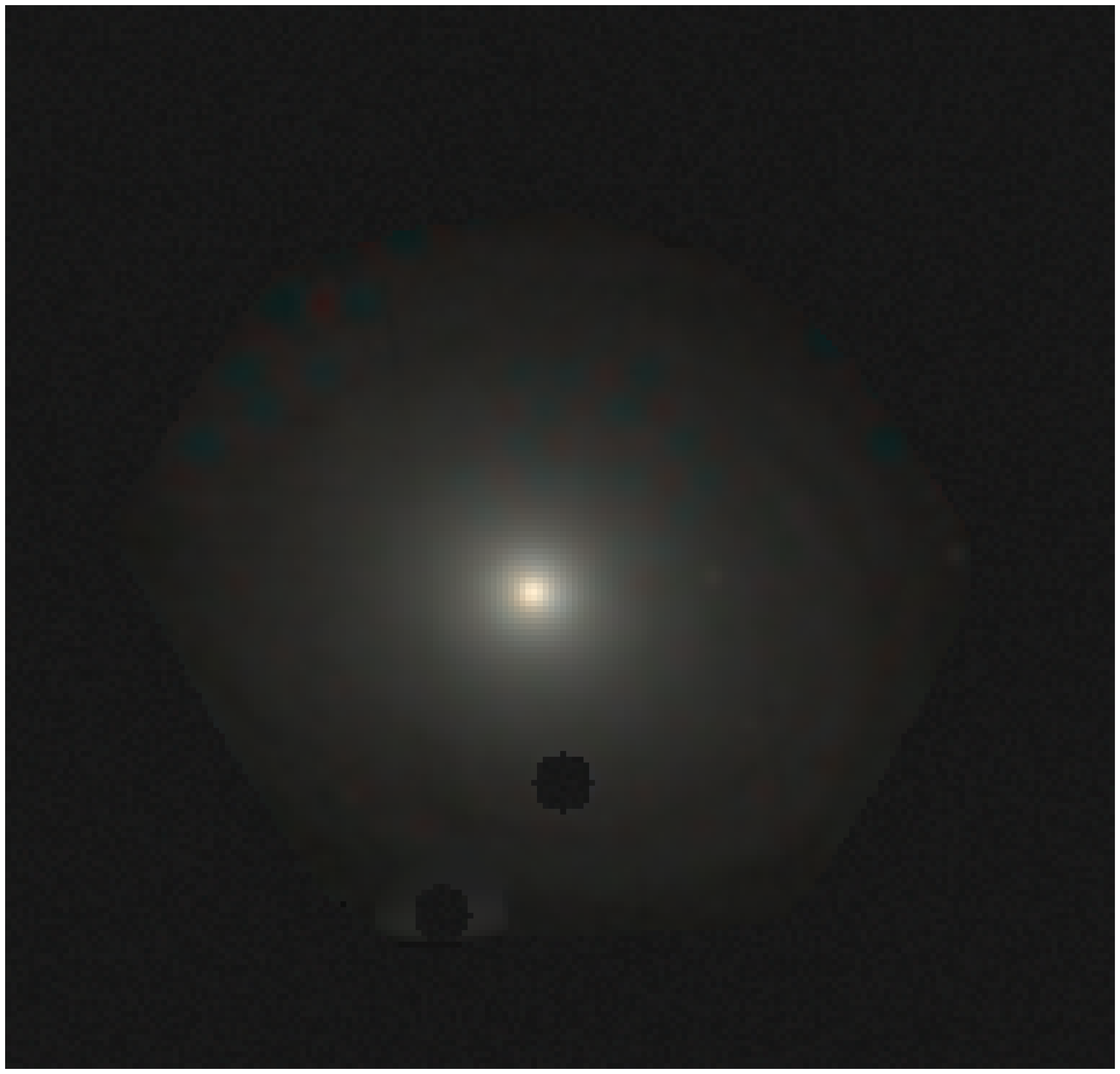} & 
        \includegraphics[width=4.2cm,height=4.2cm]{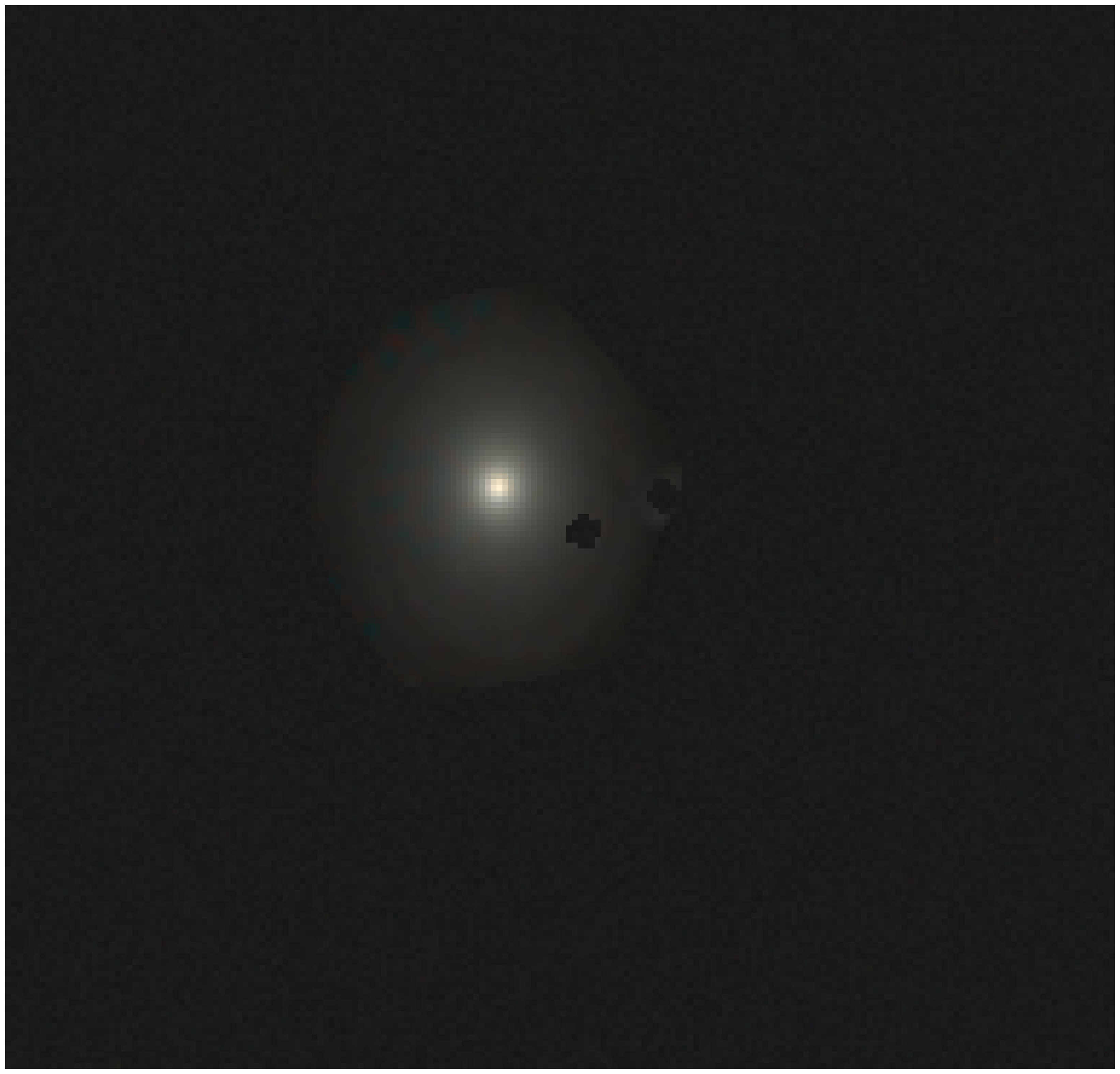}
    \end{tabular}
    \caption{\small{Comparison between original galaxies (left column) and their transformed counterparts (right column). The transformations applied preserve the physical characteristics of the galaxies while introducing variations in their appearance. For a detailed description of the transformation processes, refer to Sec. \ref{subsec:galaxy_augmentation}.}}
    \label{fig:transformation}
\end{figure}

 We provide a visual representation of these augmentations in Fig.~\ref{fig:transformation}. The original galaxies are displayed in the left column, offering a baseline for comparison. Adjacent to each, in the right column, are their corresponding transformed versions. These images illustrate the effects of the transformations, while maintaining the fundamental physical characteristics. The transformations, though diverse, are designed to be physics-preserving, ensuring that the scientific integrity of the galaxy representations remains intact.

\subsection{Dimensionality reduction of eCALIFA spectra}
\par The process of dimensionality reduction in the context of CALIFA data cubes plays a fundamental role in addressing several computational challenges. The primary objectives of this reduction strategy are twofold: firstly, to significantly speed up the training process of neural networks, and secondly, to ensure efficient data management within the limitations of available memory resources. The acceleration of the training process is a direct consequence of reducing the data size. By decreasing the number of dimensions, we effectively lessen the computational load on the system. This reduction not only makes it feasible to process larger batches of data simultaneously but also reduces the time complexity of each operation involved in the learning process. Such efficiency is particularly crucial when dealing with extensive datasets like eCALIFA, where the sheer volume of data can be a major bottleneck.
\par For the dimensionality reduction of eCALIFA spectra, we employ Principal Component Analysis \citep[PCA][]{doi:10.1080/14786440109462720}. This choice is grounded in both its efficacy and its comparative performance against other techniques, such as variational autoencoders (VAEs). According to \cite{2020AJ....160...45P}, it has been demonstrated that beyond 10 components, the reconstruction error of optical spectra remains consistent whether one employs PCA or more complex methods like VAEs. This finding underscores the efficiency of PCA in our context, where we reduce the original wavelength dimension of 1877 points to 29 principal components. This level of reduction is sufficient for reconstructing a wide range of spectra in CALIFA, encompassing star-forming regions, AGN activity, and spectra from older stellar populations. 
\par To ensure the quality and reliability of the PCA, we preselect spectra from all eCALIFA galaxies with a S/N ratio greater than 30 before conducting the PCA decomposition. This approach guarantees that the PCA components are derived from physical features rather than noise. All spectra have been normalised to the median within a window between 5600 and 5800~$\AA$ . This window was selected due to the absence of emission and absorption lines in the rest-frame spectra of galaxies, thereby not influencing the PCA decomposition. Furthermore, the edges of the spectrum are discarded, constraining the spectral range between 3739 Å and 6804~$\AA$. The final reduced 'spectral' dimension consists of these 29 PCA components, complemented by an additional component that encodes the luminosity of each spaxel within the normalisation window. It is important to note that all data cubes are aligned to the rest frame before this process is initiated. The PCA auto-spectra are shown in the appendix \ref{app:PCA} (Fig.~\ref{fig:PCA1}, and Fig.~\ref{fig:PCA2}). We also present the fourth PCA maps alongside the luminosity maps of the galaxies in Fig~\ref{fig:transformation} located in the appendix \ref{app:SNN_input} (see Figure~\ref{fig:PCA_maps}).

\par The choice of 29 PCA components effectively captures the general characteristics of the majority of spectra in eCALIFA data cubes. We achieve a reconstruction error below $3\%$ in most of the spectra in the PCA training set ($\sim98\%$). However, it should be acknowledged that this approach may encounter limitations in reconstructing certain peculiar spectra. For instance, the model might struggle to perfectly reconstruct specific features like the [OI] line or the Na I $\lambda \lambda$ $5890, 5896$ present in Low-ionisation nuclear emission-line  (LINER) galaxies. This limitation is primarily due to the lower relative abundance of such spectra in the sample. It is worth noting that varying the exact number of PCA components around 29 does not significantly impact the main findings of this study. This is because the encoded representation at this level does not distinguish between more peculiar features in the data.
\par In future work, should the sample size of galaxies be expanded, there might be an opportunity to increase the number of PCA components. Such an expansion would enable the model to capture secondary or tertiary order features more effectively, potentially enhancing the reconstruction of less common spectral features. This evolution of the model would depend on the availability and diversity of the data, which could provide a richer basis for distinguishing subtle spectral variations.

\par It is important to emphasize that the use of PCA in this study is not to create a representation space. This distinction arises because PCA, along with models such as autoencoders that aim to reconstruct input images from a latent representation, are not invariant to non-physical variations in the data. In contrast, the CL framework is specifically designed to incorporate a set of symmetries that prevent non-physical discrepancies from manifesting in the representation space. Direct application of PCA to galaxy images can result in a latent space where instrumental effects emerge as significant differences. This issue has been documented in the literature; for instance, \cite{2021ApJ...911L..33H} examined how galaxies shift within the representation space under various transformations, such as rotation or cropping of the images. When these images were projected directly from pixel space, such transformations influenced the galaxies' positions in the diagram. Similarly, \cite{2021ApJ...921..177S} demonstrated that instrumental effects present in the MaNGA survey, such as the number of fibers used - which affects the apparent size of a galaxy - or the presence of zeros in the maps, create a discernible gradient in the PCA projection. Although rotating images to a common axis can partially mitigate this effect \citep{2020MNRAS.498.4021U}, other instrumental variances, such as the apparent size of galaxies in the sky or the presence of foreground stars, cannot be reversed.
\subsection{Model architecture}
\par Our study employs a Siamese neural network (SNN) architecture, which is particularly effective for analysing and comparing pairs of related data samples \citep{2020arXiv200205709C}. The SNN consists of two identical subnetworks, each dedicated to processing a different set of data: one subnetwork processes the original galaxy maps, and the other handles the transformed versions of these galaxies. These subnetworks have the same configuration and share weights, ensuring that they process their respective inputs in a comparable manner. The purpose of using a siamese architecture in our study is to project both the original and transformed galaxy images into a lower-dimensional space (sometimes called the contrastive space). By doing so, the network captures essential features and characteristics of the galaxies, both in their original and transformed states. This is particularly important given the nature of our transformations, which are designed to be physics-preserving yet could introduce variations in appearance. 
\par In this low-dimensional space, a contrastive loss function is computed. We used the \textit{normalised cross entropy} loss \citep{10.5555/3157096.3157304} which is defined for the i-th galaxy in the set as:
\begin{equation}
L_{i} = -\log \frac{\exp(\langle z_i, z_i^+ \rangle / h)}{\sum_{k} \exp(\langle z_i, z_k \rangle / h)}
\end{equation}
where $\langle z_i,z_i^+\rangle$ is the cosine similarity between the original galaxy and its transformation pair in the contrastive space,  $\langle z_i, z_k \rangle$ is the cosine similarity for the rest of the galaxies in the batch, and $h$ is the temperature used to control the concentration of the constrastive space by weighing the cosine similarity. We have established the temperature at its standard value, $0.5$, upon noting that variations in this parameter do not significantly affect the results. 
\par The normalised cross entropy loss evaluates the similarity or dissimilarity between the pairs of projected data points (original and transformed galaxies). The aim is to minimise this loss, ensuring that the network learns to effectively distinguish between the intrinsic characteristics of the galaxies despite the transformations applied. 
\par In this work, we made used of  SNN architecture specifically customised for the analysis of eCALIFA data cubes, now dimensionally reduced to $192x184x30$, where 192x184\footnote{While the original spatial dimensions of eCALIFA data cubes are $159x151$ pixels, we have expanded them to $192x184$. This enlargement ensures that the transformation, which translates the galaxies to centre them within the cube, does not result in any part of the galaxy being lost as it extends beyond the original dimensions.} represents the spatial dimensions and 30 is the number of spectral components, i.e. the $29$ PCA features maps plus the luminosity at the normalisation window. The model is designed to efficiently handle these data cubes while extracting meaningful features for further analysis.
\par Our model's architecture is inspired by the one in \citet{2021ApJ...921..177S} but features fewer layers, as detailed in Table~\ref{table:encoder-architecture}. This design choice is influenced by the size of our training set, as model complexity needs to be proportional to the training size. The current architecture requires fitting approximately 2 million parameters. Adding an extra convolutional layer with 1024 filters and an additional dense layer of 1024 neurons would quadruple the model's parameters, an unnecessary increase given our dataset's scope. Our representation space is defined by the first dense layer following the flattening step (DENSE\_1, see Table~\ref{table:encoder-architecture}). This choice is based on findings that the initial layers of the network yield better representations than the later layers \citep{2020arXiv200205709C}.
\newpage
\par The training of the model was conducted with an early stopping mechanism set to a patience of five, closely monitoring the validation loss. The maximum batch size used was equal to the count of primary galaxies in the training set, which is 772. However, the effective number of galaxy examples used for training significantly increased, as we expanded our training set by a factor of 50 through the described augmentation techniques. This augmentation started from 10 realisations and was progressively increased until no further improvements were observed in the validation loss score. This approach ensured that each batch contained unique galaxy representations, avoiding the comparison of different realisations of the same galaxy as if they were distinct entities.
\par We utilize an NVIDIA A100-PCIE-80GB GPU for training the SNN. The model approximately requires 10 minutes for training, completing 50 epochs.
\section{Results}\label{sec:results}
\subsection{2D projection maps}\label{subsec:2Dprojection}
\par In the representation space of our analysis, galaxies are described by vector of 512 dimensions. To facilitate practical visualisation, this high-dimensional data is projected onto a 2-dimensional plane. This projection is achieved using the Uniform Manifold Approximation and Projection (UMAP) algorithm \citep[UMAP,][]{2018arXiv180203426M}. The UMAP is calculated solely using the training set, and then applied to project both the training and validation sets onto this 2D space. In the UMAP, the training set and validation set are distinguished with circles and stars, respectively. This approach is important to ensure that our model is not overfitting. This will be explored in more detail in Sect.~\ref{subsec:dist}.Galaxies are colour-coded based on their global or integrated properties, drawing from the table catalogue provided by \cite{2024RMxAA..60...41S}.
\begin{figure*}[!ht]
    \centering
    \begin{tabular}{cc}
        \includegraphics[width=9.2cm,height=8.2cm]{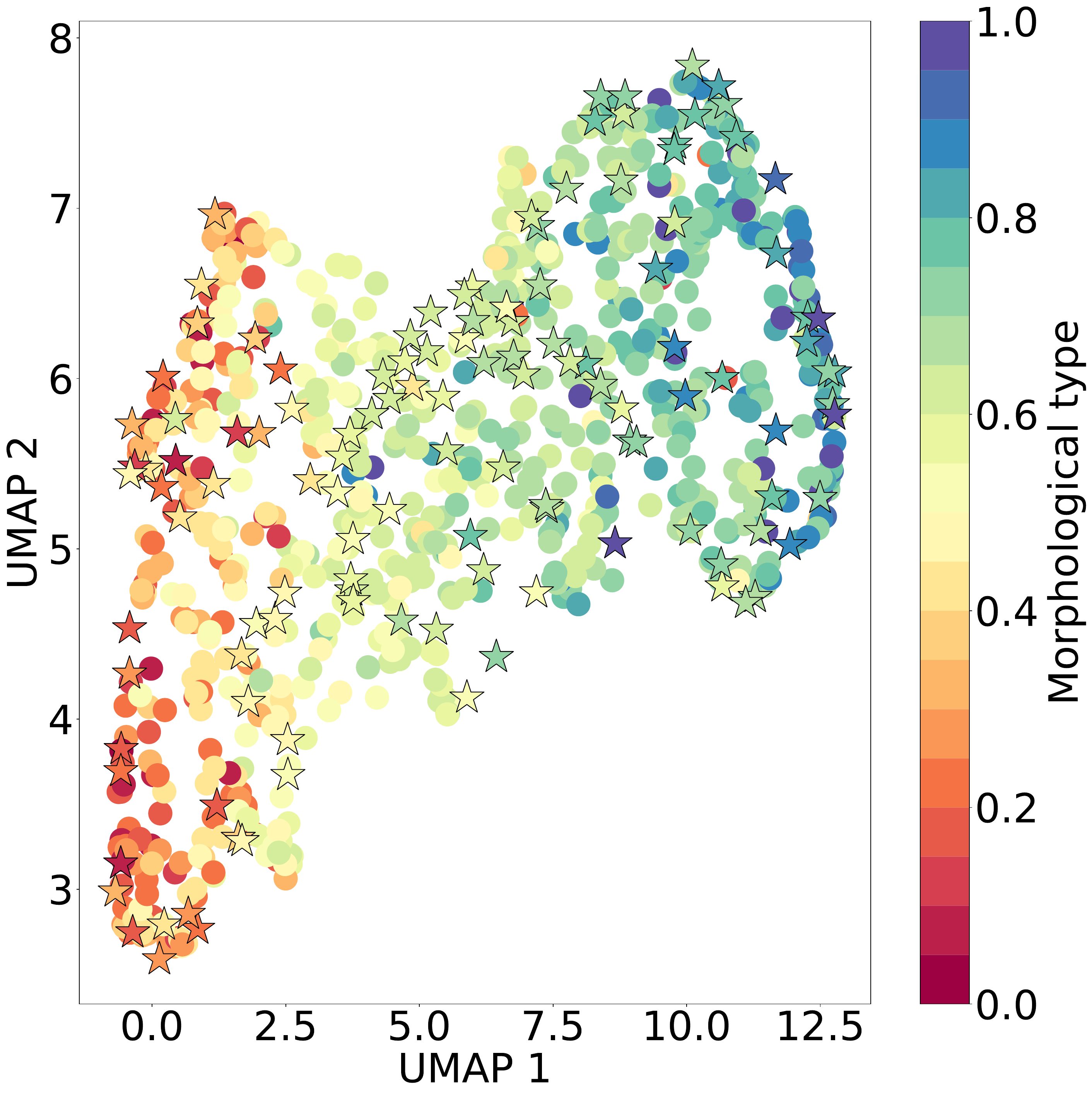} & 
        \includegraphics[width=9.2cm,height=8.2cm]{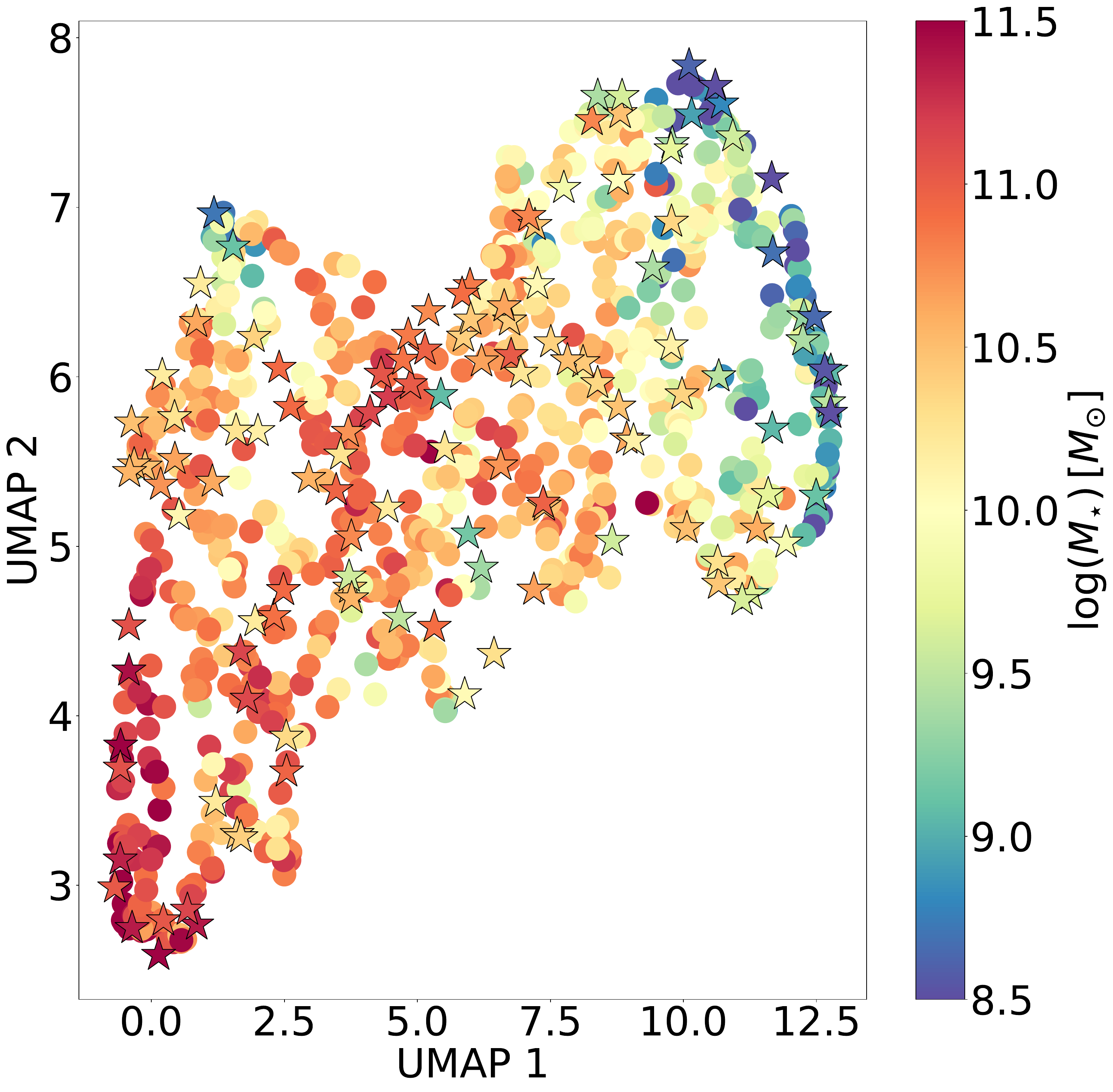} \\   
        \includegraphics[width=9.2cm,height=8.2cm]{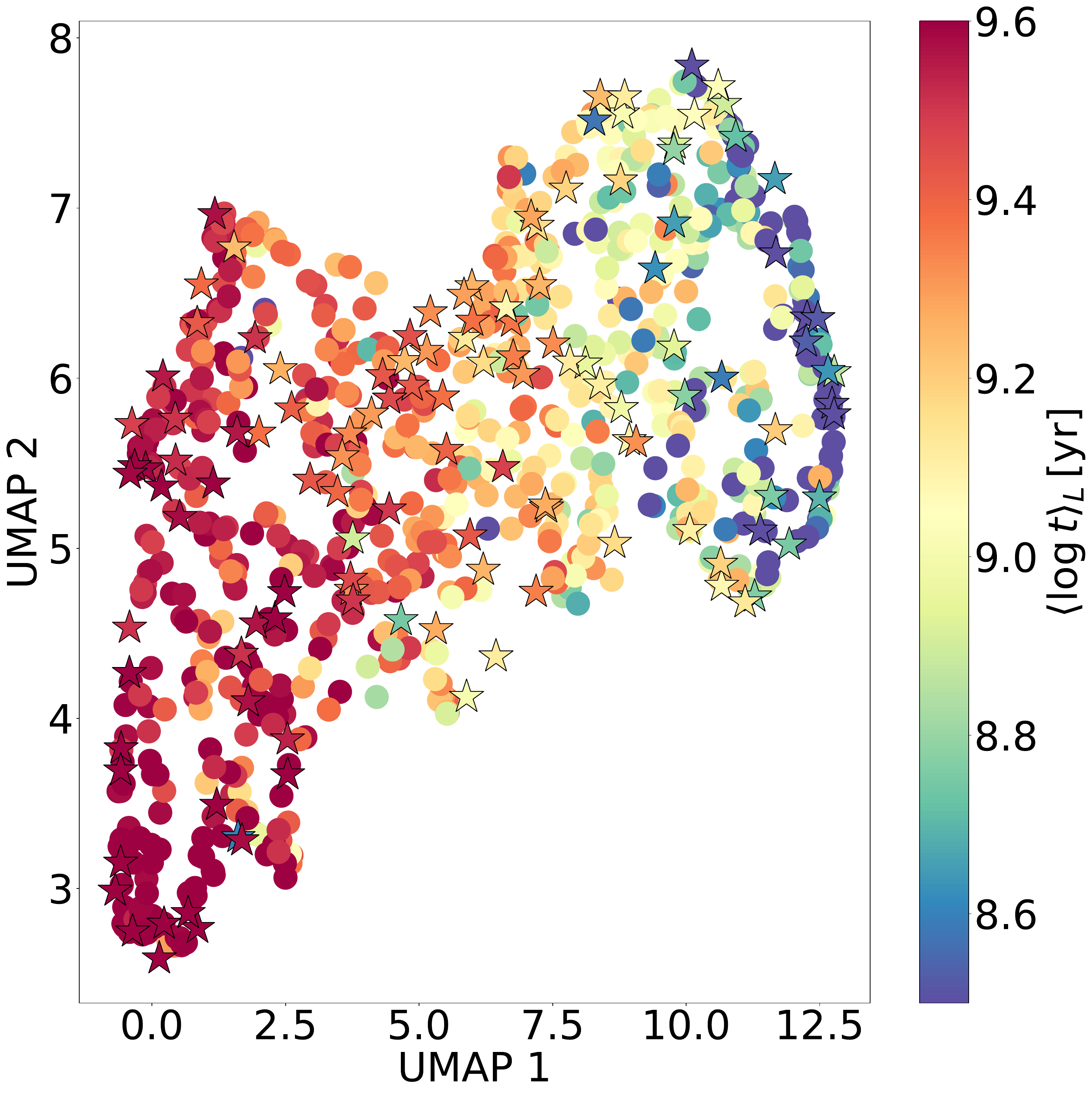} &
        \includegraphics[width=9.2cm,height=8.2cm]{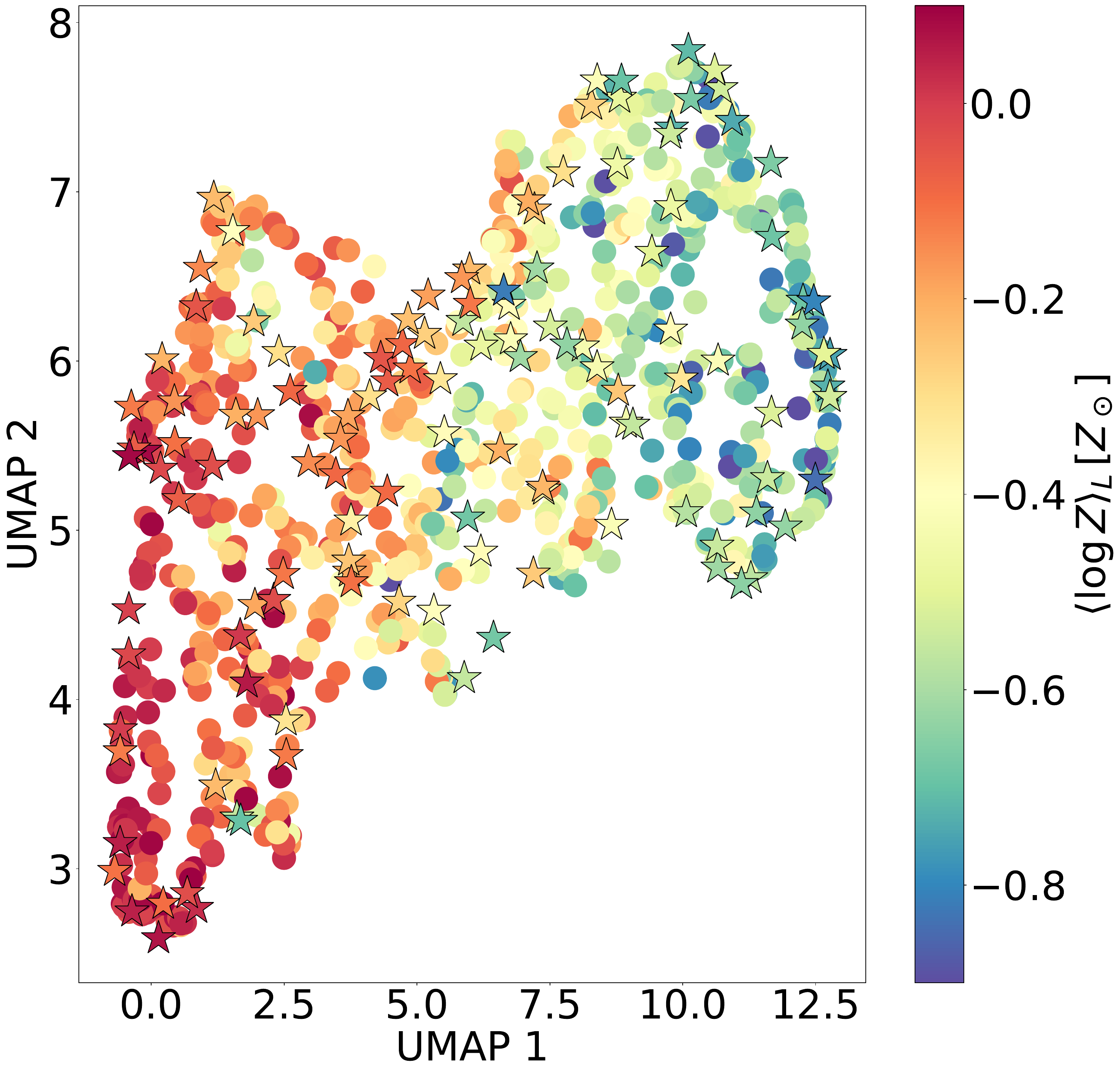} 
    \end{tabular}
    \caption{\small{UMAP projection of embedded space, colour-coded by galaxy attributes. Top-left: Morphological types, with early types in red and late types in blue (refer to the text for the colour-coding scheme of different morphologies). Top-right: Stellar mass. Bottom-left: Mean luminosity-weighted age of the stellar population. Bottom-right: Metallicity luminosity-weighted of the stellar population. Circles represent the training set, while stars denote the validation sample}}
    \label{fig:SSP_morph}
\end{figure*}
\par In Fig.~\ref{fig:SSP_morph}, the UMAP projection is colour-coded by galaxy morphology, stellar mass, and characteristics of the stellar population, such as luminosity-weighted age, and metallicity. The morphological spectrum includes ellipticals (E0 to E7), spirals (S0, Sa, Sab, Sb, Sbc, Sc, Scd,Sd, Sdm), irregulars (Irr), and blue compact dwarfs (BCD) as distinct entities, each identified with a specific colour. A notable gradient is observed: early-type galaxies display older, metal-rich stellar populations (SP), whereas late-type galaxies are younger and less metal-rich. Stellar mass, a crucial parameter, reveals a gradient segregating low-mass ellipticals from their larger counterparts. This trend is mirrored in spiral galaxies. 
\begin{figure*}[!ht]
    \centering
    \begin{tabular}{ccc}
        \includegraphics[width=9.2cm,height=8.2cm]{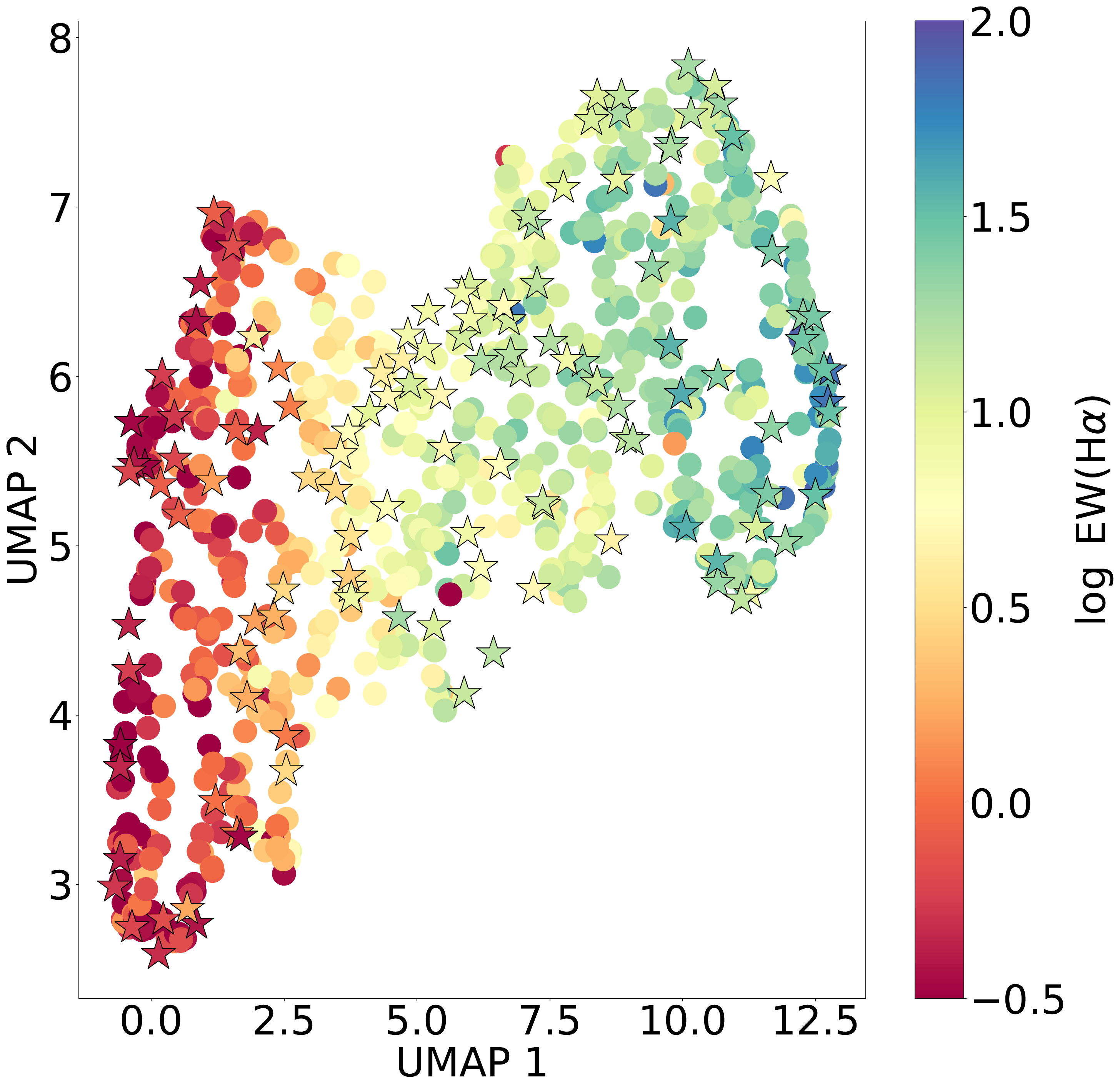} & 
        \includegraphics[width=9.2cm,height=8.2cm]{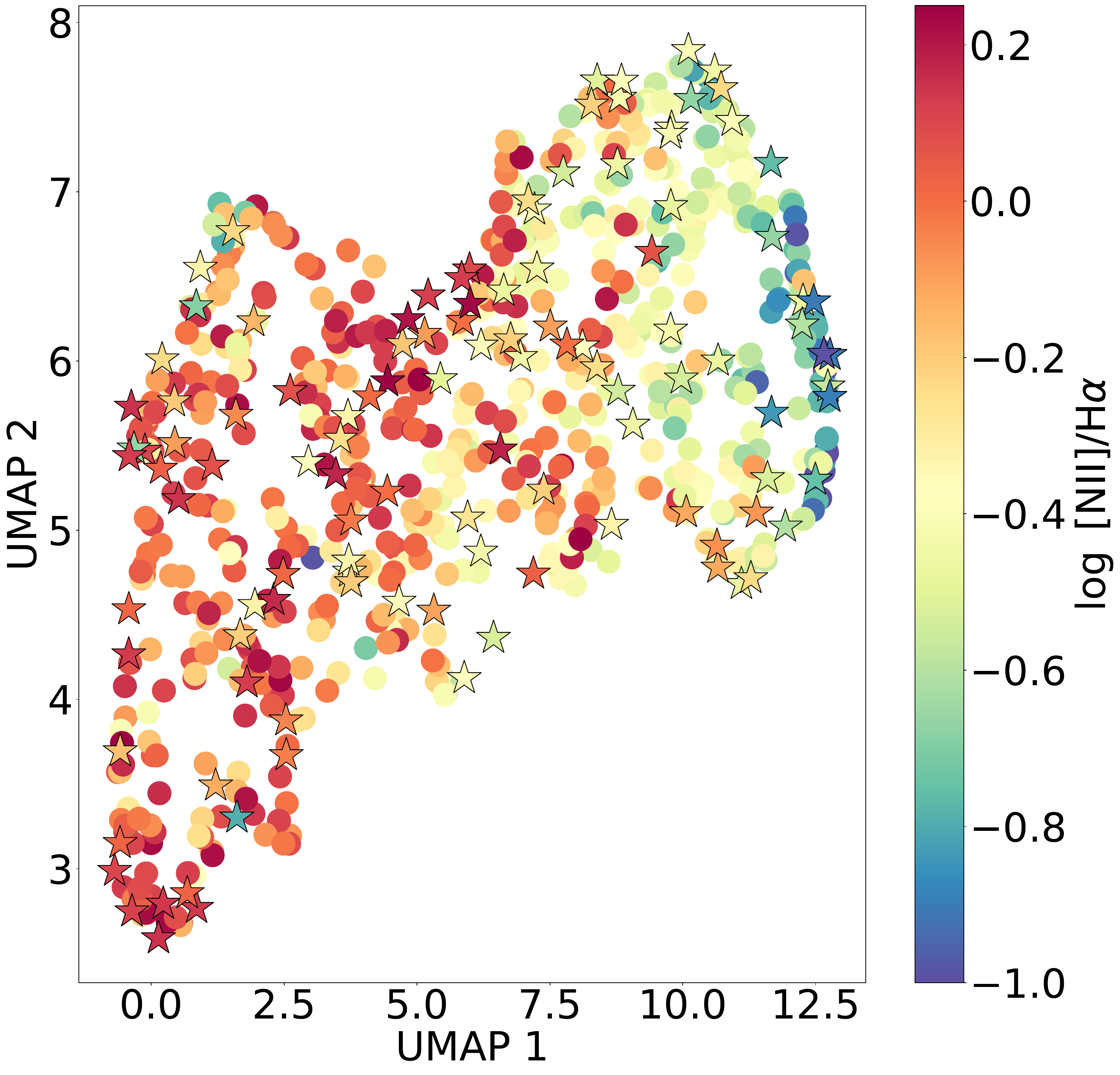} \\   
        \includegraphics[width=9.2cm,height=8.2cm]{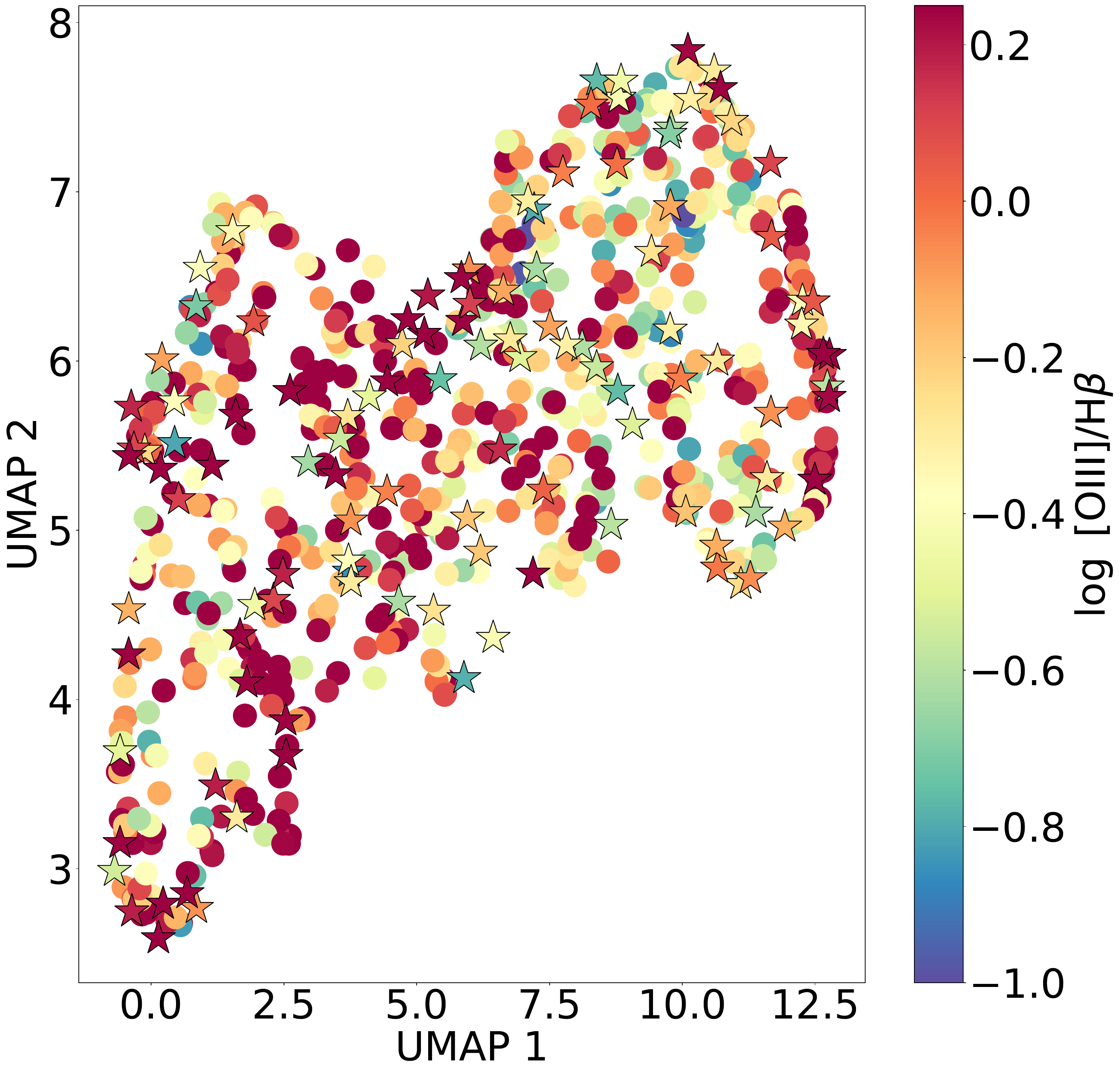} &
        \multicolumn{2}{l}{\includegraphics[width=7.2cm,height=8.2cm]{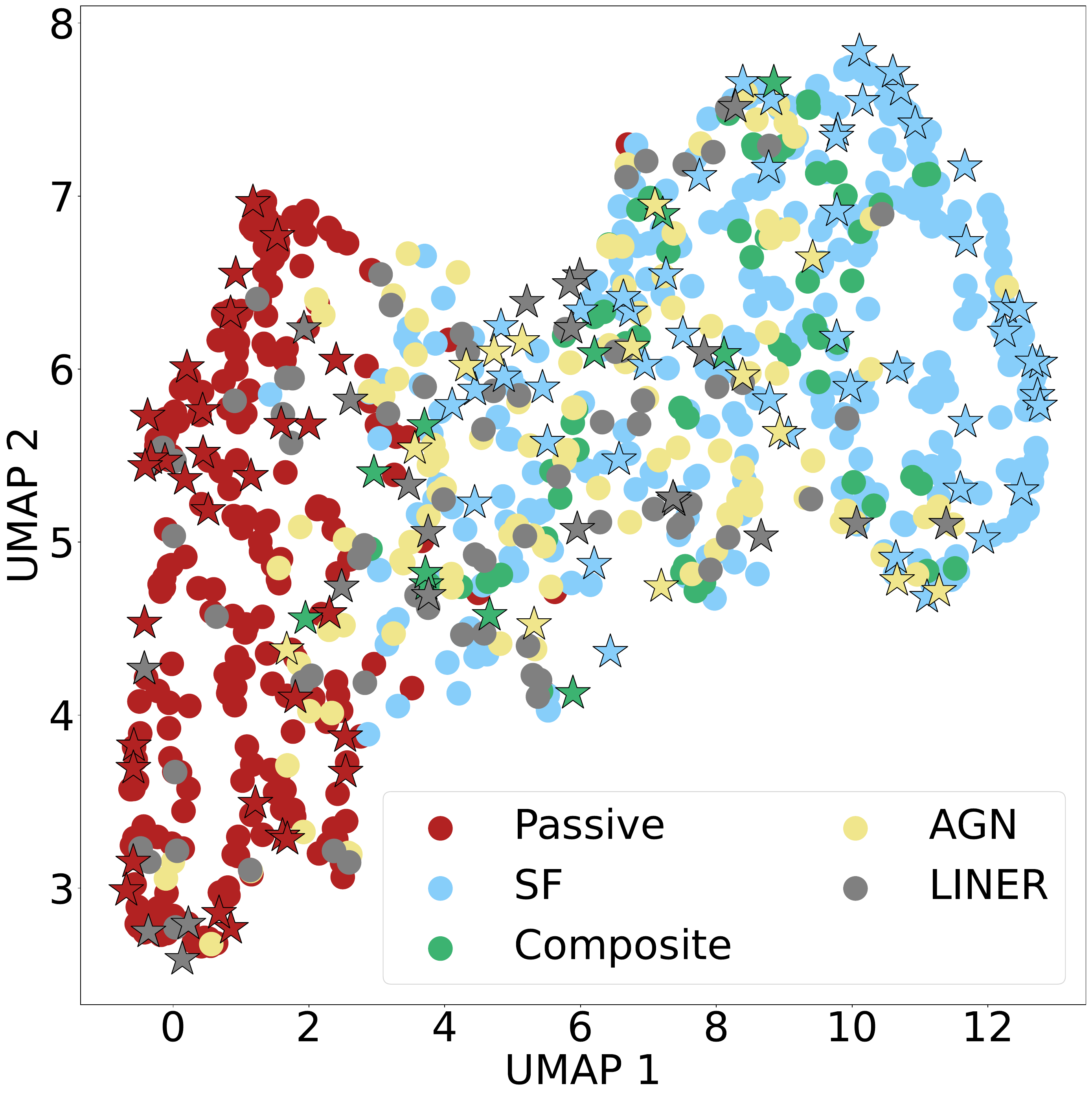}} 
    \end{tabular}
    \caption{UMAP projection of embedded space, colour-coded by galaxy emission line properties and ionisation mechanisms. Top-right: Logarithm of the integrated EW of $H\alpha$. Top-left: Logarithm of the central [NII]/H$\alpha$ ratio. Bottom-left: Logarithm of the central [OIII]/H$\beta$ ratio. Bottom-right: Classification of galaxies according to their primary ionisation mechanism, categorised into AGN, Star-forming, LINERs, Passive, and Composite galaxies. Circles represent the training set, while stars denote the validation sample. }
    \label{fig:EL}
\end{figure*}
\par The UMAP projection, colour-coded according to emission line-derived properties as shown in Fig.~\ref{fig:EL}, displays a distinct gradient in the integrated equivalent width (EW) of H$\alpha$. This gradient is reflective of the specific star formation rate (sSFR) in galaxies, thereby enabling the SNN to distinguish galaxies based on this characteristic. Shifting attention to the galactic nucleus of the galaxy, a gradient is also observed in the [NII]/H$\alpha$ ratio within galaxy centres, related to the gas phase metallicity of the nucleus. However, no gradient is evident for the [OIII]/H$\beta$ ratio, suggesting the ionisation state of the nucleus might not be the primary differentiator in this 2D projection. The bottom right of Fig.~\ref{fig:EL} categories galaxies in five types: active galaxy nucleus (AGN), star forming (SF), LINER, composite, and passive galaxies. In order to achieve this classification, we follow the WHAN diagram \citep{2011MNRAS.413.1687C} with central emission line values. Composite galaxies are defined as the ones between the \cite{2003MNRAS.346.1055K} and the \cite{2001ApJ...556..121K} curves in the WHAN diagram. In order to consider a galaxy as passive, we require an absence of recent star formation throughout the galaxy, not just centrally. Galaxies with any H$\alpha$-detected formation are considered star-forming (SF), even if they have quenched bulges. Figure~\ref{fig:EL} reveals a distinction between passive and SF galaxies, but AGN, SF, LINER, and composite galaxies overlap in this projection. Nevertheless, it is clear that AGN, composites, and LINERs are more common in S0, Sa, or Sb types, while Irr, Sc, Sd, and BCD galaxies are predominantly SF.

\begin{figure*}[!ht]
    \centering
    \begin{tabular}{cc}
        \includegraphics[width=9.2cm,height=8.2cm]{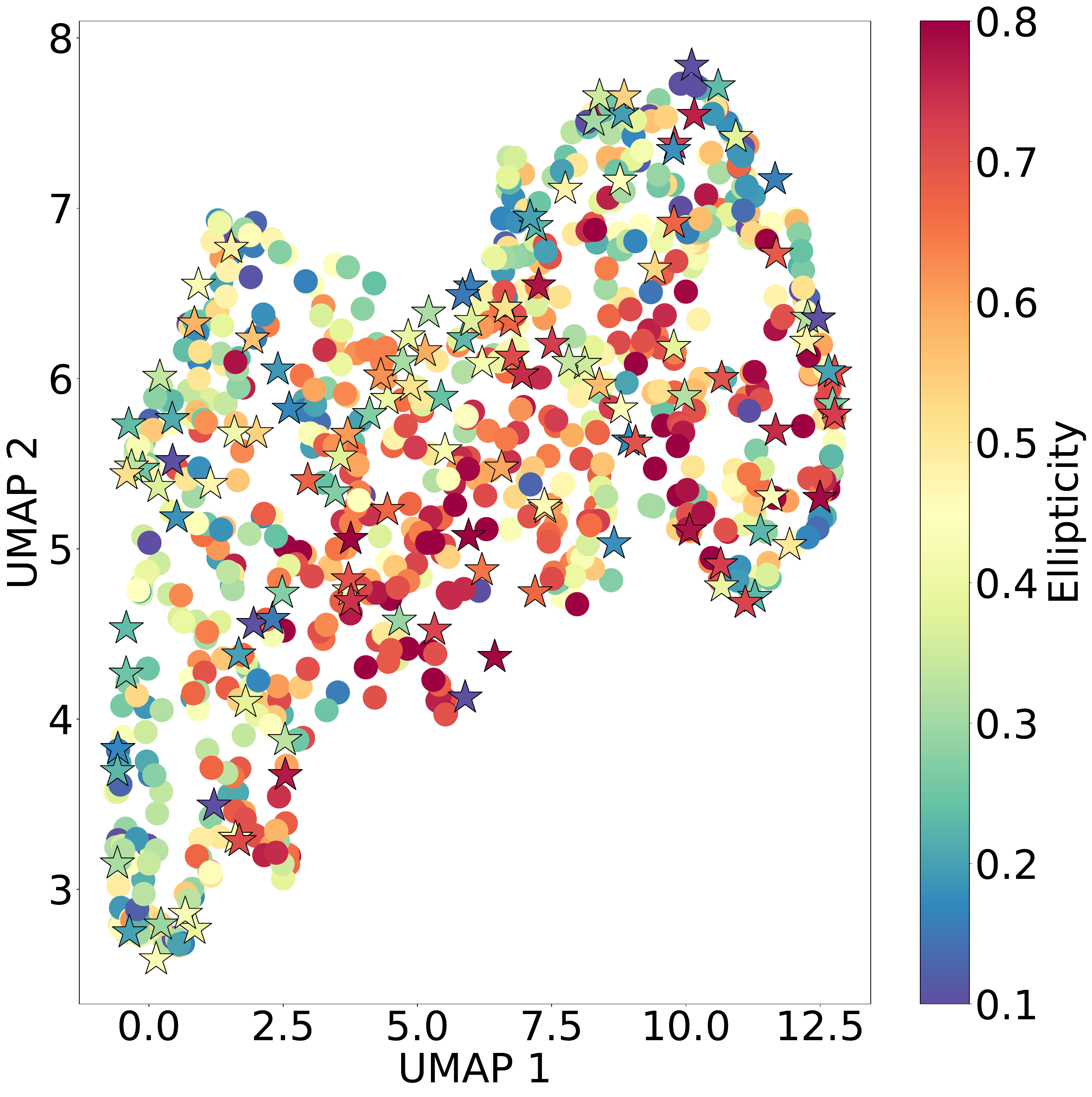} & 
        \includegraphics[width=9.2cm,height=8.2cm]{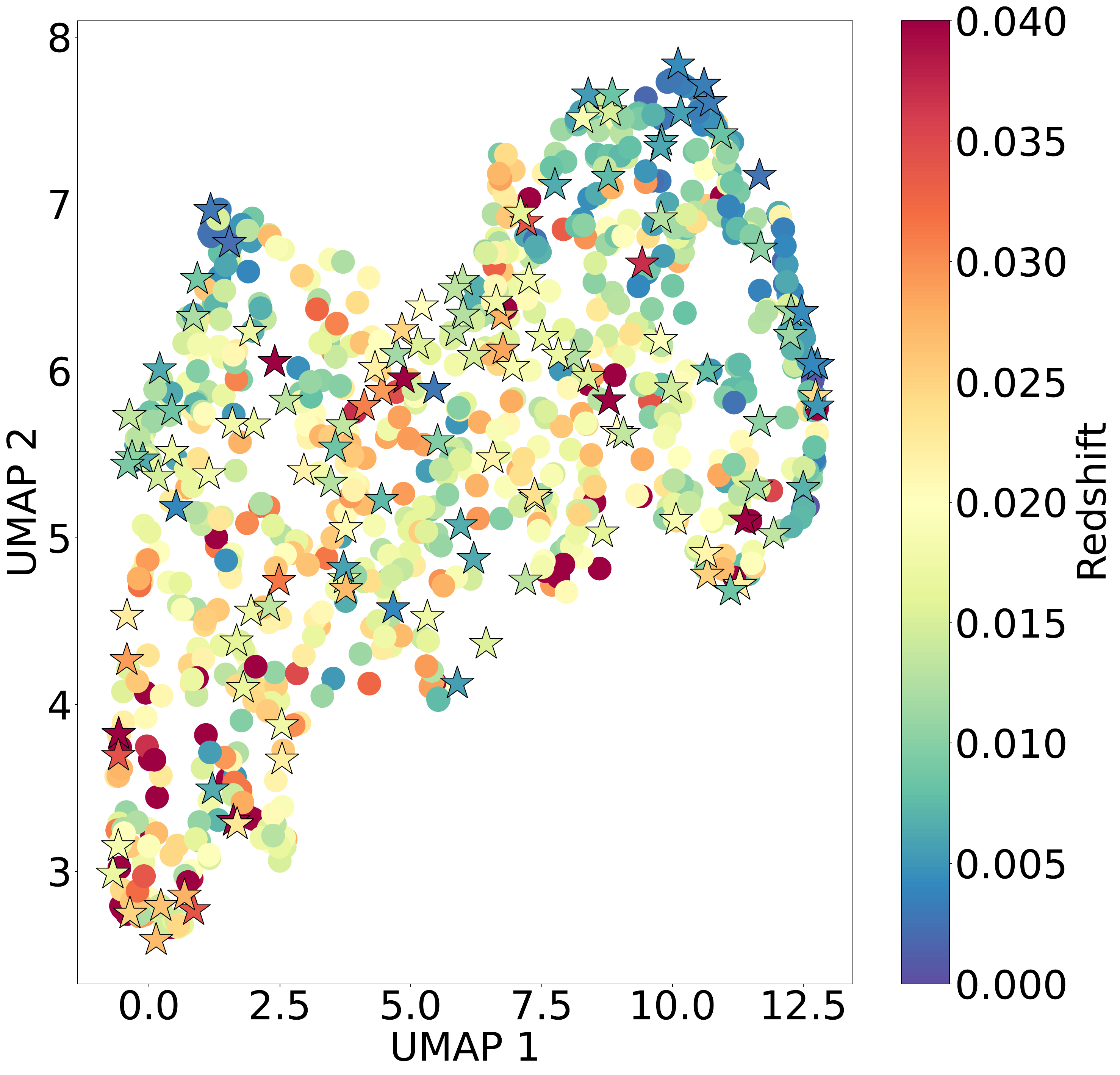} 
    \end{tabular}
    \caption{\small{UMAP projection of embedded space, featuring two primary galaxy characteristics. Left: Redshift distribution of the sample. Right: Ellipticity of galaxies, reflecting the shape and orientation of the galaxies within the sample. Circles represent the training set, while stars denote the validation sample.}}
    \label{fig:z_el}
\end{figure*}

\par In Fig.~\ref{fig:z_el}, UMAP projections are colour-coded for galaxy ellipticity and redshift. Intriguingly, despite not accounting for galactic inclination, an ellipticity gradient is unexpectedly absent. This omission might indicate that other factors, such as stellar mass or morphology, are more influential in differentiating galaxies in this representation. For redshift, while no strong gradient is evident, there is some clustering, likely due to the selection process of the observed galaxies. This includes low-mass ellipticals, Irr galaxies, and BCD, primarily observed at lower redshifts, a reflection of observational constraints and scheduling.
\par
\begin{figure}[!ht]
    \centering
    \includegraphics[width=9.2cm,height=8.2cm]{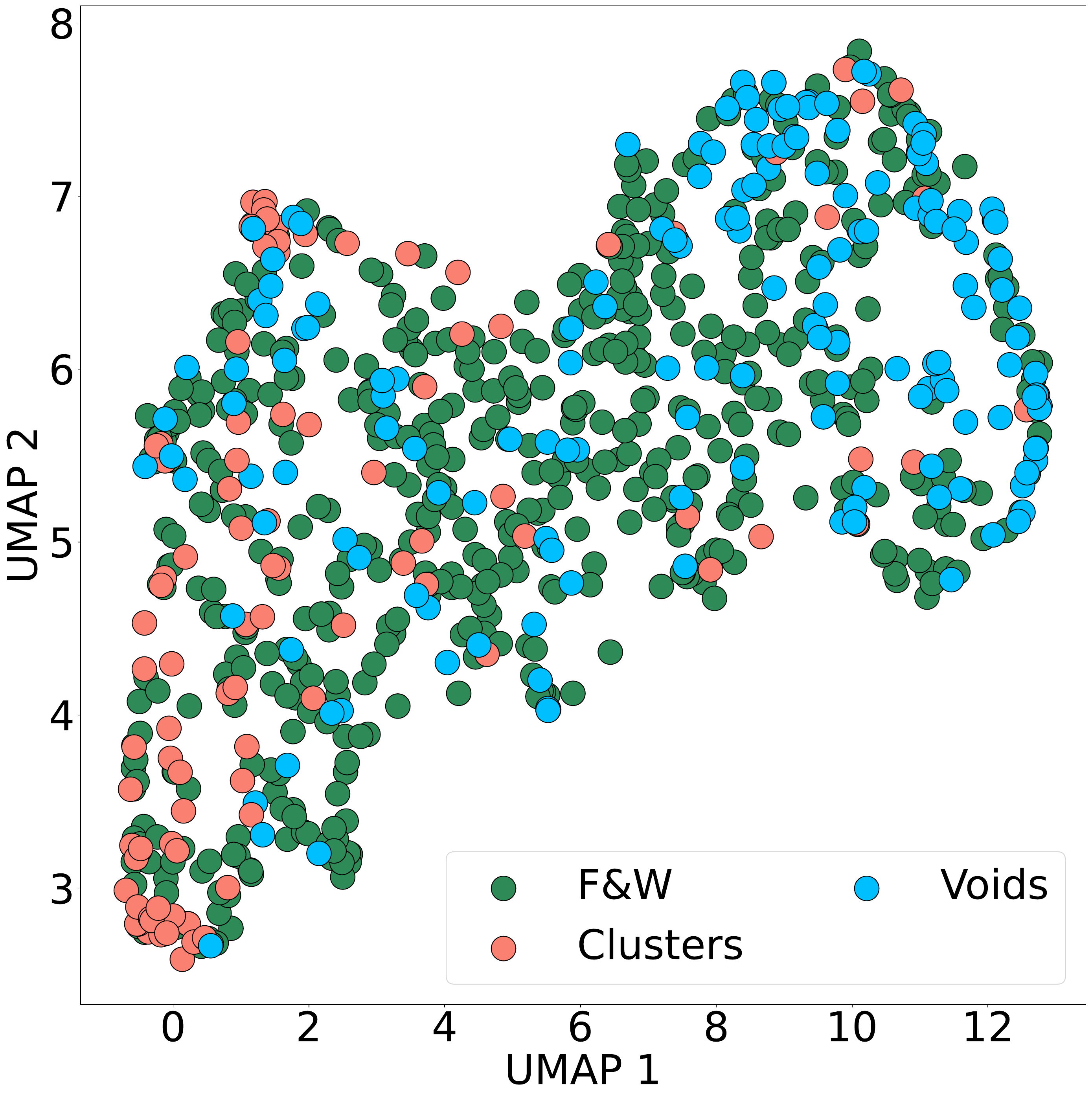} 
    \caption{\small{Color-coded UMAP projections of galaxies into different environments, including clusters, filaments and walls (F$\&$W), and voids.}}
    \label{fig:enviroment}
\end{figure}

 In Fig.~\ref{fig:enviroment} we colour-coded the UMAP projections to delineate various galactic environments. This visualisation facilitates the differentiation of galaxies within clusters, those situated in the filaments and walls (F$\&$W), and those in very low density environments, namely voids. The categorisation of these environments was achieved through the integration of data from multiple catalogues. To identify galaxies in clusters, we amalgamated catalogues from \cite{2017A&A...602A.100T}, which utilised SDSS galaxies, with those from \cite{2017MNRAS.470.2982L}. The latter combined data from four extensive redshift surveys: the Two Micron All-Sky Redshift Survey \citep[2MRS,][]{2012ApJS..199...26H}, the Six-degree Field Galaxy Survey\citep[6dFGS,][]{2004MNRAS.355..747J}, the SDSS, and the Two-degree Field Galaxy Redshift Survey \citep[2dFGRS,][]{2001MNRAS.328.1039C}, creating a comprehensive catalogue of galaxy groups in the low-redshift Universe. A galaxy is deemed to be in a cluster if it appears in a group with a minimum of 30 members in any of these catalogues. For galaxies in very low density environments, we utilised the catalogue from \cite{2012MNRAS.421..926P}. A specific criterion was established to classify a galaxy as a void galaxy, requiring its location to be within 100$\%$ of the effective radius of its respective void. Galaxies found in cluster catalogues with fewer than 30 members or outside the void radius are classified as F$\&$W galaxies. eCALIFA encompasses 95 galaxies in clusters, 162 in voids, 559 in F$\&$W, and 76 that were not categorised in any of the mentioned catalogues, thus remaining unclassified and excluded from Fig.~\ref{fig:enviroment}. 
\par It is important to note that this classification represents a preliminary attempt to categorise galaxies environment in eCALIFA. Our aim in this paper is not to provide a thorough analysis of environmental influences on eCALIFA galaxies, but rather to offer initial insights into how these factors impact galaxy distribution in our representation space. It should be acknowledged that galaxies situated in walls and filaments in our study could still belong to medium or small groups, which have distinct evolutionary impacts compared to isolated galaxies \citep[see e.g.][]{2022A&A...666A..84G}. Moreover, the 100$\%$ effective radius criterion might include galaxies that are actually on the fringes of voids, adjacent to filament and wall structures where the density approaches the universe's mean density. Bearing this in mind, observational data from the figure indicate that galaxies in clusters predominantly comprise early-type galaxies, aligning with previous expectations, while void galaxies more frequently exhibit the latest morphological types. We explore this pattern further in Sect.~\ref{subsec:dist}.
\subsection{Clustering }\label{subsec:clustering}
\par In this section, we perform clustering in the 512-dimensional where eCALIFA galaxies have been projected in order to explore patterns of similarity among galaxies and to investigate the correlation of these clusters with various physical properties. We employ the k-Means clustering algorithm with the cosine distance, exploring cluster solutions ranging from two to four. Our approach does not seek to determine the optimal number of clusters, a process typically involving metrics such as the elbow method, silhouette scores, or the Davies-Bouldin index, each with its own underlying assumptions that can lead to varying conclusions about the 'optimal' cluster count. Instead, we attempt to offer an insight into how these clusters align with certain physical characteristics identified as significant in the preceding sections of our analysis.
\par
\begin{figure*}[]
    \centering
    \begin{tabular}{ccc}
        \includegraphics[width=6.2cm,height=5.5cm]{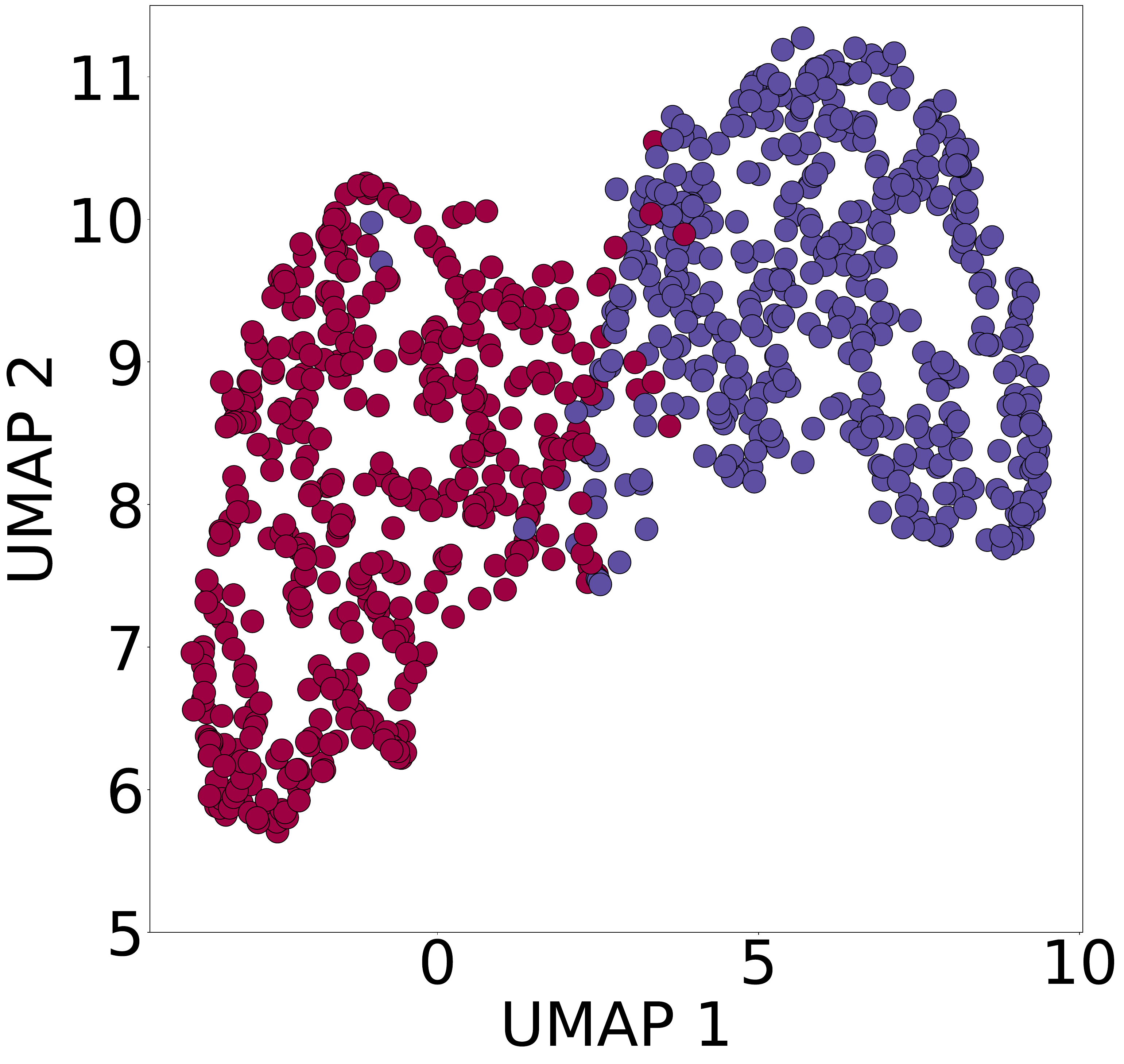} & 
    \includegraphics[width=5cm,height=5.5cm]{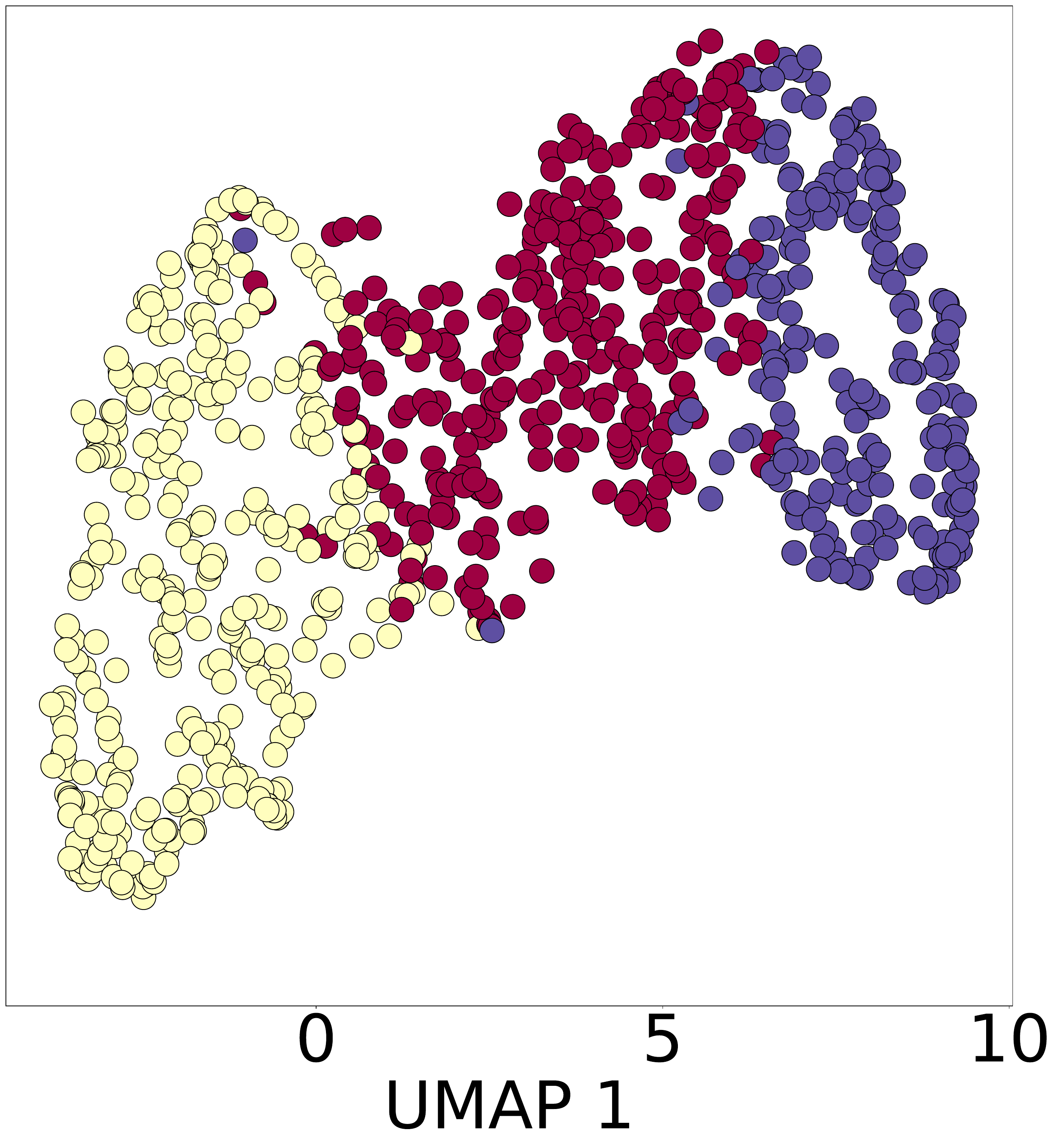} &
        \includegraphics[width=5cm,height=5.5cm]{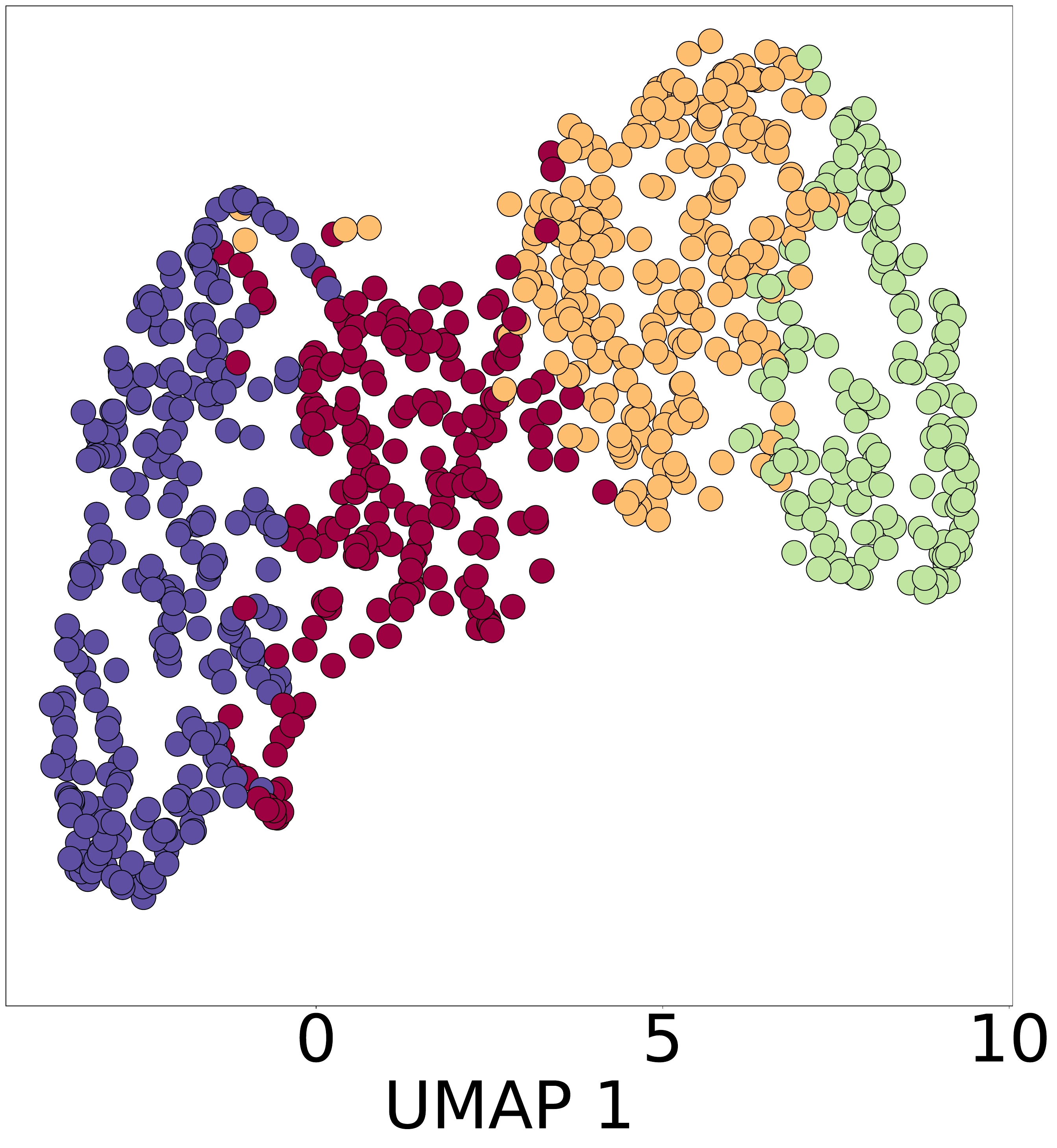}   

        \\
        \includegraphics[width=6.2cm,height=5.5cm]{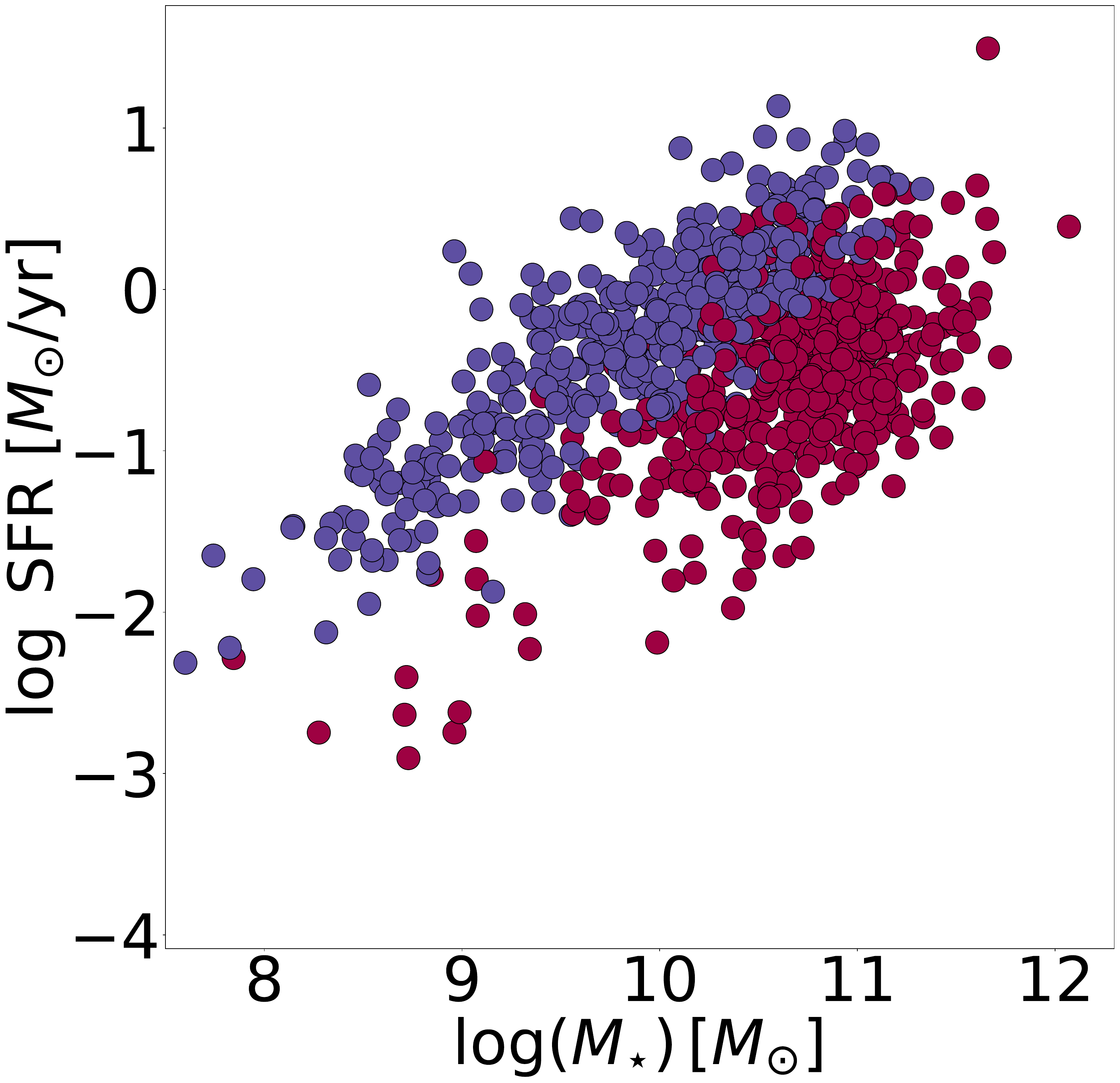} & 
        \includegraphics[width=5cm,height=5.5cm]{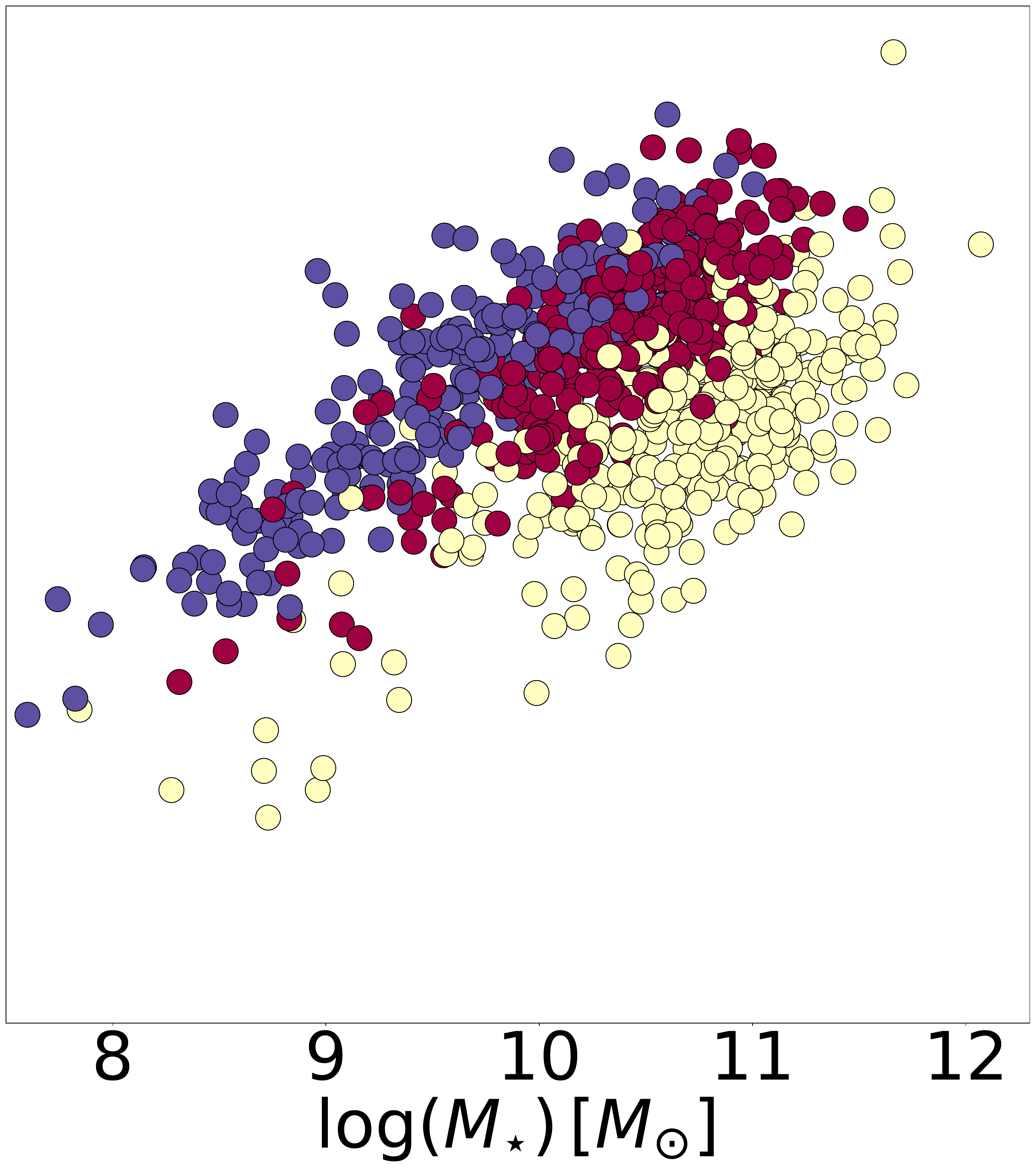} &
        \includegraphics[width=5cm,height=5.5cm]{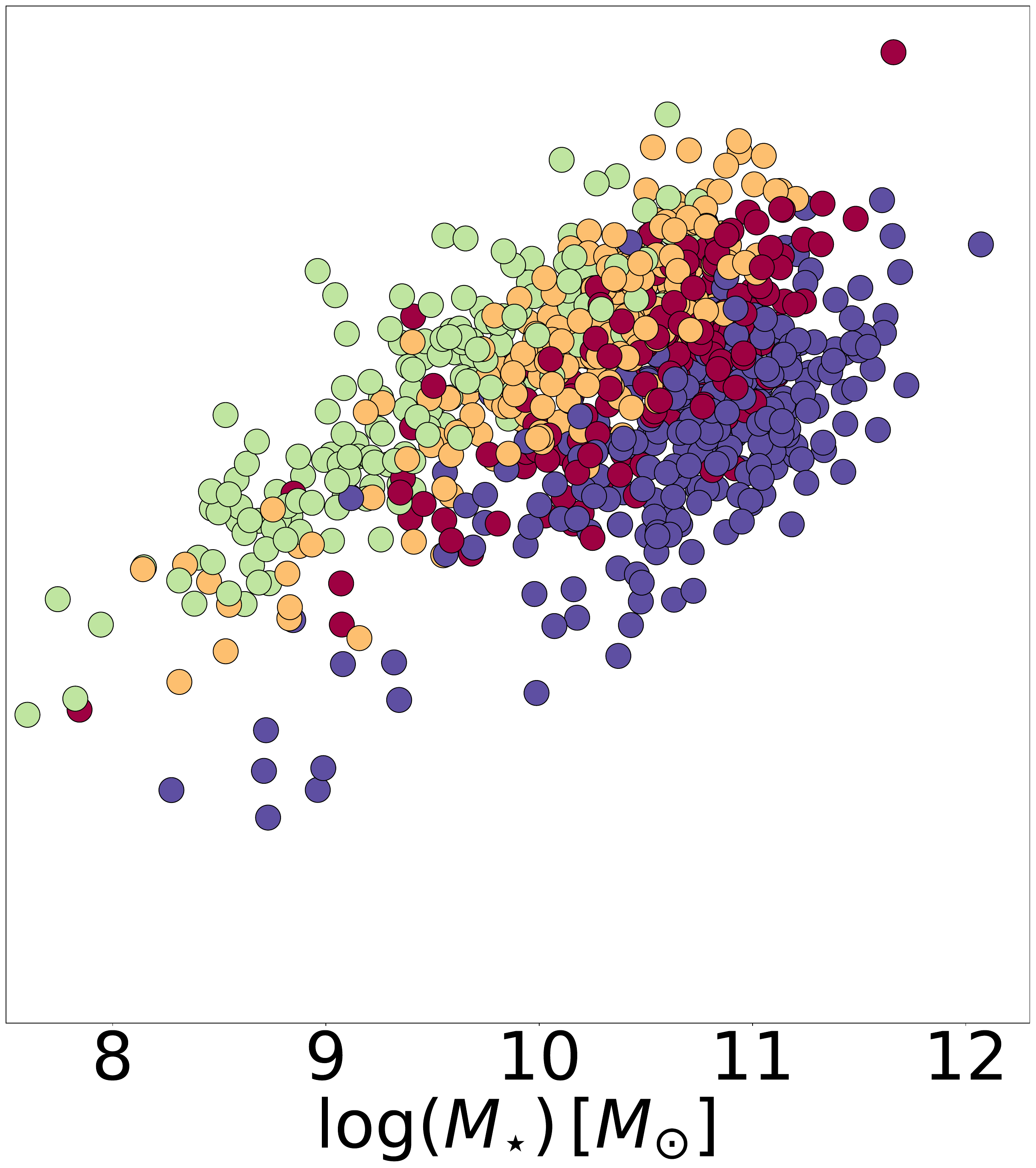}  

        \\
        \includegraphics[width=6.2cm,height=5.5cm]{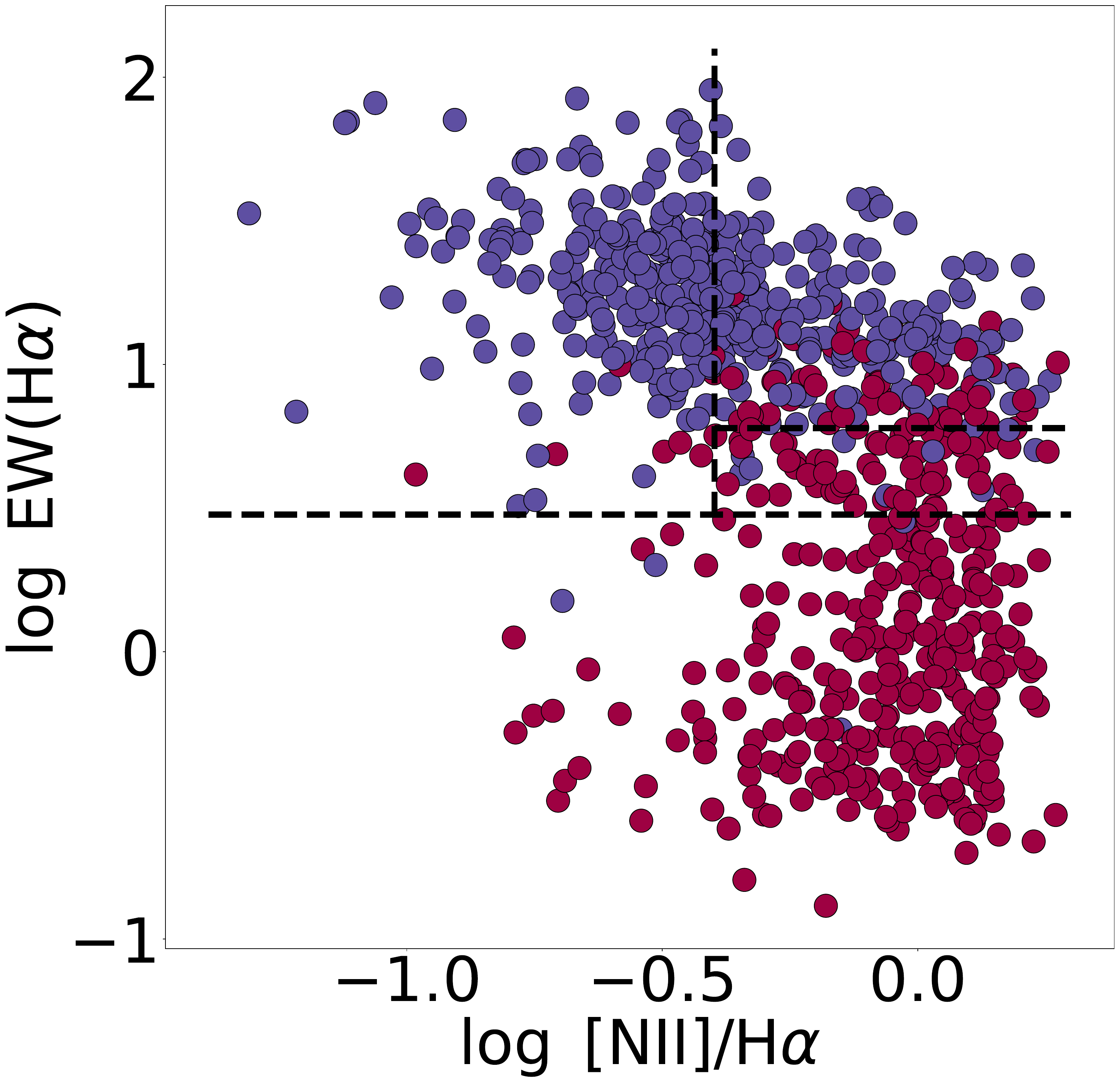} & 
        \includegraphics[width=5cm,height=5.5cm]{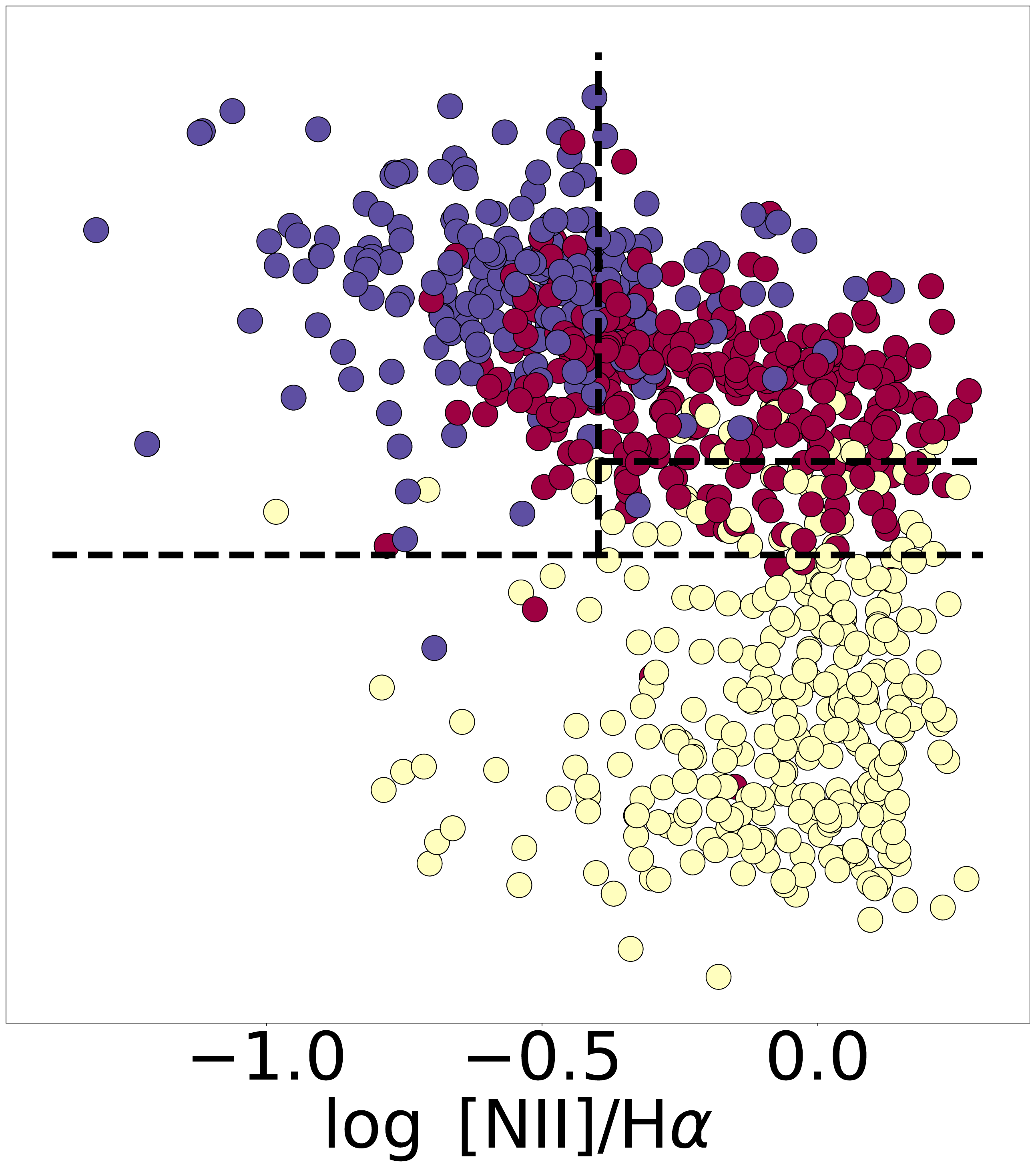} &
        \includegraphics[width=5cm,height=5.5cm]{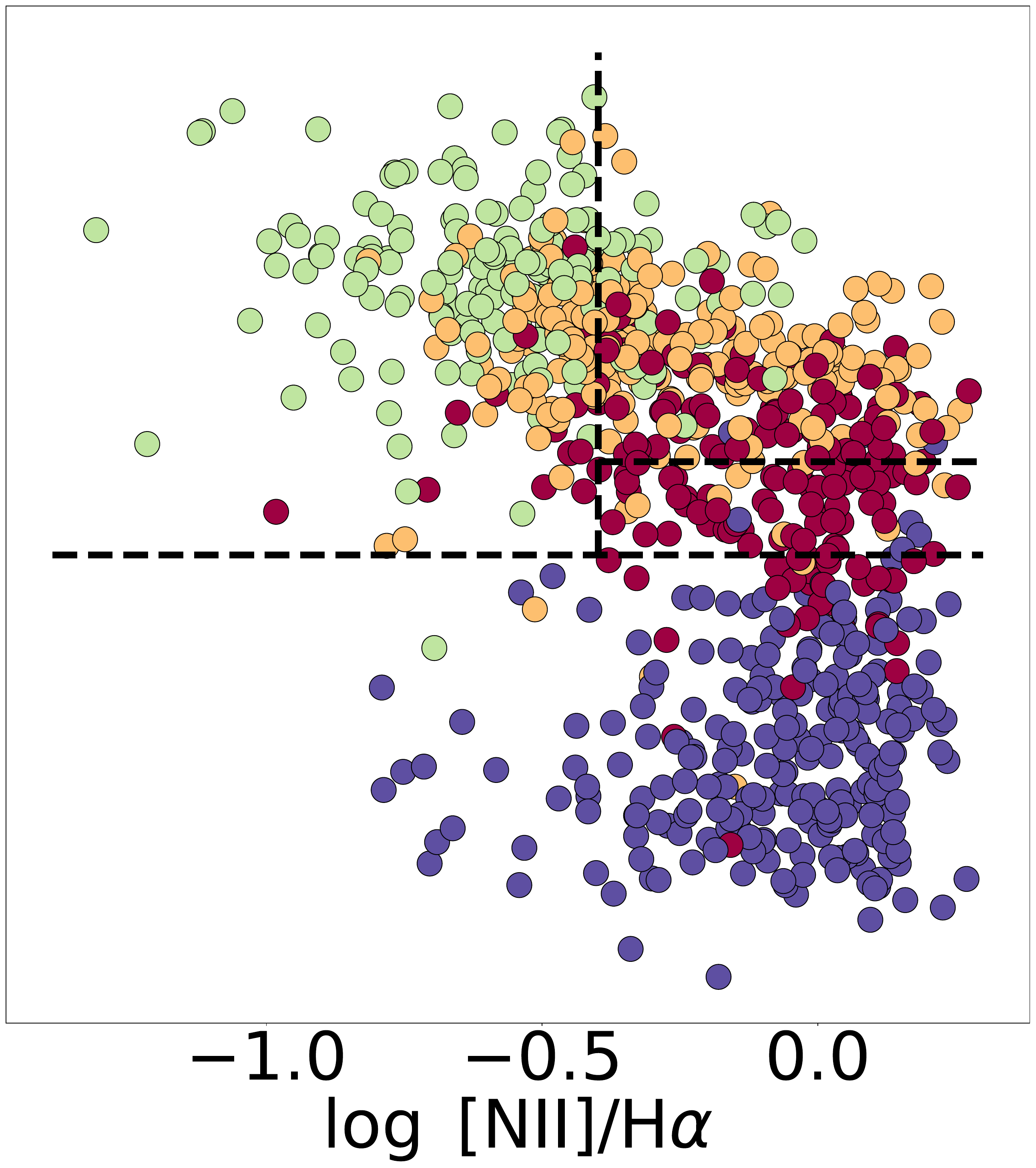}          
        \\
        \includegraphics[width=6.2cm,height=5.1cm]{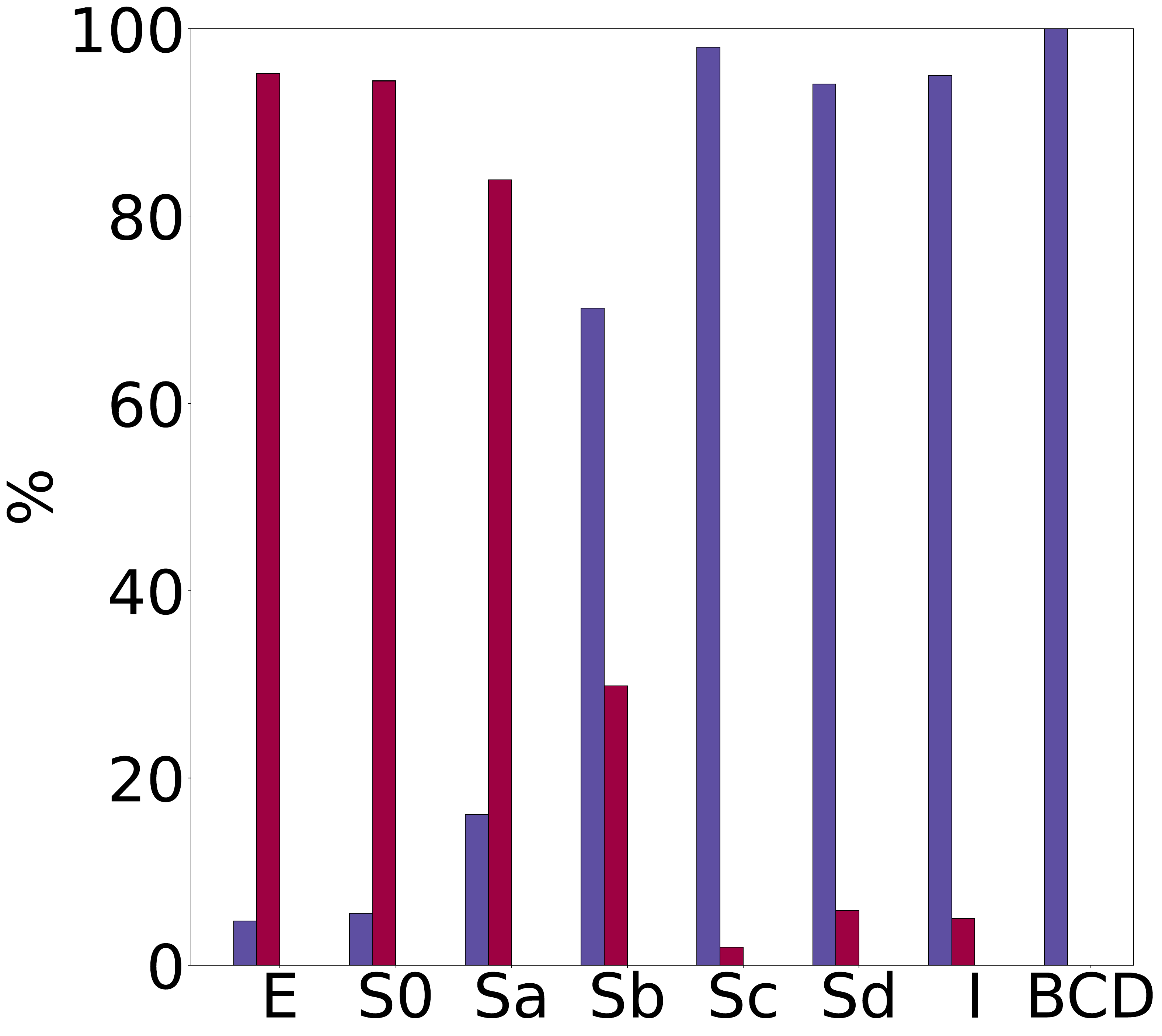} & 
        \includegraphics[width=5cm,height=5cm]{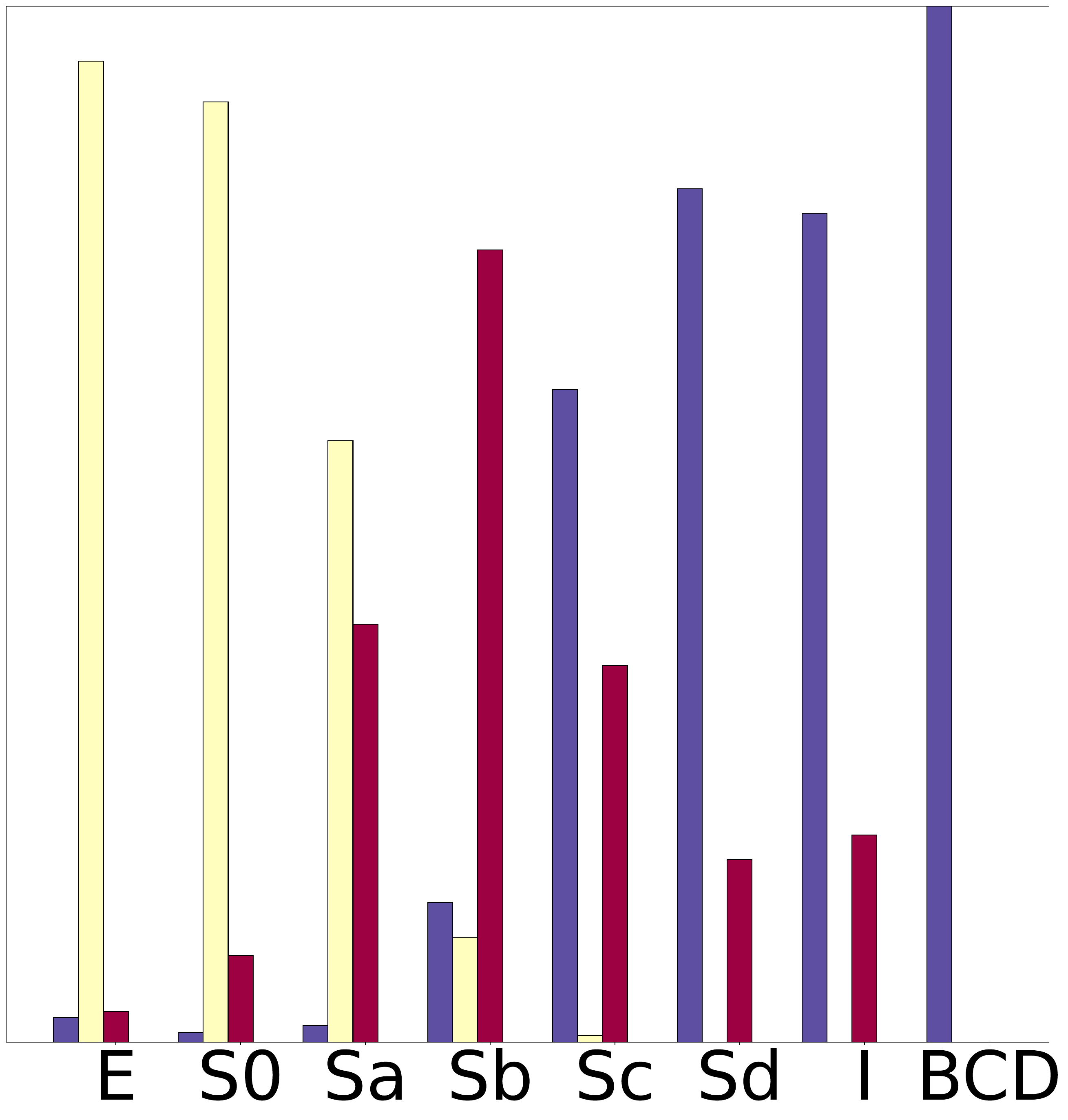} &
        \includegraphics[width=5cm,height=5cm]{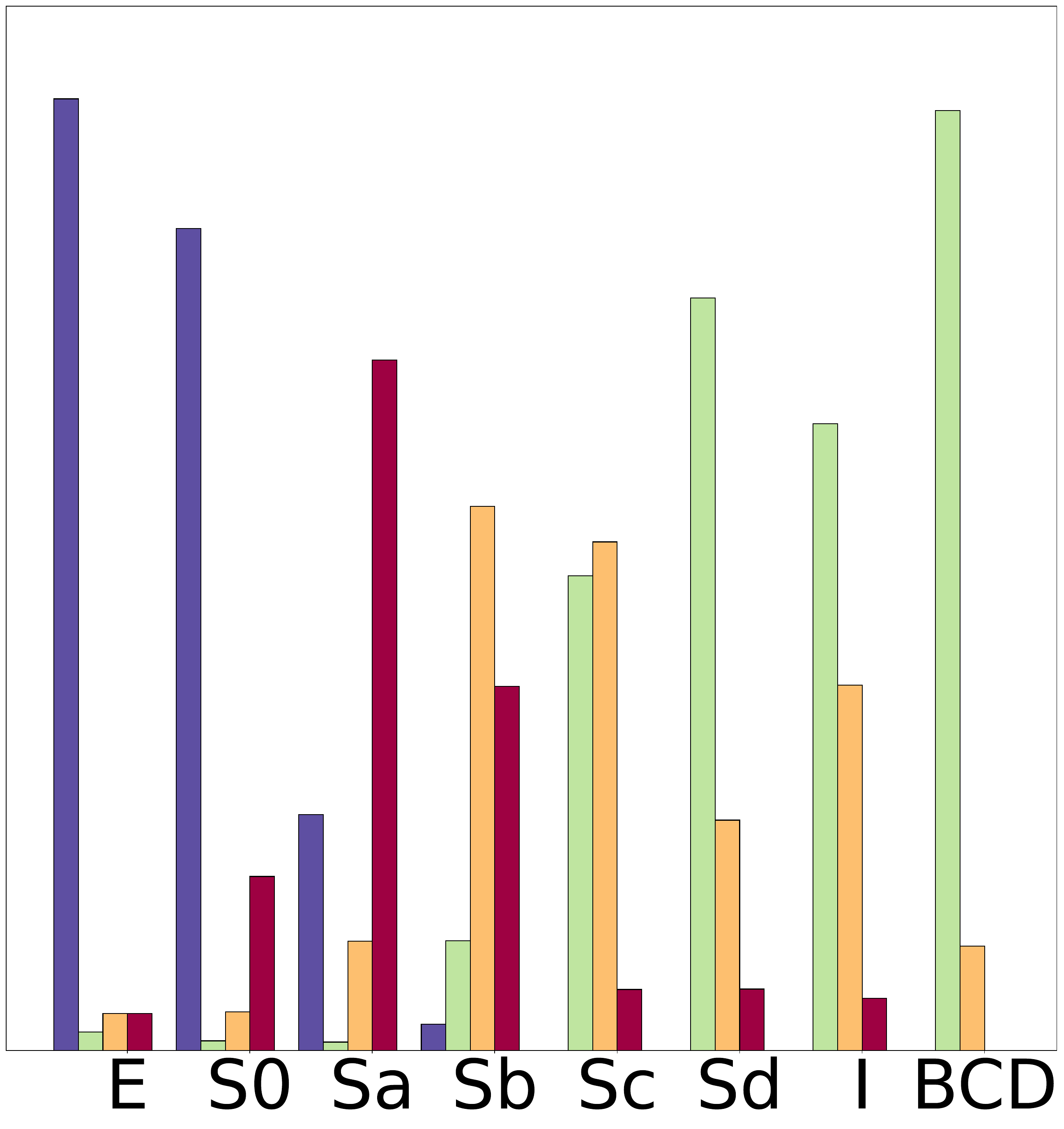}          
        \end{tabular} 
    \caption{\small{Visualisation of galaxy clustering and properties from the eCALIFA sample. Each column in the figure represents a different number of clusters, with the number increasing from two to four as we move from left to right. The rows are organised as follows: the first row presents the UMAP projection of galaxies, colour-coded to differentiate the clusters. The second column illustrates the SFMS for each clustering scenario, showing the SFR derived from stellar populations. The third column features the WHAN diagram for each set of clusters. In the last row we include a histogram showing the morphological distribution within each cluster, expressed as a percentage relative to the total number of each morphological type in the eCALIFA sample.}}
    \label{fig:cluster}
\end{figure*}

 In Fig. \ref{fig:cluster}, we provide a comprehensive visual exploration of galaxy clustering in our study. Each column of the figure corresponds to an increasing number of clusters, ranging from two to four, and showcases four distinct components: a UMAP projection, the Star Formation Main Sequence (SFMS),  the WHAN diagram, and the morphology composition of each cluster. The UMAP projection in each row is colour-coded to distinguish different clusters, offering an intuitive understanding of their distribution in the reduced two-dimensional space.

Next to the UMAP projections are visual representations of key astrophysical aspects. The SFMS plots provide insights into the SFR, derived from stellar populations, across different clusters. This is complemented by the WHAN diagrams, which are instrumental in unravelling the primary ionisation mechanisms at play within these galaxies.

\par The last row of Fig.~\ref{fig:cluster} show an histogram that displays the morphological distribution within each cluster. This histogram illustrates the percentage of each morphological type, relative to its total representation in the eCALIFA sample, that is found within a given cluster. This addition provides a more nuanced view of the composition of each cluster, emphasising the prevalence of various galaxy morphologies in relation to the overall dataset. 
\par Collectively, these visual elements in Fig.~\ref{fig:cluster} offer a multi-faceted perspective, combining spatial distribution, morphological composition, star formation activity, and ionisation characteristics to present a holistic view of galaxy clustering as observed in the eCALIFA sample. In the 2-cluster scenario, we observe a natural division into the classic red sequence and blue cloud, with a notable correspondence between red galaxies and early-type morphologies and vice versa. When we expand to three clusters, a transition population emerges, potentially linked to 'green valley' galaxies. These galaxies are characterised by high mass and low SFR, yet they still fall within the SFMS \citep{2014MNRAS.440..889S,2022MNRAS.512.3566N}. A significant proportion of this group comprises AGN and LINER galaxies, and it encompasses about 80$\%$ of the Sb galaxies in our sample.
\newpage
\par The situation becomes more complex to interpret with four clusters. We identify a cluster of quenched/passive galaxies, predominantly of E and S0 morphologies. Another cluster is mainly composed of Sa and Sb galaxies with low specific SFR (sSFR), followed by a cluster characterised by moderate sSFR spiral galaxies, primarily of Sb and Sc types, and a fourth cluster with very high sSFR and late-type morphologies, including Irregular galaxies and Sdm types. 
\par In our study, we also conducted the clustering experiment using the Gaussian Mixture Model (GMM) algorithm, aiming to cross-verify our results obtained from the k-Means approach. The outcomes were strikingly similar for cluster groupings of two to four, which underlines the robustness of our clustering analysis. However, a notable divergence was observed upon adding a fifth cluster; specifically, when the clustering algorithms were fitted with five clusters, the GMM presented a different cluster composition than that of k-Means.
\par This discrepancy in the five-cluster model suggests several possibilities. Primarily, it might indicate that a configuration of five clusters could be overly granular for effectively representing the galaxy data. This level of segmentation might lead to clusters that are more a byproduct of the algorithm's intrinsic tendencies rather than a true reflection of distinct galaxy groupings. The difference could also be attributed to the inherent algorithmic nature of k-Means and GMM. While k-Means excels in identifying spherical clusters, GMM is adept at recognizing more complex, elliptical shapes. This fundamental distinction could be driving the differing results in more nuanced clustering scenarios, like with five clusters. Such an outcome accentuates the critical role of algorithm selection and suggests a cautious approach when interpreting clustering results, especially as the number of clusters increases.
\par Concluding this section, it is essential to underscore that our primary objective was not to establish definitive categories of galaxies. Instead, our focus was on garnering a deeper understanding of the diverse properties and behaviours of galaxies within the eCALIFA sample. Through our clustering analysis, we have demonstrated that the representation space crafted via CL is meaningful and effectively segregates galaxies based on their intrinsic properties.
\par This analysis has shown significant correlations between the physical attributes of galaxies and their distribution in the representation space. By examining the clusters across different configurations, we have gained insights into how galaxies group based on various characteristics such as SFR, ionisation mechanisms, and morphological types. These findings provide a deeper appreciation of the complex nature of galaxies and their evolutionary stages, as observed in the eCALIFA dataset. Furthermore, this exploration has validated the efficacy of our SNN-based approach in creating a representation space that is not only discriminative but also insightful for galaxy evolution research.
\subsection{Comparing distributions of galaxy population}\label{subsec:dist}
\par In the field of galaxy evolution, a fundamental question often arises: do distinct galaxy samples exhibit divergent physical properties? This inquiry extends to exploring environmental influences on SFR quenching \citep[e.g.][]{2021MNRAS.502.4457G,2023A&A...680A.111D}, the impact of AGN on galactic evolution \citep[e.g.][]{2023A&A...675A.137M}, the correlation between galaxy size and the stellar content of galaxies \citep{2019A&A...631A.158D}, the result of the merger state on stellar and ionized gas kinematic \citep[e.g.]{2015A&A...582A..21B,2023MNRAS.526.2863M}, etc. Traditionally, addressing these questions involves analysing and contrasting the physical properties of each galaxy population to determine the influence of each physical variable. While this approach offers clear interpretations based on physical characteristics, it is not without limitations, notably susceptibility to model degeneracies and assumptions. 
\par With  contrastive learning we can investigate if two population of galaxies have the same observation properties in a data-driven manner under the assumption that the constrastive space is sufficiently descriptive to distinguish the phenomenon under study. This method is based on the premise that if two galaxy populations have intrinsic physical differences, they should occupy distinct positions within this space. As galaxies exist within a 512-dimensional manifold, we need to examine the distribution of data points in this high-dimensional space for each population and compare them. One approach consists on comparing the ten closest neighbours for each distribution in order to discern whether a set of populations are distinguishable based on their positioning in this representation space \citep{2024ApJ...961...51V}. 
\par
\begin{figure*}[!ht]
    \centering
    \begin{tabular}{ccc}
    \includegraphics[width=6.cm,height=5.5cm]{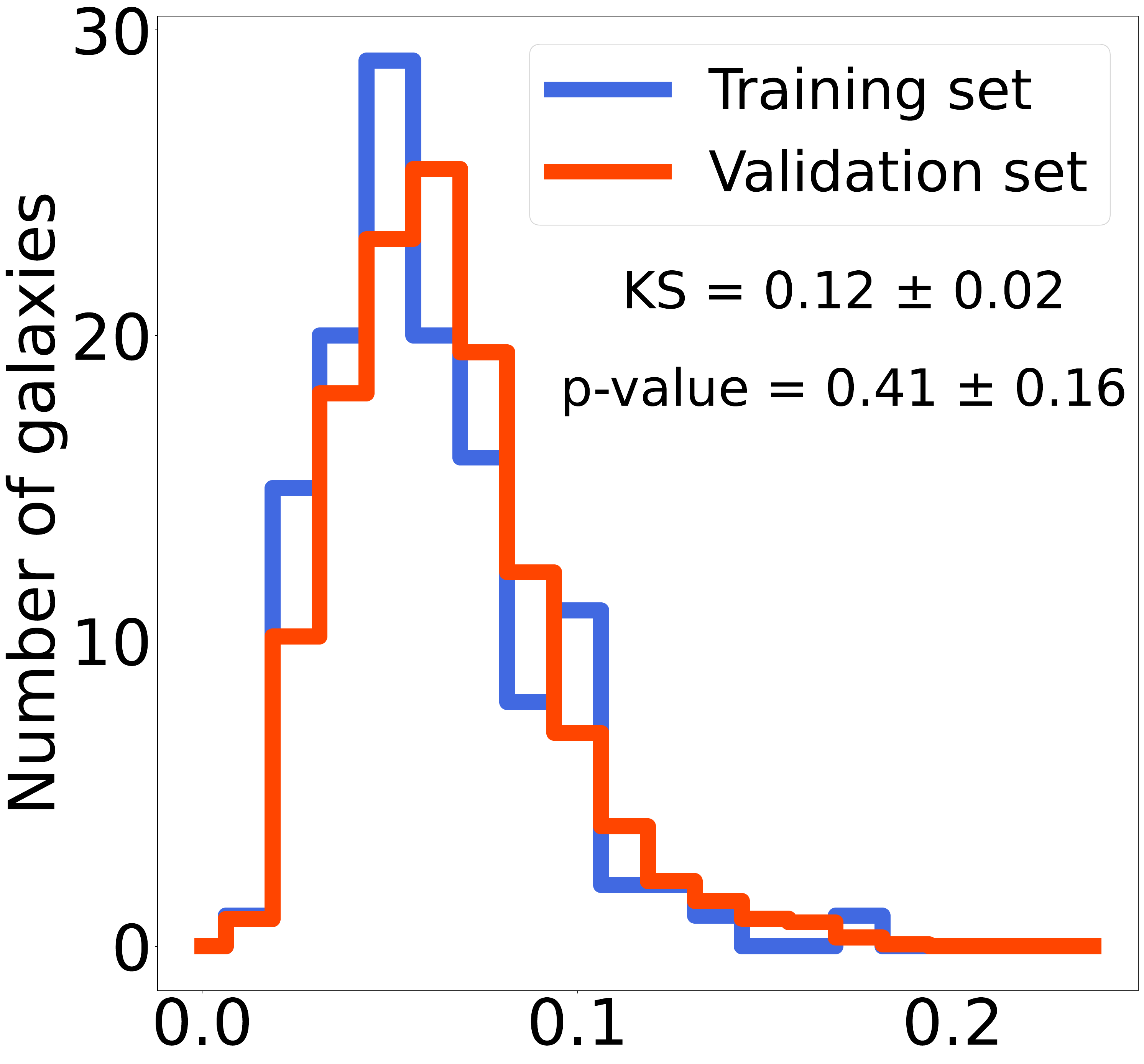} &    
    \includegraphics[width=6.cm,height=5.5cm]{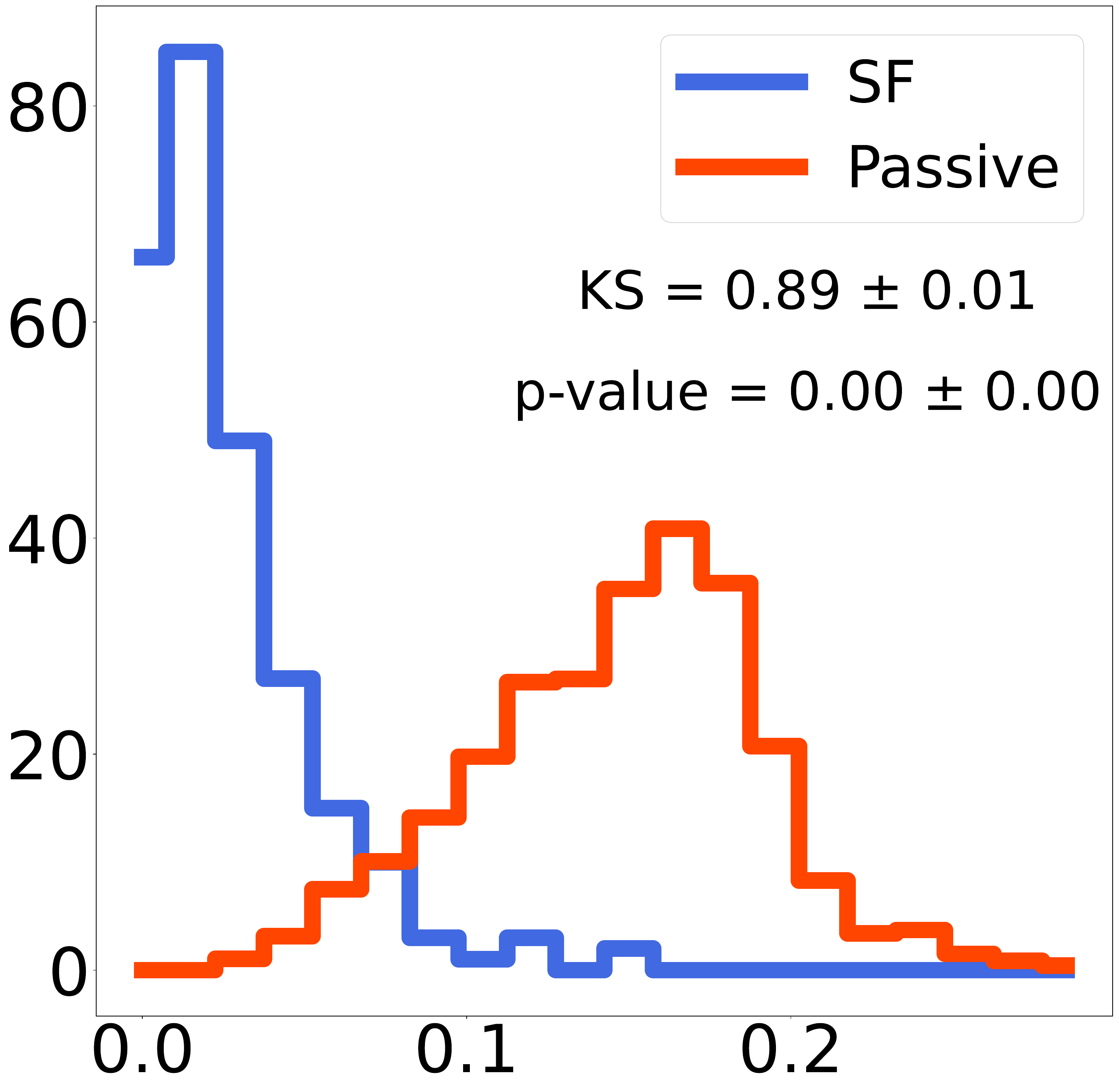} &
    \includegraphics[width=6.cm,height=5.5cm]{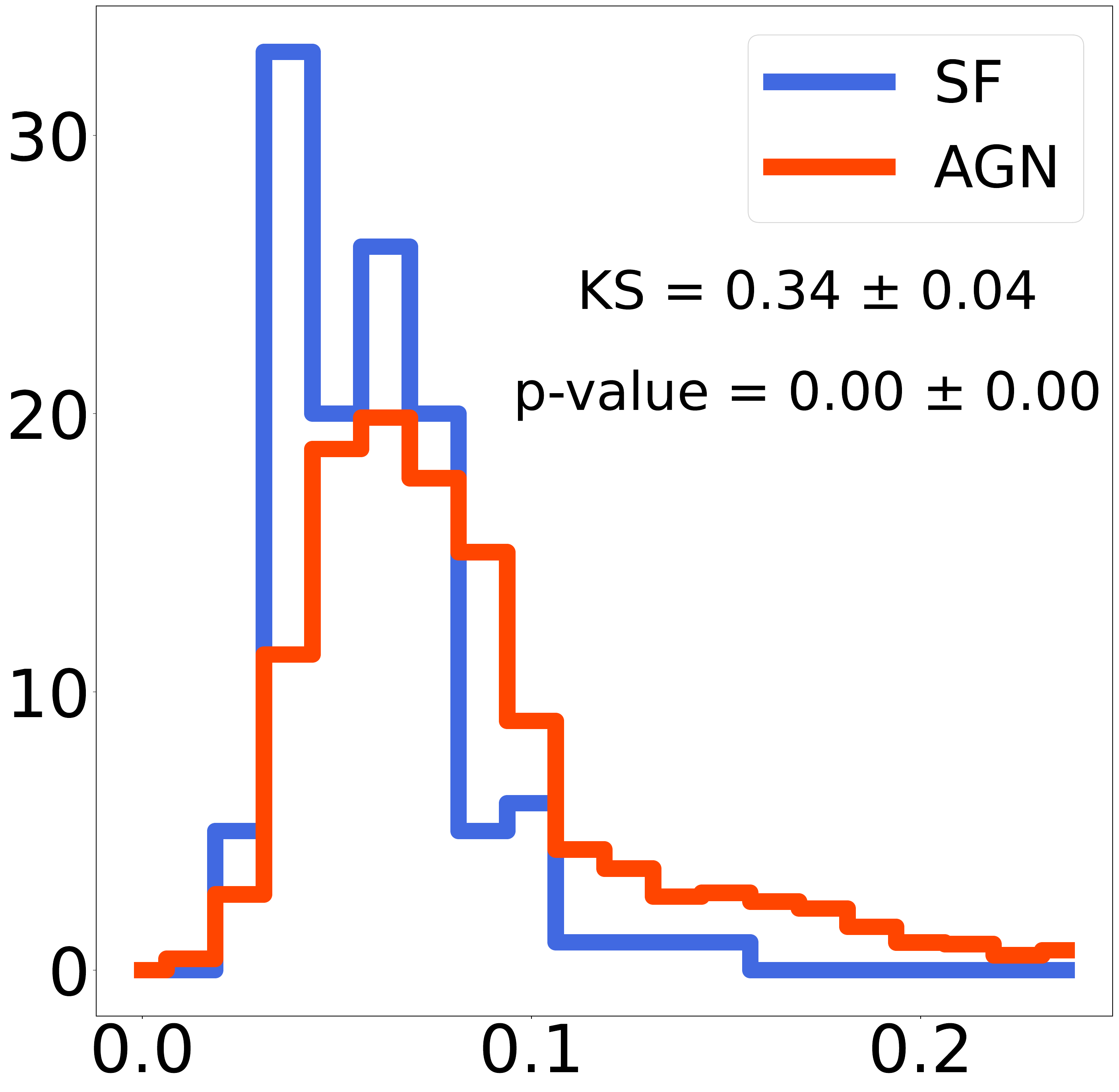} \\
    
    \includegraphics[width=6.cm,height=5.5cm]{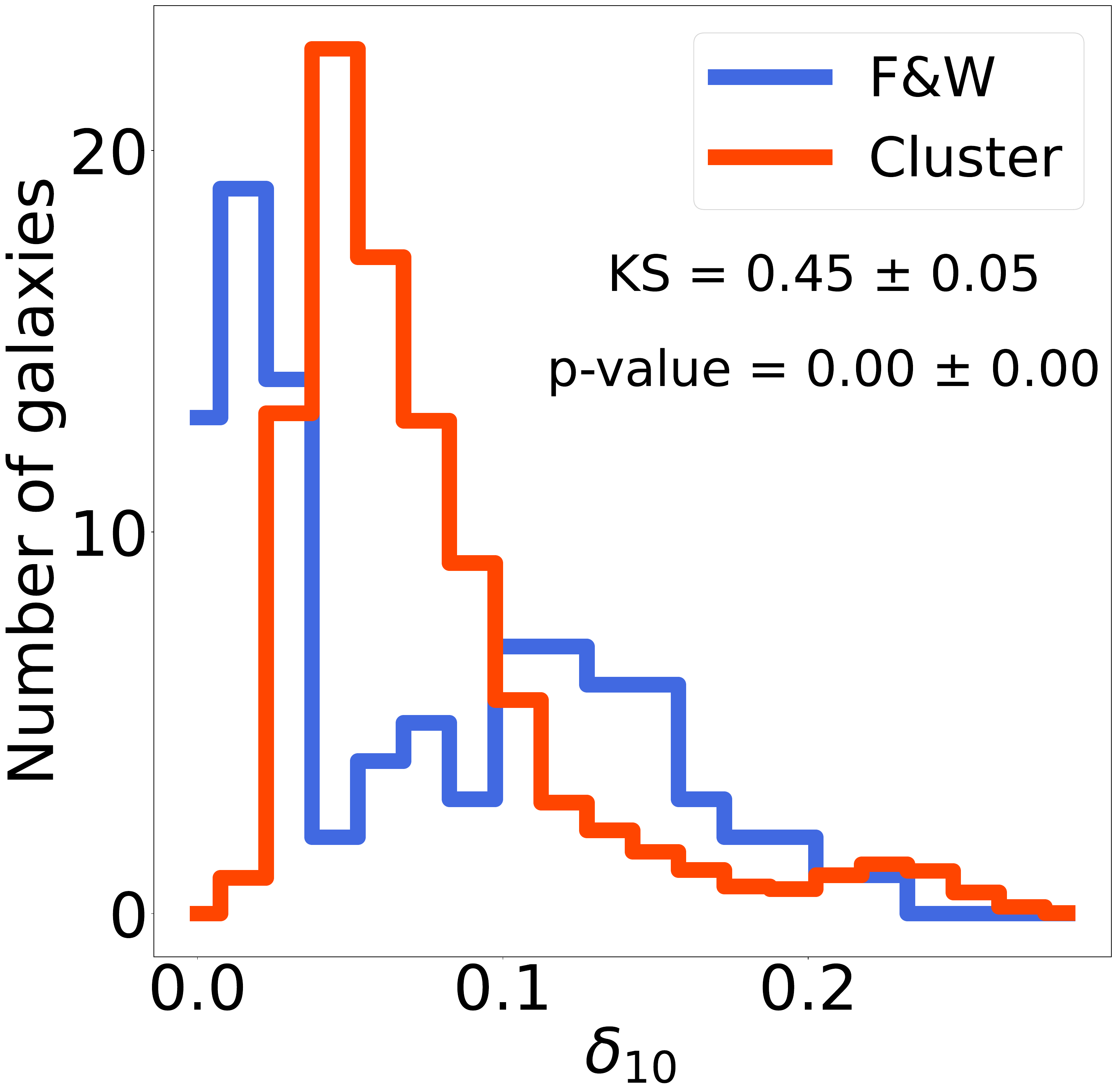} &
    \includegraphics[width=6.cm,height=5.5cm]{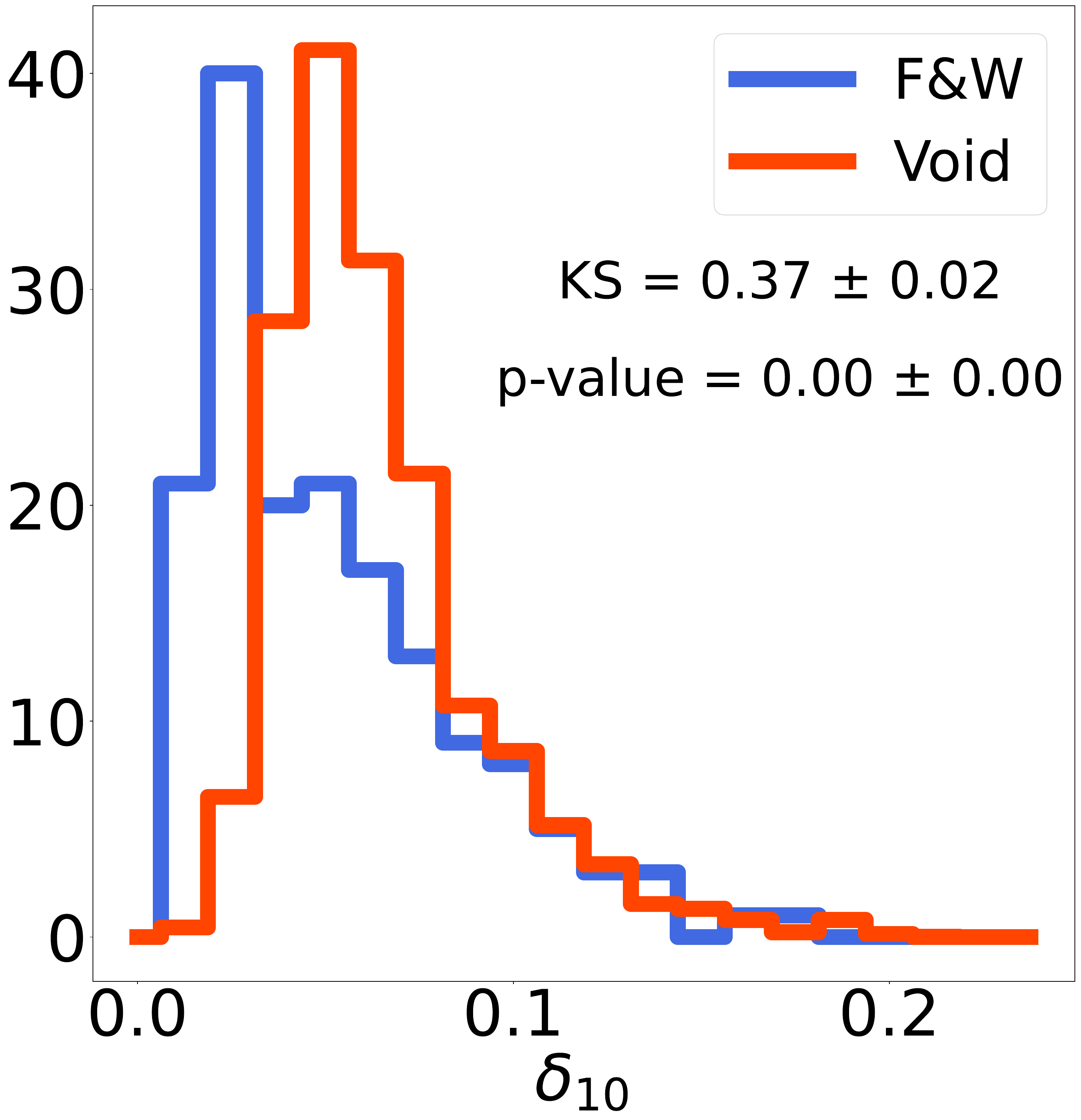} & 
    \includegraphics[width=6.cm,height=5.5cm]{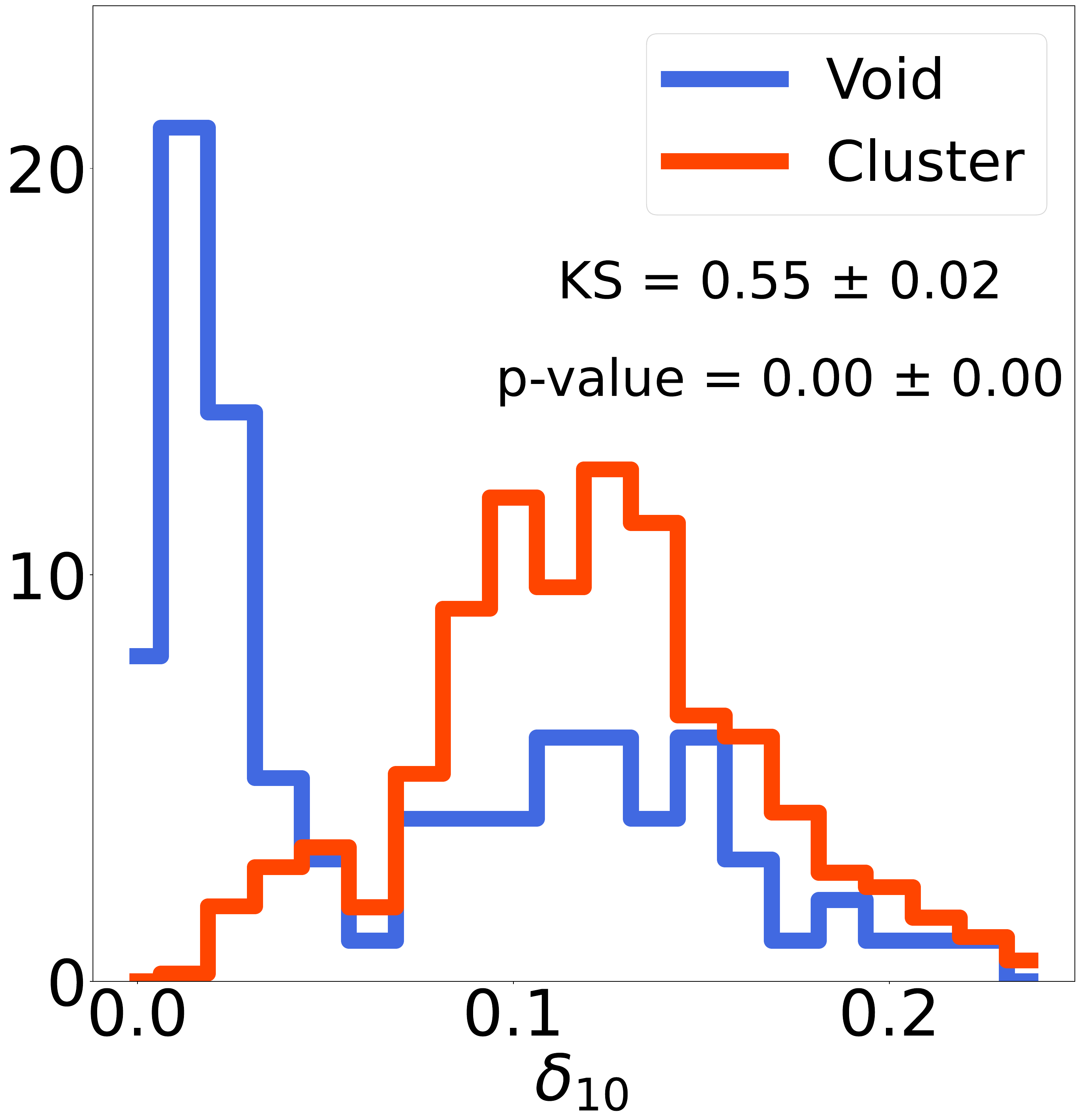}
    
    \end{tabular} 
    \caption{\small{Comparative histograms of the 10th nearest neighbour distances for different galaxy populations. Top left: Training vs. validation set. Top middle: Passive vs. SF galaxies. Top right: AGN vs. SF galaxies. Bottom left: F$\&$W galaxies vs. galaxies in clusters. Bottom middle: F$\&$W galaxies vs. void galaxies. Bottom right: Void galaxies vs. galaxies in clusters.}}
    \label{fig:dist_hist}

\end{figure*}

 To ascertain the efficacy of our methodology, we first compare the galaxy distributions in both the validation and training sets. The construction of the validation set ensured it mirrored the morphological type proportions found in the training set. The UMAP projection suggested that both sets occupy similar regions, implying that, barring overfitting, our methodology should reveal no significant differences between these populations in the representation space. This hypothesis was tested by calculating the cosine distances to the ten nearest neighbours from the training set for each galaxy in both the validation and training sets. Given the larger size of the training set, multiple iterations were performed, each involving a random selection of 126 galaxies for comparison with the validation set, followed by averaging the results. A pronounced discrepancy in these distance distributions would suggest an uneven statistical representation across the two categories of galaxies. For a quantitative evaluation of this disparity, we applied the Kolmogorov–Smirnov test. This non-parametric technique assesses whether two one-dimensional probability distributions significantly diverge. It compares empirical distribution functions of both samples and computes a p-value, reflecting the likelihood that any observed differences arise under the null hypothesis, which posits no disparity in distributions. As shown in the top left panel of Fig. \ref{fig:dist_hist}, the distributions closely align, and a p-value exceeding 0.05 – the conventional threshold for null hypothesis rejection – strongly indicates that the galaxy distributions in the validation and training sets are sourced from an identical underlying distribution.
\par We have now expanded our analysis to encompass a comparative study of the distributions of SF and passive galaxies. The UMAP projection distinctly delineates the varying spatial regions occupied by each galaxy type. Notably, SF is often suppressed in elliptical galaxies, while it occurs more frequently in spiral galaxies. This divergence is prominently illustrated in the histograms displayed in the top middle of Fig. \ref{fig:dist_hist}, underscored by a KS value approaching unity, and a p-value of zero. These results serve dual purposes: firstly, as a validation of the SNN capability to effectively project eCALIFA data cubes into a lower-dimensional space, albeit preliminarily; and secondly, as a methodological approach to examine the distribution of distances within the representation space using nearest neighbours to identify similarities among galaxy populations. 
\par Our focus now shifts to more ambiguous scenarios. In examining the distributions of AGN and SF galaxies, we observe an overlap in the UMAP projection, with an exception noted in the latest morphological types where AGN presence is unlikely.  However, this apparent convergence might be attributed to the limitations of projection and may not accurately reflect the true nature of the distributions in the representation space.  The findings, as indicated in the top left panel of Figure~\ref{fig:dist_hist}, suggest distinct distributions for the two galaxy types. Yet, the resulting KS value is lower compared to the previous case, where the distributions exhibited greater disparity. Subsequent analysis will demonstrate that when morphology and mass distribution are harmonised, the differences between AGN and SF galaxy distributions in the contrastive space diminish.
\par In this stage of our investigation, we focused on the comparative analysis of galaxies situated in various environments, as outlined in Sect.~\ref{subsec:2Dprojection}. Consistent with the observations in Fig.~\ref{fig:enviroment}, the spatial distribution of galaxies in our representational framework exhibits notable distinctions across cluster, void, and F$\&$W environments. The most pronounced contrasts emerge between cluster and void environments, which is an anticipated outcome given their diametrically opposed characteristics—high-density versus low-density settings. Although F$\&$W galaxies exhibit discernible differences when compared to those in clusters, these variances are less marked when contrasted with void galaxies (see bottom panels of Fig.~\ref{fig:dist_hist})
\par During this stage of our analysis, it becomes pertinent to question whether the observed differences among specific galaxy populations arise from variations in the distribution of galaxy types, considering aspects like mass and morphology, which would lead to distinct placements in the representation space. Alternatively, we must consider if other intrinsic galaxy characteristics could cause these populations to exhibit varied distributions. To put it simply, if we compare AGN galaxies with SF galaxies of similar mass and morphology, will they still be distinguishable in the representation space? To address this question, we have identified galaxy twins pairs of galaxies that have comparable mass (within a 0.2 dex range) and identical morphological types across the groups being compared. This approach is applied to contrast SF and AGN galaxies, as well as to compare galaxies in F$\&$W versus void environments and cluster galaxies with those in F$\&$W settings. The number of pair twins are 99, 136, and 92, respectively. 
\par
\begin{figure*}[!ht]
    \centering
    \begin{tabular}{ccc}
    \includegraphics[width=6.cm,height=5.5cm]{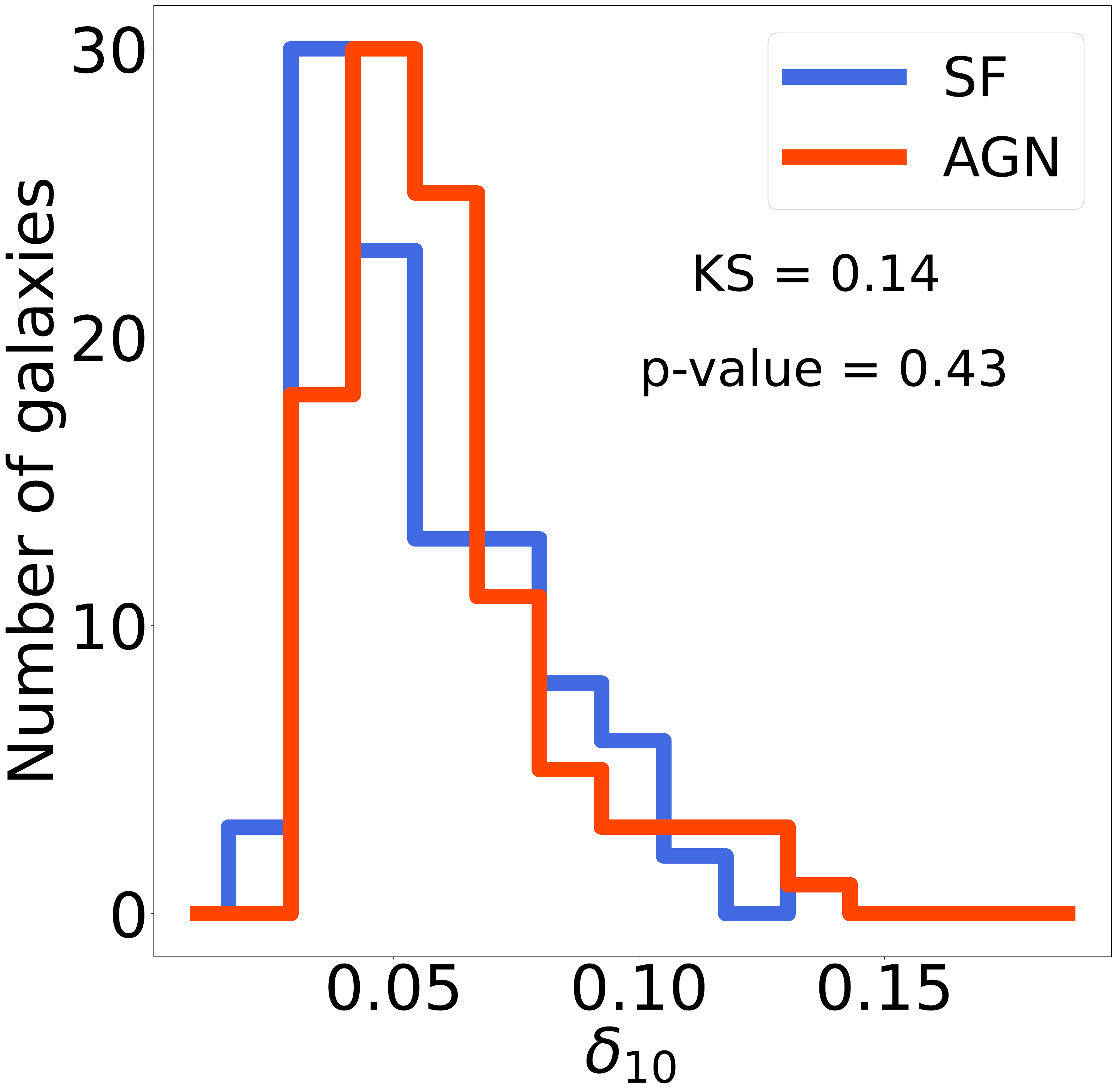} &    
    \includegraphics[width=6.cm,height=5.5cm]{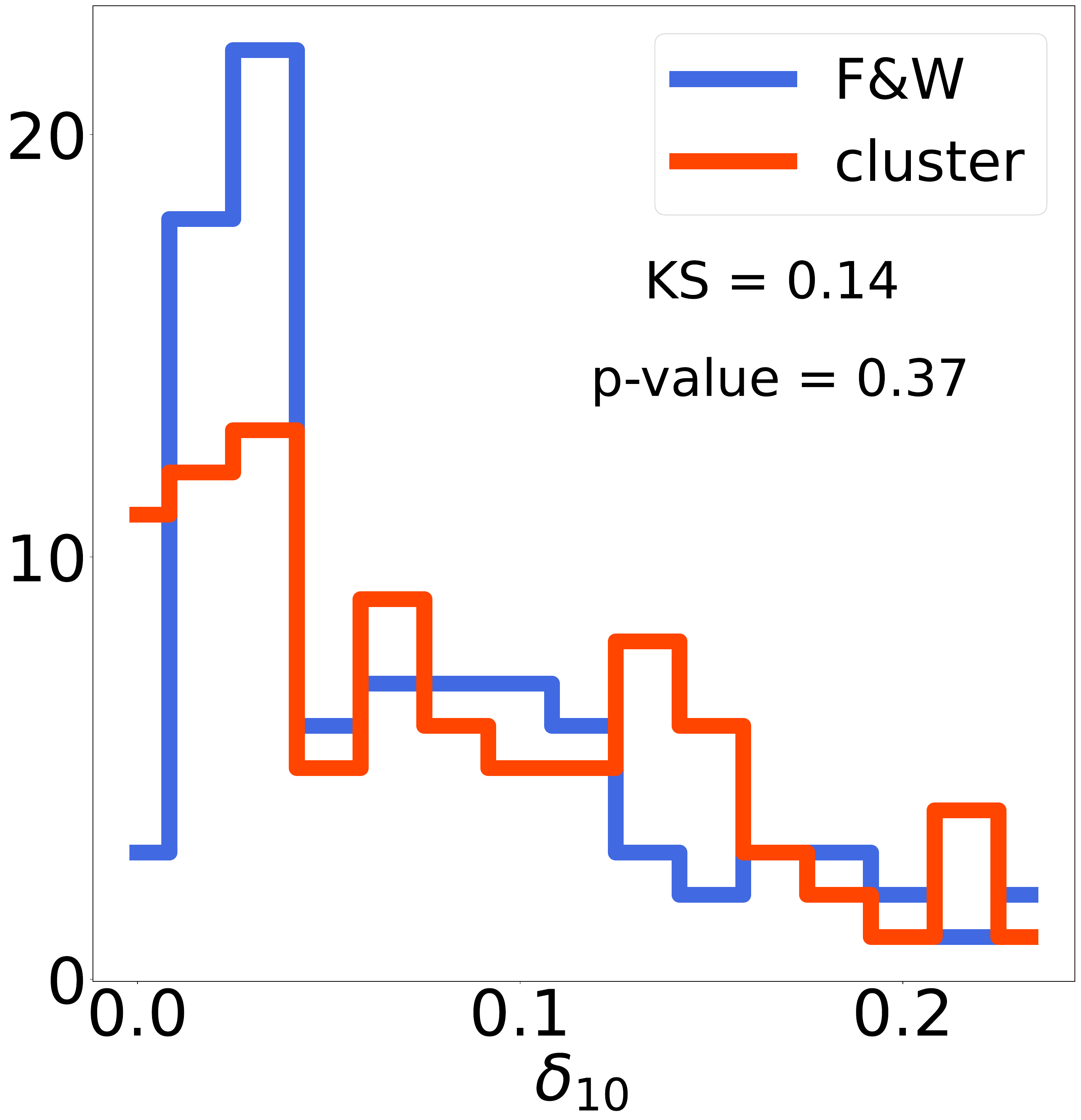} &
    \includegraphics[width=6.cm,height=5.5cm]{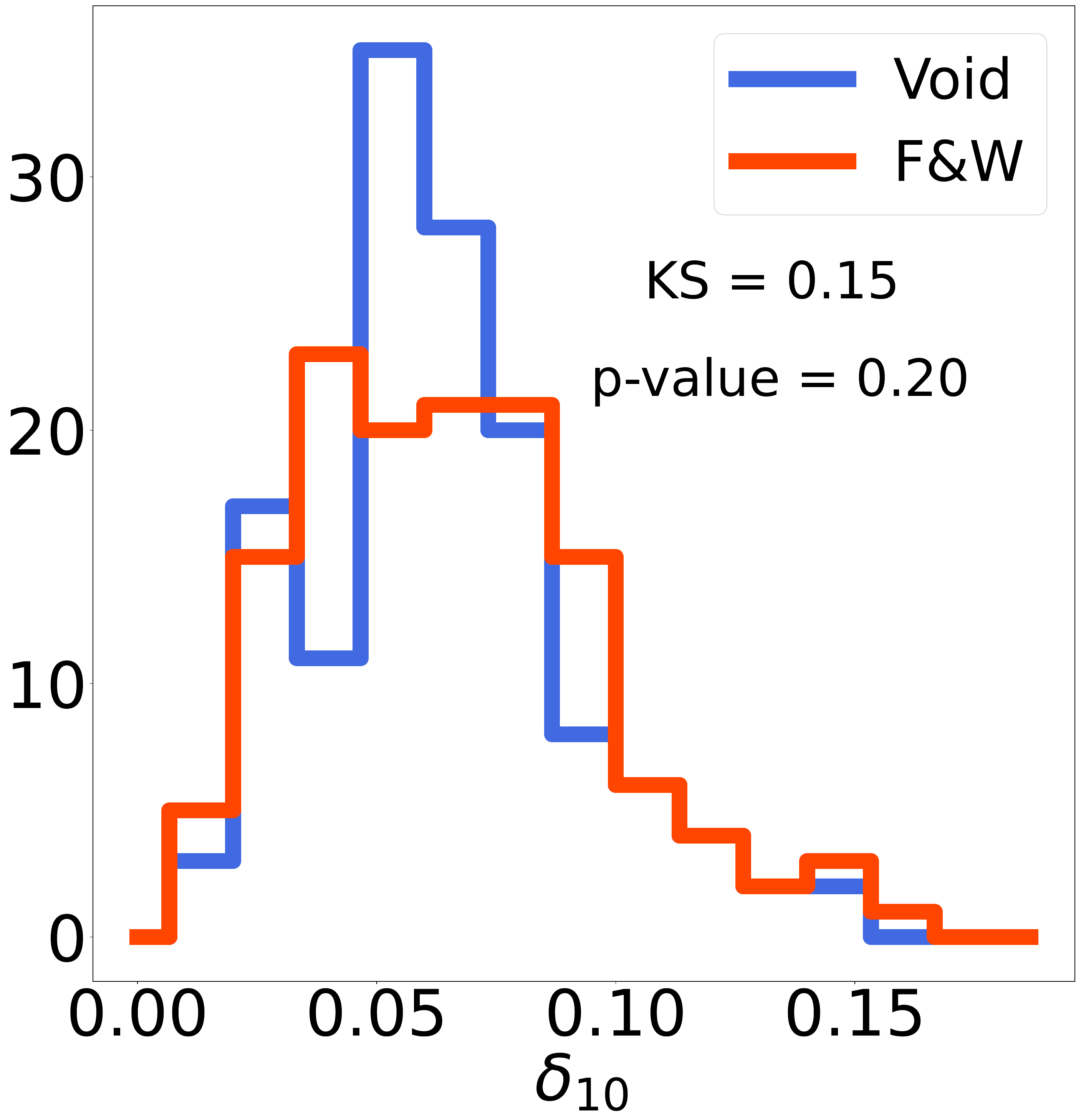}     
    \end{tabular} 
    \caption{\small{Comparative histograms of the 10th nearest neighbor distances for different galaxy populations composed of twin galaxies, i.e. similar stellar mass (within 0.2 dex difference), and same morphological type. Left: AGN vs. SF galaxies. Middle : F$\&$W galaxies vs. galaxies in clusters. Right: Void galaxies vs. F$\&$W galaxies. }}
    \label{fig:dist_hist_twin}

\end{figure*}

 The results presented in Fig.~\ref{fig:dist_hist_twin} reveal that galaxy populations, when observed in the representation space, do not exhibit statistically significant intrinsic differences once they are matched for mass and morphology. This observation holds true for comparisons between AGN and SF galaxies, galaxies in F$\&$W versus those in voids, and galaxies in clusters versus those in voids. Focusing on the first comparison, i.e., SF versus AGN galaxies, it becomes evident from the standpoint of similarity measures that an AGN galaxy is fundamentally different from a SF galaxy, even when the AGN is hosted in a galaxy with the same morphology and mass. The lack of significant differences between these two galaxy populations in the representation space might initially suggest that the projection made by SNN is not capturing the subtle differences, such as the few pixels occupied by the AGN in the eCALIFA data cubes, with mass and morphology dominating the projection. 

\section{Discussion}\label{sec:discussion}

\subsection{Galaxy properties and the representation space} 
\par CL has emerged as a tool to build proto-foundation models in astronomy, providing a framework for the construction of representation spaces in large datasets, predominantly composed of unlabelled data. This methodology facilitates a set of downstream tasks, such as the morphological classification of astronomical sources \citep[see e.g.][]{2024RASTI...3...19S}. However, our research primarily aims to delve deeper into the characteristics of these representation spaces, with a specific focus on reinterpreting galaxy properties within eCALIFA data cubes. Applying CL to modern IFU datasets presents significant challenges due to the inherent complexity of the observations, which encompass both images and spectra. Nevertheless, this approach offers the distinct advantage of preserving morphological information across various wavelengths while simultaneously presenting spectral features in a spatial context. This dual perspective enables the concurrent study of galaxy morphology and spectral characteristics. Traditional analysis of IFU data tends to address these aspects in isolation, typically conducting spaxel-by-spaxel examinations of the stellar population or ionised gas to generate two-dimensional maps of physical properties. Morphology is then considered as a singular parameter to categorise samples and explore the two-dimensional variations of these properties as a function of morphological classification. 
\begin{figure}[!ht]
    \centering
    \includegraphics[width=9.2cm,height=7cm]{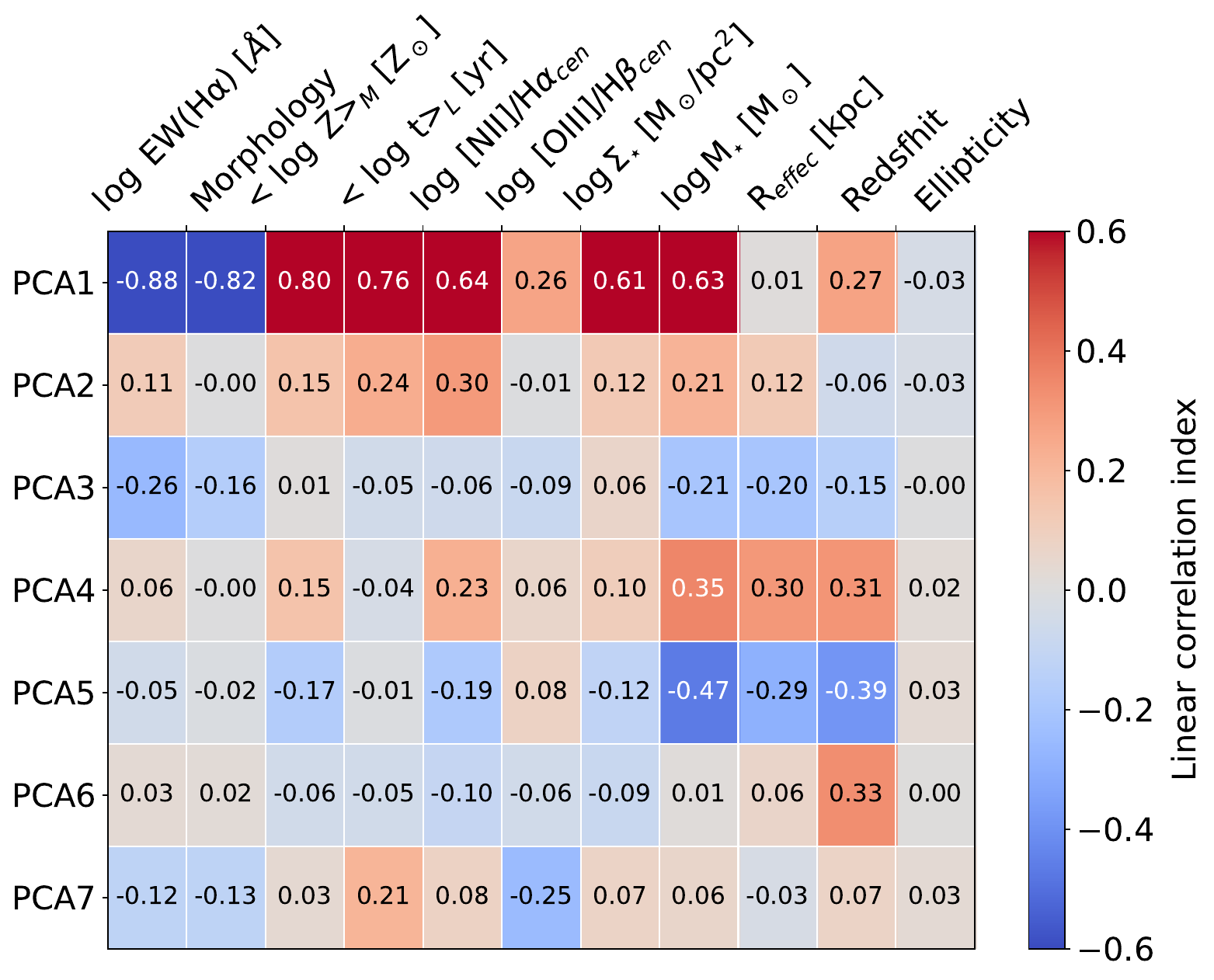} 
    \caption{\small{Linear pair wise correlations between the first four PCA of the representation space and the physical properties. Cells are colour-coded according to the values of the Pearson’s correlation statistic.}}
    \label{fig:correlation}
\end{figure}
\par The analyses conducted in Sect.~\ref{subsec:2Dprojection} and~\ref{subsec:clustering} have already demonstrated the linkage between the representation space and a range of physical properties, including morphology, H$\alpha$ EW, stellar metallicity, and stellar mass. To quantify better these relationships, PCA was applied to the representation space, followed by the computation of correlation coefficients between each PCA component and various physical properties. The initial seven components accounted for 90$\%$ of the variance, highlighting the redundancy within the representation space. As shown in Fig.~\ref{fig:correlation}, the H$\alpha$ EW and morphology exhibit the strongest correlations, succeeded by stellar metallicity and the luminosity-weighted age. The [NII]/H$\alpha$ ratio in galaxy centres displays a significant correlation, whereas no correlation is observed with the [OIII]/H$\beta$ ratio. This disparity is likely attributable to the [NII]/H$\alpha$ ratio's association with other variables, such as H$\alpha$ EW or stellar mass, rather than a direct correlation with the gas phase metallicity of the galaxy's centre and the representation space. Correlations with stellar mass and stellar surface density are also noted but are less pronounced compared to those with morphology or H$\alpha$ EW. Lastly, minimal correlations are identified with the galaxy's effective radius, redshift, and ellipticity, indicating a weaker relationship with these parameters.
\par At the heart of the CL framework lies a series of transformations designed to ensure data invariance. Among these, we consciously chose not to implement a transformation to account for the inclination of galaxies, a decision that stemmed from the necessity to accurately model the three-dimensional distribution of gas and dust, alongside establishing parameters for stellar and gas extinction. However, our analysis reveals that galaxy inclination has a minimal impact on their representation within the designated space. This observation diverges from findings reported by \cite{2021ApJ...911L..33H}, where SDSS galaxies viewed face-on and edge-on occupy distinct regions in similar analyses. The lack of observed differentiation in our study might indicate that spectral characteristics play a more significant role than morphological features in this context, attributes not captured in the SDSS's ugriz images. Our examination, as illustrated in Fig.~\ref{fig:correlation}, does not conclusively determine whether the SNN prioritise specific physical properties over others. Instead, it assesses the presence of correlations between each property within the representation space. Given the close relationship between the EW of H$\alpha$ and galaxy morphology, it is conceivable that observed correlations with morphology could indirectly reflect those with the EW, or vice versa. This complexity underscores the need for further investigation.
\par Our findings indicate that the ionisation state of a galaxy's nucleus does not serve as a distinguishing feature in the representation space, which likely results from multiple factors. Firstly, it is important to recognise a primary constraint of this study: the relatively limited size of the training dataset. This limitation restricts the diversity of galaxy populations represented, particularly when considering variables such as stellar mass, morphology, and ionisation states under various ionisation conditions. Consequently, the SNN may lack the requisite data to identify subtler distinctions beyond the most prominent features. Additionally, AGNs are localised to the nuclei of galaxies, affecting only a small portion of the spaxels within the eCALIFA data cubes. Given the current design of the transformations, these nuanced differences might be overlooked as the focus is on the galaxy as an entirety. Exploring alternative SNN architectures such as visual transformers \citep[see e.g.][]{2020arXiv201011929D,2023arXiv230108243A} or incorporating additional symmetries—such as uniform spectra across the galaxy's centre—could potentially enhance the segregation of these populations. However, such investigations fall beyond the scope of this paper, which primarily aims to evaluate the applicability of the well established CL framework to eCALIFA data cubes.
\subsection{Galaxy populations}
The bimodal distribution of galaxy populations is a well-documented phenomenon observed both in the local universe and at intermediate redshift \citep[see e.g.][]{2001AJ....122.1861S,2004ApJ...600..681B,2009ARA&A..47..159B,2019A&A...631A.156D,2021A&A...649A..79G,2022A&A...661A..99M}. This bi-modality has undergone thorough examination from various analytical perspectives. These investigations have confirmed the presence of primarily two distinct categories of galaxies. The first, known as the blue cloud, encompasses predominantly spiral galaxies. These galaxies are characterised by their gas-rich composition, lower masses, metal poor, and elevated rates of star formation activity. Conversely, the second category, the red sequence, consists of elliptical galaxies. These are notable for their significant mass, metal abundance, and the incorporation of very ancient stellar populations, alongside markedly suppressed star formation rates. 
\par The cluster analysis delineated in Sect.~\ref{subsec:clustering} demonstrates that the observed bimodality arises naturally from the analysis of eCALIFA data cubes in an unsupervised, data-driven manner. This outcome is particularly noteworthy considering that the SNN was trained without any explicit physical information. The emergence of two principal galaxy populations was facilitated solely through the implementation of transformations designed to mitigate the network's focus on observational disparities among galaxies, thereby allowing these populations to manifest based on the inherent qualities of the data. Moreover, the application of clustering within the representation space using three clusters revealed the existence of an intermediate population. This group, characterised by moderate SFR, masses, and morphological types that bridge the gap between the aforementioned primary populations, aligns with the characteristics typically associated with green valley galaxies. This re-discovery of the star formation main sequence corroborates findings previously established by \cite{2021ApJ...921..177S} through models trained on stellar population and kinematic maps derived from the analysis of MaNGA data cubes. However, our analysis uniquely achieves this insight by directly examining eCALIFA data cubes, without the reliance on physically informed maps provided to the SNN.
\par We investigate now the impact of excluding direct mass information on the training of a SNN  by omitting the primary map that encodes galaxy luminosity from the input. Instead, the network is trained exclusively on PCA maps, excluding the one representing the luminosity at the normalisation window. Subsequently, we employ the k-Means clustering algorithm to delineate galaxy populations in the representation space and assess the overlap in clusters when luminosity data is integrated. The outcomes are summarised in Table~ \ref{tab:cluster_results}. This illustration emphasises the fraction of galaxies that are consistently categorised together, irrespective of whether luminosity information is included in the SNN training process. Our analysis reveals that galaxy luminosity exerts minimal influence on distinguishing between galaxy populations. Instead, the spectral shape information, as deciphered through PCA decomposition maps, emerges as the paramount factor in constructing the representation space. This observation is consistent with the results presented by \cite{2021ApJ...921..177S}, who investigated cluster memberships using various input maps for SNN training. Their research highlights the critical importance of kinematic maps for precise identification of the galaxy main sequence. While acknowledging the significant, though lesser, influence of stellar population attributes such as age and metallicity in the detailed characterisation of galaxy features, the study found that excluding V-band luminosity from the SNN training inputs did not markedly affect the clustering outcomes. This finding accentuates the relatively minor role of luminosity profiles in distinguishing galaxy populations when spectral data is accessible. 
\begin{table}[h!]
\caption{Galaxy Clustering Consistency: Luminosity Impact}
\centering
\begin{tabular}{@{}ccc@{}}
\toprule
Clusters & Common Members (\%) & Different Members (\%) \\ \midrule
2 & 92.8 & 7.2 \\
3 & 84.2 & 15.8 \\
4 & 90.4 & 9.6 \\  \bottomrule
\end{tabular}
\tablefoot{Comparative analysis of galaxy clustering with and without luminosity information. The table summarises the alignment of clusters formed using a k-Means algorithm, indicating the percentage of galaxies that remain consistently grouped when luminosity is excluded from the training of the SNN.}
\label{tab:cluster_results}
\end{table}

\subsection{Galaxy environment}
In this study, we utilised the representation space to examine the distinctions among galaxies based on their large-scale structure setup, specifically focusing on galaxies in clusters, F$\&$W, and voids. It is crucial to note that the eCALIFA sample employed in this research is not tailored for in-depth analyses of galaxy environments. Rather, its purpose is to provide a comprehensive overview of galaxy types present in the nearby universe. Consequently, galaxies found in clusters and voids, which are less prevalent in our sample and may be biased towards certain types, do not accurately reflect the actual proportion of each galaxy type within their respective environments. Bearing this limitation in mind, we observed that the distribution of galaxies within the representation space varies significantly among those situated in voids, clusters, and F$\&$W. This variation is attributed to differences in mass and morphology distributions across the samples. Our findings align with previous research indicating that galaxies in voids tend to be less massive, bluer, and exhibit later-type morphologies \citep[see, e.g.,][]{2015ApJ...810..108M,2012MNRAS.426.3041H}. Conversely, galaxies located in clusters are more massive, display redder colours, and are predominantly composed of early-type morphologies \citep{2010ApJ...721..193P,2014MNRAS.440..889S}.
\par In comparing samples with matched mass and morphology, galaxies exhibit a consistent distribution within the representation space. This observation suggests that, once mass and morphology are controlled for, there may not be inherent differences in the properties of galaxies across various environments. However, it is possible that the SNN may lack the sensitivity to detect subtle variations present in the 2D distribution of stellar populations or gas profiles. Indeed, the recent study by \cite{2024arXiv240410823C} compares a set of CALIFA galaxies located in F$\&$W environments with galaxies observed in voids by the CAVITY survey (I. Pérez in prep.). This research reveals that, despite matching for morphology and stellar mass, void galaxies exhibit a lower stellar mass surface density, younger stellar populations at the outskirts, and marginally higher specific sSFR compared to their CALIFA F$\&$W galaxy counterparts.
\par  Our goal is not to draw definitive conclusions about the environmental effects on nearby galaxies, which would necessitate a more systematic investigation. Instead, we aim to demonstrate the potential of CL to approach the issue from a novel, data-driven perspective. In the future, we plan to enhance our dataset by including galaxies observed by the CAVITY survey, which utilises the same integral field spectrograph instrument than CALIFA galaxies. This expansion will not only augment the training set size but also improve the training of the SNN, enabling a more detailed analysis of environmental influences on galaxy evolution.

\subsection{SNN as a galaxy tokenizer}
\par In this study, considerable effort was dedicated to exploring the characteristics of the representation space generated by the SNN when applied to eCALIFA data cubes. We have demonstrated that this representation space offers valuable insights into galaxy properties, allowing for a data-driven examination of galaxy populations. However, the specific SNN designed in this work is currently confined to eCALIFA data cubes. The challenge in extending this methodology simultaneously to other IFU surveys, such as MaNGA \citep{2015ApJ...798....7B}, MUSE-Wide \citep{2019A&A...624A.141U}, or forthcoming instruments like WEAVE \citep{2023MNRAS.tmp..715J}, lies in the varying spectral and spatial resolutions these surveys present. Direct application of the same SNN model to different IFU datasets is hindered by these discrepancies. Nonetheless, integrating data from multiple surveys would be advantageous, as it could create a more extensive training dataset, thereby enhancing the quality of the representation space. Such integration would also mitigate the instrumental and observational biases inherent in each survey. Unfortunately, the current absence of a sufficient cross-match between these surveys obstructs the generation of paired data across different datasets, a crucial step in overcoming the limitations posed by disparate data cube dimensions. Homogenising data cubes to a uniform spatial and spectral resolution does not seem a promising solution, as it necessitates a reduction to the lowest spatial and spectral resolution and the narrowest wavelength coverage, compromising the integrity of the data.
\par For these reasons, we believe that the SNN might serve effectively as a tokenizer prior to training more complex models. To elucidate, in the field of natural language processing, a tokenizer is a tool that standardises language by converting text into a structured format that a model can understand \citep{2013arXiv1301.3781M}. It breaks down text into units, such as words or phrases, standardising language, meaning, and expressions, which facilitates the uniform processing of text data across various documents. Similarly, the SNN might be employed individually for each IFU dataset to generate a preliminary compressed representation of the data. Through this method, each galaxy in the training set would be represented by a vector of uniform dimensionality, despite the projections being executed by slightly differing functions within each SNN. By capturing the most pertinent features of the data, the SNN efficiently condenses the most relevant physical information present in the data cubes.
\par Subsequently, it would become feasible to train transformer-based models on these compressed vector representations, akin to the methodology in Large Language Models  where subsequent tokens are predicted based on preceding ones. During this phase, the model, in an unsupervised manner, would learn to generate a new embedding space that is common across all surveys. This approach offers a promising pathway to amalgamate data from multiple surveys, thereby laying the groundwork for more robust foundation models for galaxy analysis. This strategy would not only enhance the comprehensiveness of the data analysis but would also significantly improve the predictive capabilities of subsequent models by leveraging a unified representation of diverse datasets.

\section{Summary and conclusion}\label{sec:conclusions}
\par In this paper, we presented a comprehensive analysis of the eCALIFA dataset through the lens of CL. Specifically, we implemented a SNN architecture to derive insightful representations from eCALIFA galaxies, marking, to our knowledge, the first application of CL directly to IFU data cubes rather than to pre-processed physically informed maps. Through the use of PCA for the reduction of spectral dimensionality and the incorporation of data augmentation strategies, we significantly improved the learning process. This approach enabled the effective training of the SNN, facilitating the identification of galactic features within a relatively small training set, in contrast to the requirements of more conventional ML algorithms.
\par Our research highlights the significant potential of CL in the field of astrophysics as a useful instrument for classifying galaxies and reevaluating their intrinsic physics, complementing the physical insights derived from spectral modelling. The embedding generated by our model categorise galaxies based on their physical attributes. We observed correlations, in descending order of strength, with the EW of H$\alpha$, galaxy morphology, stellar metallicity, luminosity-weighted age, stellar surface mass density, the [NII]/H$\alpha$ ratio, and stellar mass. 
\par Nevertheless, no correlation was identified with the [OIII]/H$\beta$ ratio, indicating that the primary ionisation mechanisms at the centres of galaxies do not significantly influence the SNN projection, thus failing to distinctly separate SF and active AGN galaxies based on this criterion. Similarly, no correlations were detected with the effective radius of galaxies or their ellipticity, which is unexpected considering no specific orientation adjustments were applied. This anomaly may be suggest that spectral features predominantly influence the SNN projection. A marginal correlation with redshift was noted, attributed to the detection of dwarf galaxies, which are only observed with adequate S/N in the eCALIFA sample at the lowest redshift bins. 
\par We examine the distribution of galaxies as a function of their environment, specifically comparing galaxies located in the F$\&$W, voids, and clusters. Our findings reveal that galaxies exhibit distinct distribution patterns across large-scale structure of varying densities, with the most significant differences observed between galaxies in clusters and those in voids. However, these differences diminish when samples are normalised for stellar mass and morphology. Further investigation is required to determine whether these observations are merely artefacts of the SNN projection, which accounts only for first-order characteristics such as stellar mass or morphology, or if galaxies indeed appear observationally identical across clusters, voids, and F$\&$W when matched for mass and morphology. The forthcoming release of the CAVITY survey promises to deliver an extensive dataset of void galaxies, utilising the same instrument as employed by CALIFA. This harmonisation of methodologies facilitates its seamless integration into our training datasets, thereby augmenting our comprehension of galactic distributions within various environmental contexts.
\par Unsupervised clustering within the representation space effectively distinguishes the SFMS, segregating the blue cloud from the red sequence. When searching for three clusters, a population consistent with the green valley is discernible, characterised by moderate SFR, intermediate morphologies, and a higher proportion of AGN galaxies. The membership of these clusters remains largely unchanged when the luminosity maps of galaxies are excluded from the training of the SNN.
\par Future research efforts will focus on expanding our dataset, potentially through the inclusion of data from additional IFU surveys, such as MaNGA, MUSE-Wide, or WEAVE. This enlargement will facilitate the development of a more comprehensive and robust embedding.
\section*{Acknowledgements}
G.M.S., R.G.B., R.G.D., L.A.D.G., A.M.C., and J.R.M. acknowledge financial support from the State Agency for Research of the Spanish MCIU through the "Center of Excellence Severo Ochoa" award to the Instituto de Astrof\'\i sica de Andaluc\'\i a  (CEX2021-001131-S) and to PID2019-109067GB-I00, and PID2022- 141755NB-I00. S.F.S. thanks the PAPIIT-DGAPA AG100622 project and CONACYT grant CF19-39578. We also thank the anonymous referee for many useful comments and suggestions.

\bibliographystyle{aa}
\bibliography{aa}

\appendix
\section{Neural network architecture}\label{app:architecture}
In Table~\ref{table:encoder-architecture} we show the architecture of the SNN employed in this work. 
\begin{table*}[!ht]
\caption{Siamese Neural Network Encoder Architecture}
\centering
\begin{tabular}{llll}
\hline
\\
\textbf{LAYER (TYPE)} & \textbf{OUTPUT SHAPE} & \textbf{PARAMETERS} & \textbf{CONNECTED TO} \\ \\
\hline
\\
INPUTLAYER & (192, 184, 30) & 0 & - \\
CONV\_2D\_1 & (192, 184, 64) & 48064 & INPUTLAYER \\
Nf $=$ 64, ks $=$ (5x5) & && \\
MAX\_POOLING\_2D\_1 & (96, 92, 64) & 0 & CONV\_2D\_1 \\
BATCHNORM\_1 & (96, 92, 64) & 256 & MAX\_POOLING\_2D\_1 \\
CONV\_2D\_2 & (96, 92, 128) & 73856 & BATCHNORM\_1 \\
Nf $=$ 128, ks $=$ (3x3) &  &  & \\
MAX\_POOLING\_2D\_2 & (48, 46, 128) & 0 & CONV\_2D\_2 \\
BATCHNORM\_2 & (48, 46, 128) & 512 & MAX\_POOLING\_2D\_1 \\
CONV\_2D\_3 & (48, 46, 256) & 295168 & BATCHNORM\_2 \\
Nf $=$ 256, ks $=$ (3x3) & &  & \\
MAX\_POOLING\_2D\_3 & (24, 23, 256) & 0 & CONV\_2D\_3 \\
BATCHNORM\_3 & (96, 92, 256) & 1024 & MAX\_POOLING\_2D\_3 \\
CONV\_2D\_4 & (24, 23, 512) & 1180160 & BATCHNORM\_3 \\
Nf $=$ 512, ks $=$ (3x3) & &  & \\
MAX\_POOLING\_2D\_4 & (12, 12, 512) & 0 & CONV\_2D\_4 \\
BATCHNORM\_4 & (12, 12, 512) & 2048 & MAX\_POOLING\_2D\_3 \\
GLOBAL\_MAX\_POOL\_2D & (512) & 0 & BATCHNORM\_4 \\
DENSE\_1 & (512) & 262656 & GLOBAL\_MAX\_POOL\_2D \\
BATCHNORM\_4 & (512) & 2048 & DENSE\_1 \\
DENSE\_2 & (128) & 65664 & BATCHNORM\_4 \\
BATCHNORM\_5 & (128) & 512 & DENSE\_2 \\
DENSE\_3 & (64) & 8256 & BATCHNORM\_5 \\
BATCHNORM\_6 & (64) & 256 & DENSE\_3 \\ \\
\hline
\multicolumn{4}{c}{Total params:  1,940,480} \\
\multicolumn{4}{c}{Trainable params: 1,937,152} \\
\multicolumn{4}{c}{Non-trainable params: 3,328} \\
\hline
\end{tabular}
\tablefoot{Encoder Architecture of our Siamese Neural Network model. Nf, and ks stand for number of filters, and kernel size. DENSE\_1 layer is the representation space of eCALIFA galaxies. This vector is then projected via fully connected layers to the constrastive space (BATCHNORM\_6) where the loss function is computed.}
\label{table:encoder-architecture}
\end{table*}

\section{PCA autospectra}\label{app:PCA}
\begin{figure*}[!ht]
    \centering
    \setlength{\tabcolsep}{0pt} 

    \begin{tabular}{ccc}
        \includegraphics[width=6.6cm,height=4.cm]{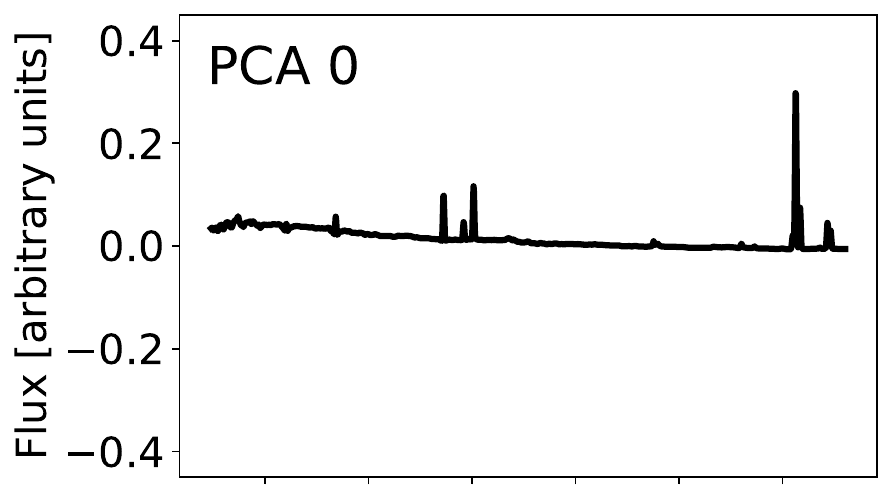} & 
        \includegraphics[width=5.8cm,height=4.cm]{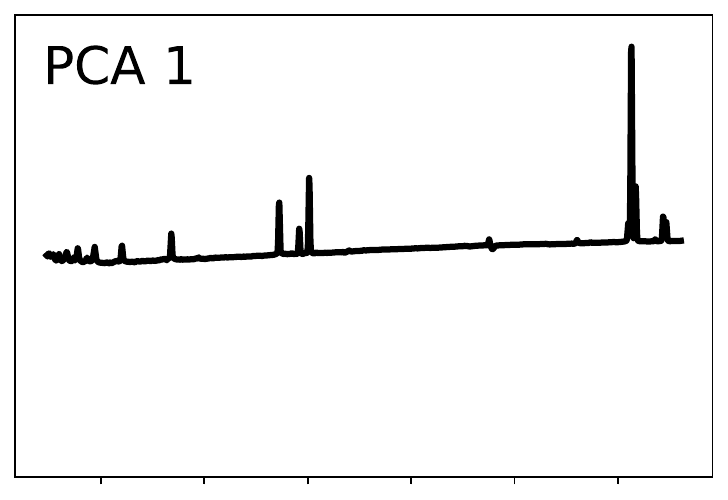} &
        \includegraphics[width=5.8cm,height=4.cm]{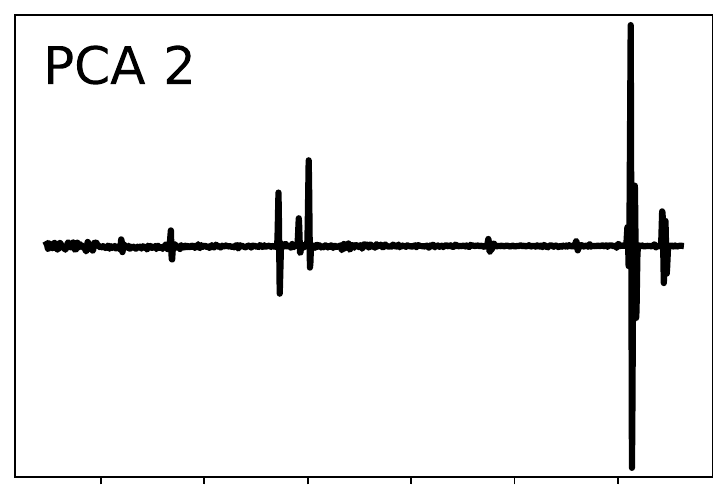} \\
        \\[-0.66cm]
        \includegraphics[width=6.6cm,height=4.cm]{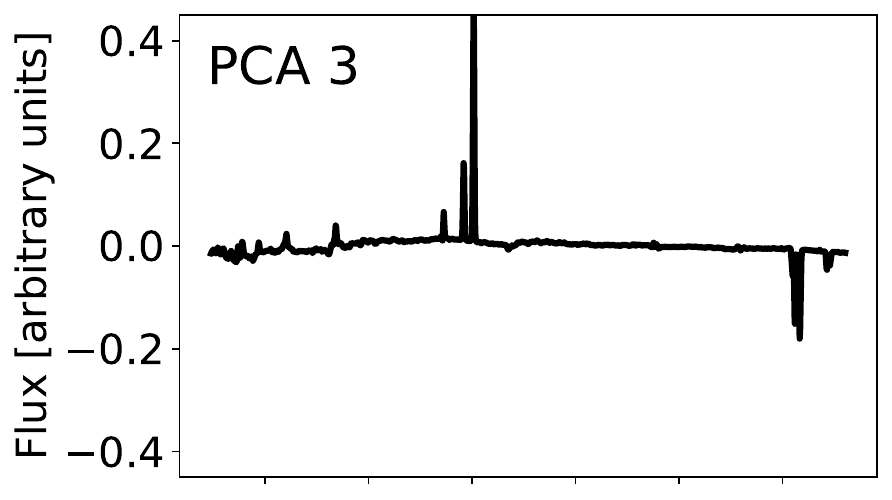} & 
        \includegraphics[width=5.8cm,height=4.cm]{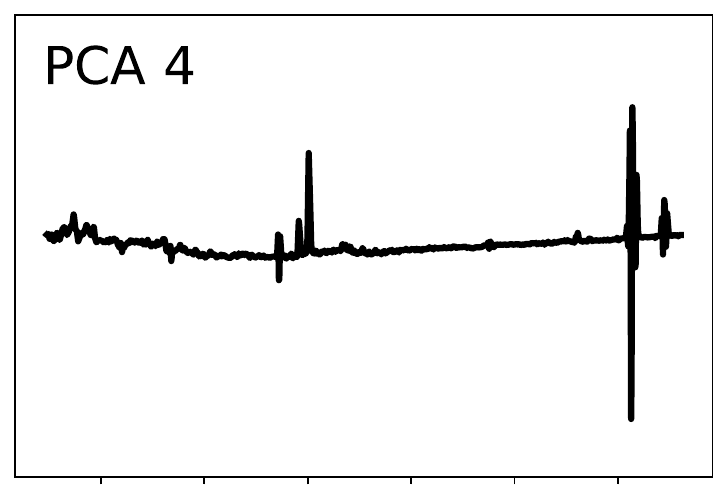} &
        \includegraphics[width=5.8cm,height=4.cm]{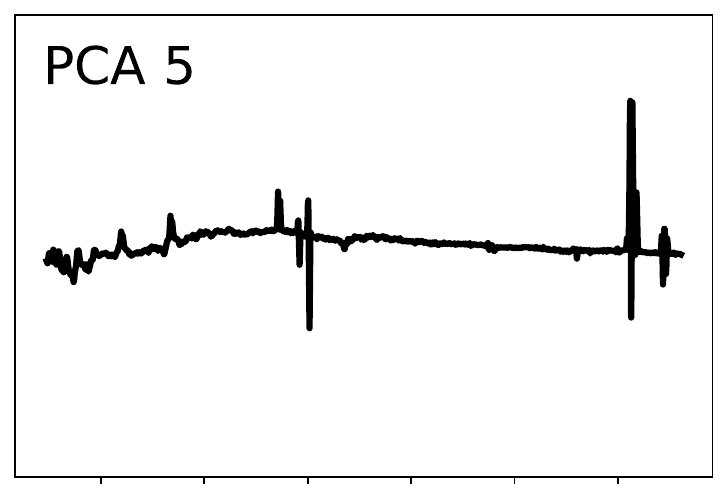} \\
        \\[-0.66cm]
        \includegraphics[width=6.6cm,height=4.cm]{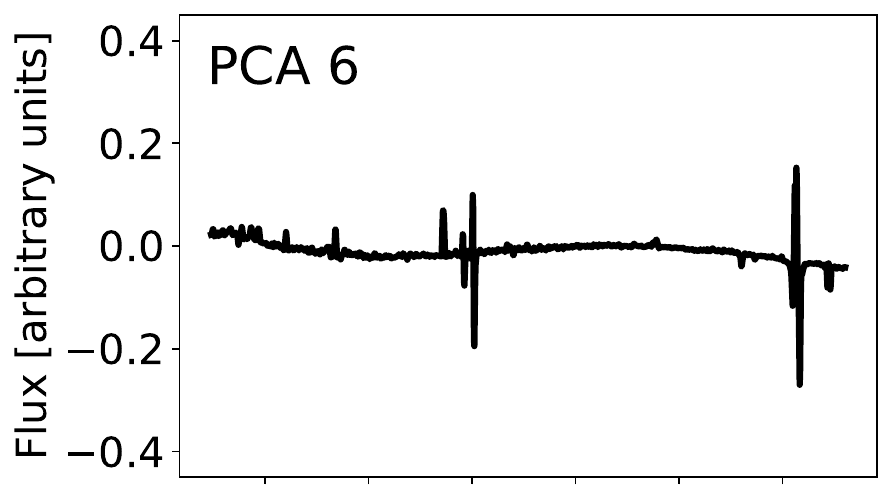} & 
        \includegraphics[width=5.8cm,height=4.cm]{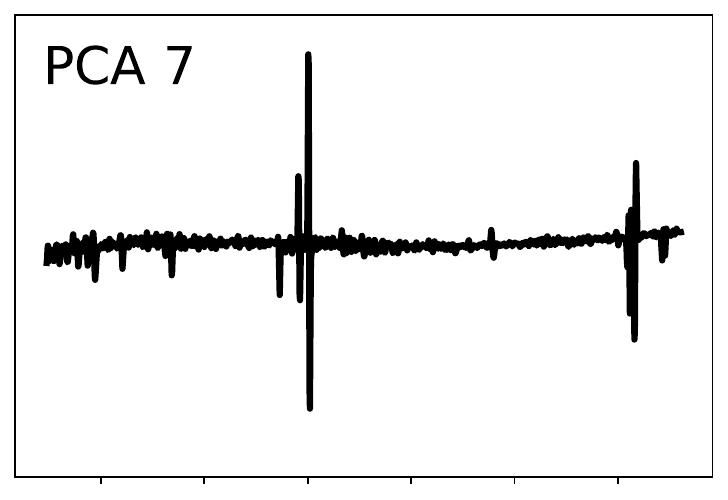} &
        \includegraphics[width=5.8cm,height=4.cm]{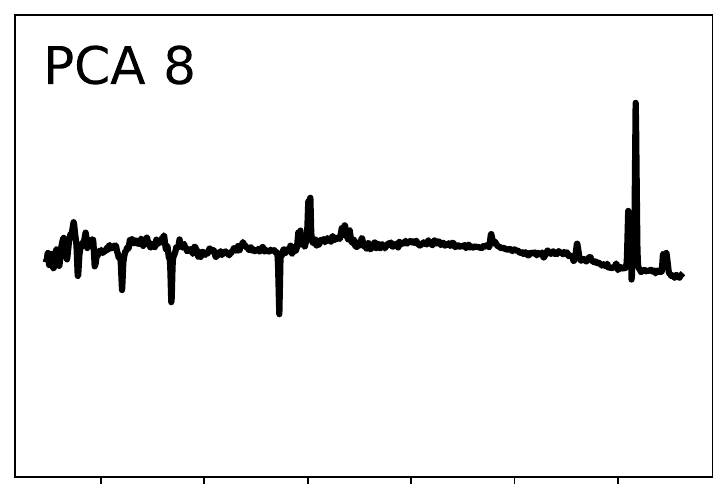} \\
        \\[-0.66cm]
        \includegraphics[width=6.6cm,height=4.cm]{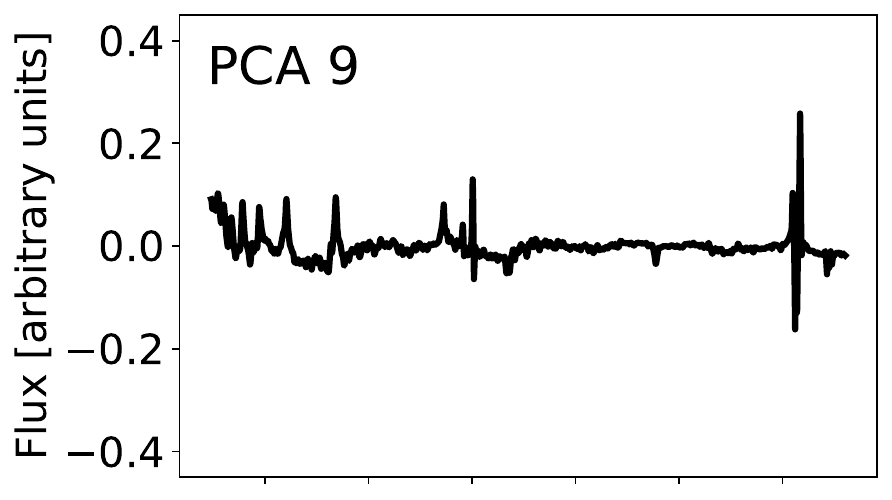} & 
        \includegraphics[width=5.8cm,height=4.cm]{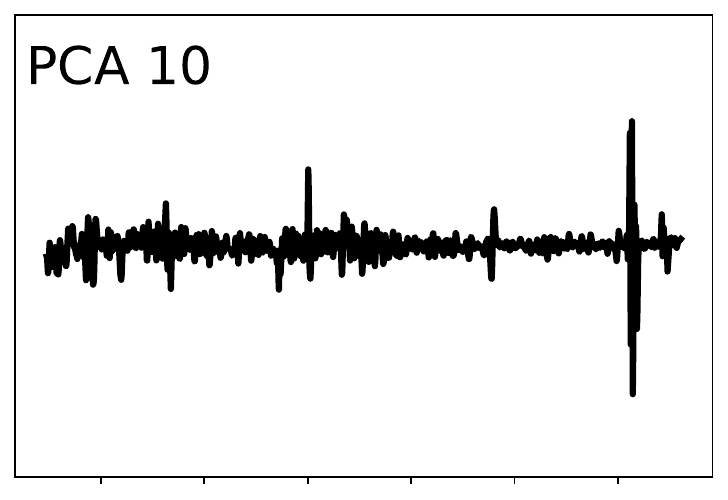} &
        \includegraphics[width=5.8cm,height=4.cm]{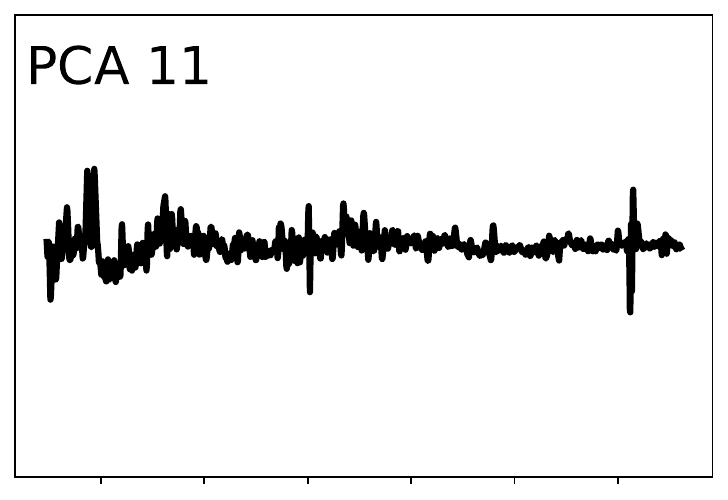} \\
        \\[-0.66cm]
        \includegraphics[width=6.6cm,height=5cm]{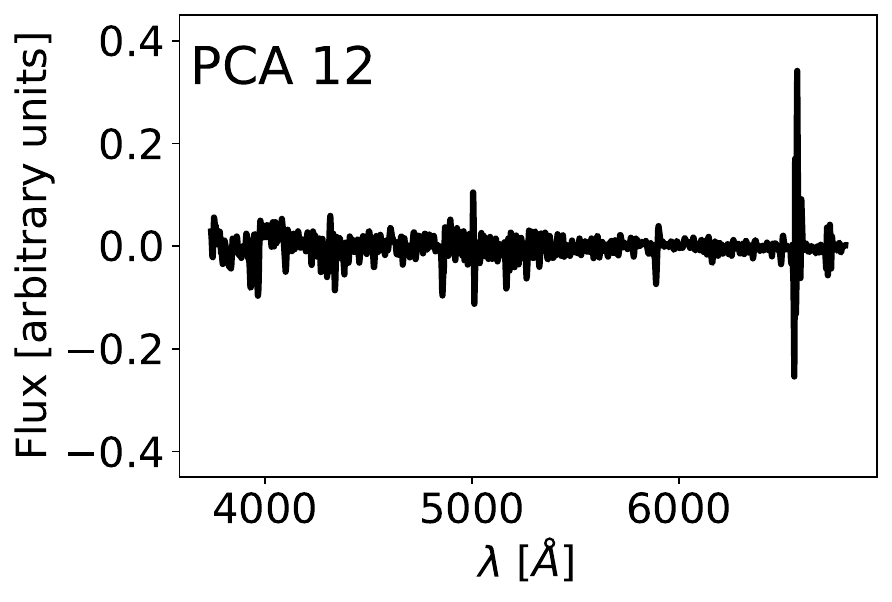} & 
        \includegraphics[width=5.8cm,height=5cm]{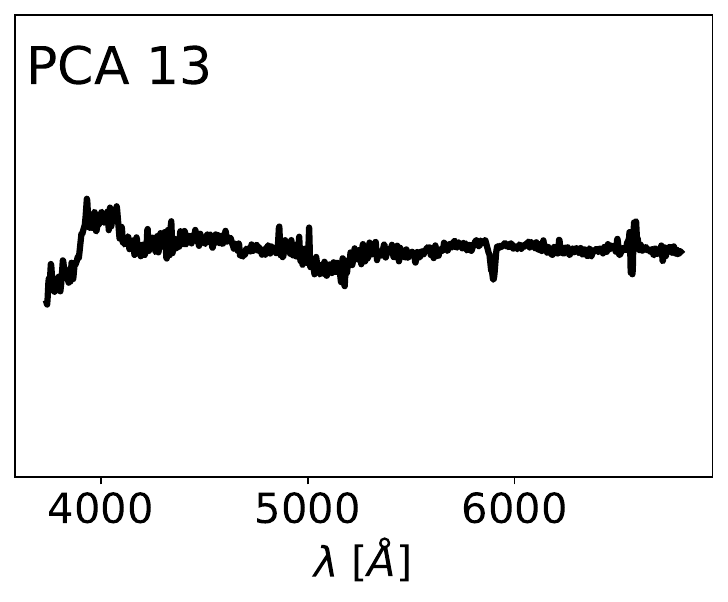} &
        \includegraphics[width=5.8cm,height=5cm]{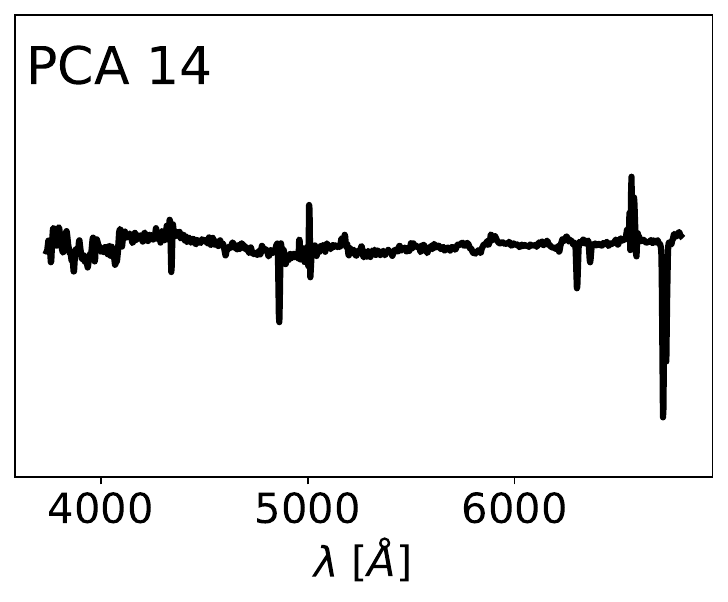} \\
        
        \end{tabular} 
    \caption{\small{First 14 PCA components used to reduce the dimensionality of eCALIFA spectra.}}
    \label{fig:PCA1}
\end{figure*}

In Figures~\ref{fig:PCA1} and \ref{fig:PCA1}, we present the PCA components utilised to reduce the dimensionality of the eCALIFA spectral data. Although it is not straightforward to discern which specific information each PCA component captures—given that this method relies on a mathematical transformation of the data rather than a physically motivated approach—the first two components predominantly capture the shape of the spectra and the intensity of the emission lines. Notably, PCA 2 appears to be a purely kinematic component, as a discernible shift in lines suggests it likely encodes the galaxy's rotation pattern. PCA 3, on the other hand, moderates the relationship between the [OIII] $\lambda\lambda$ $4959, 5006$ and the [NII] doublet, thus influencing the gas phase metallicity of each spaxel by rendering the gas more metal-poor as this coefficient increases. In PCA 4, we observe kinematic features exclusively in the Balmer series, suggesting it modulates the differential kinematics of collisional emission lines relative to those of the Balmer series. As we examine higher-order PCA components, it becomes increasingly challenging to unravel the underlying physical effects due to the interplay of multiple phenomena. These components may simply adjust minor variations in the spectra with a purely mathematical intent, aiming to encapsulate the diversity of spectra found in the eCALIFA data.

\begin{figure*}[!ht]
    \centering
    \setlength{\tabcolsep}{0pt} 

    \begin{tabular}{ccc}
        \includegraphics[width=6.6cm,height=4.cm]{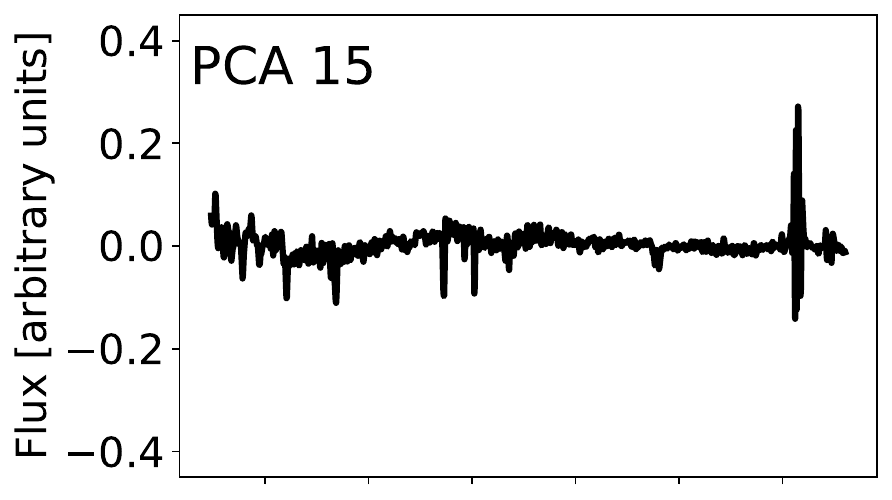} & 
        \includegraphics[width=5.8cm,height=4.cm]{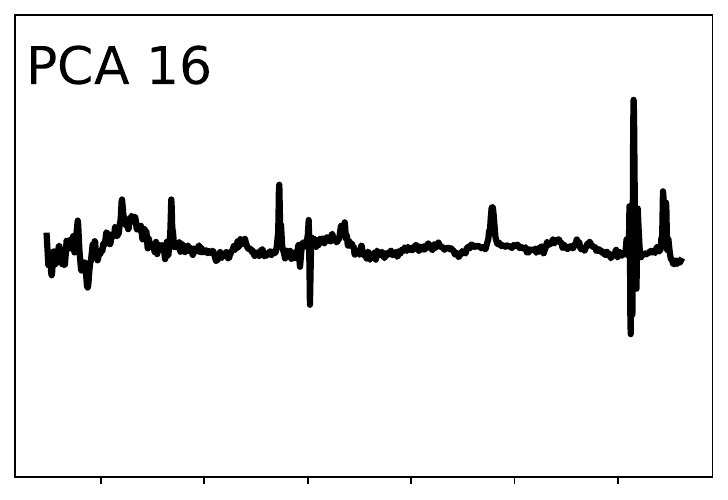} &
        \includegraphics[width=5.8cm,height=4.cm]{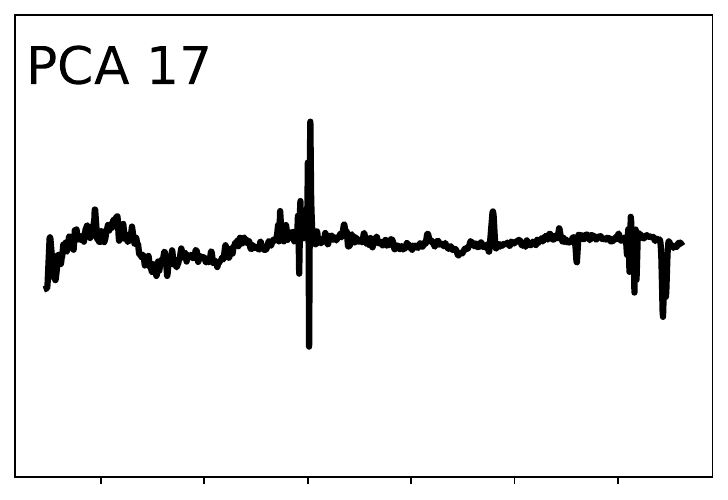} \\
        \\[-0.66cm]

        \includegraphics[width=6.6cm,height=4.cm]{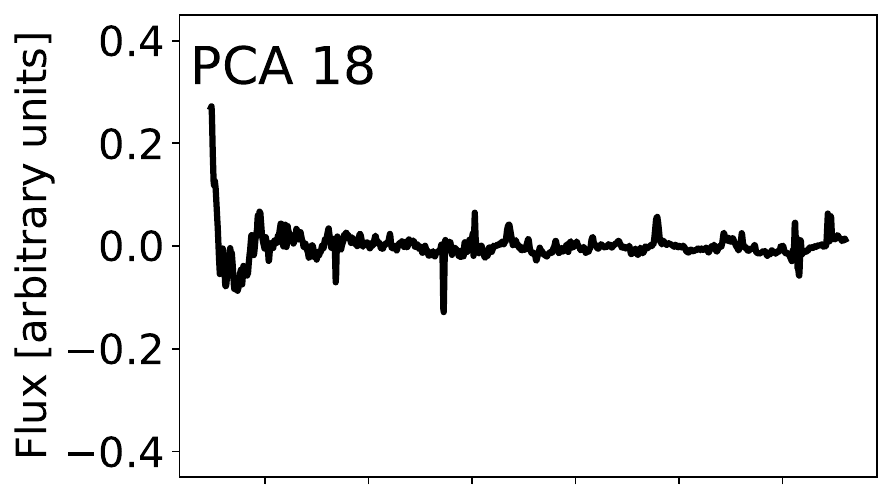} & 
        \includegraphics[width=5.8cm,height=4.cm]{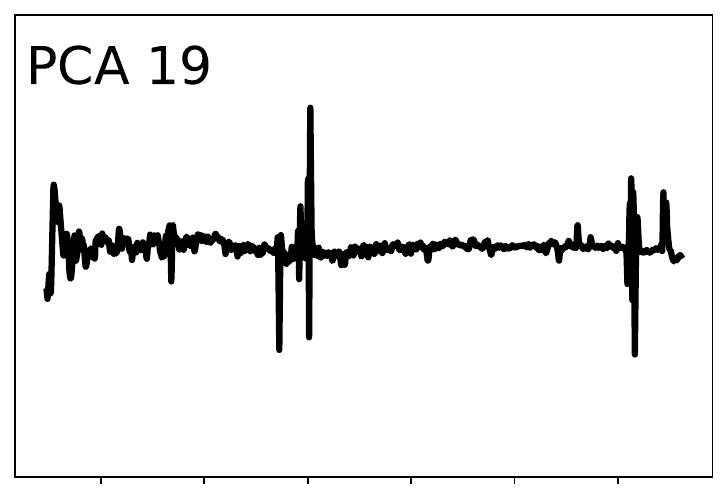} &
        \includegraphics[width=5.8cm,height=4.cm]{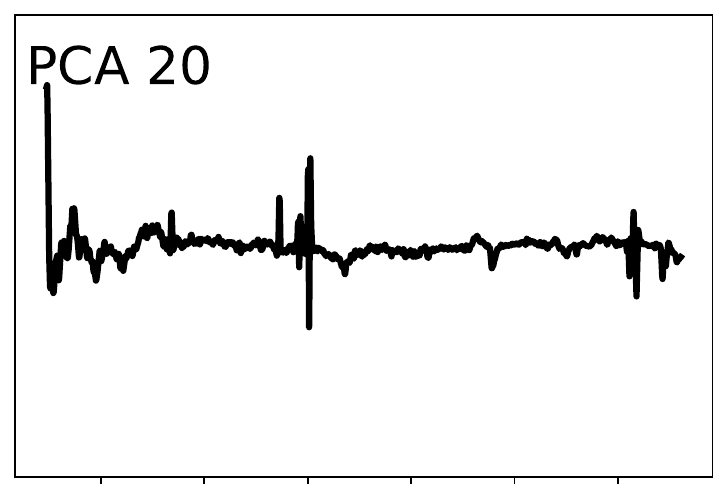} \\
        \\[-0.66cm]

        \includegraphics[width=6.6cm,height=4.cm]{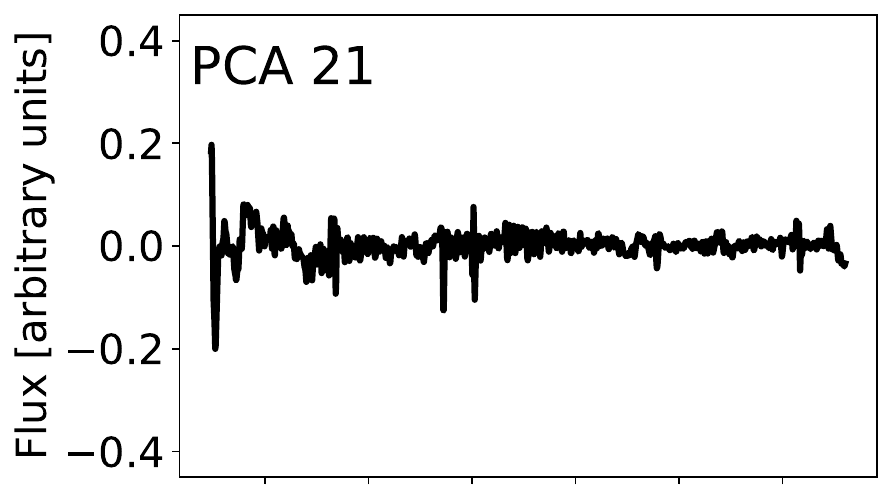} & 
        \includegraphics[width=5.8cm,height=4.cm]{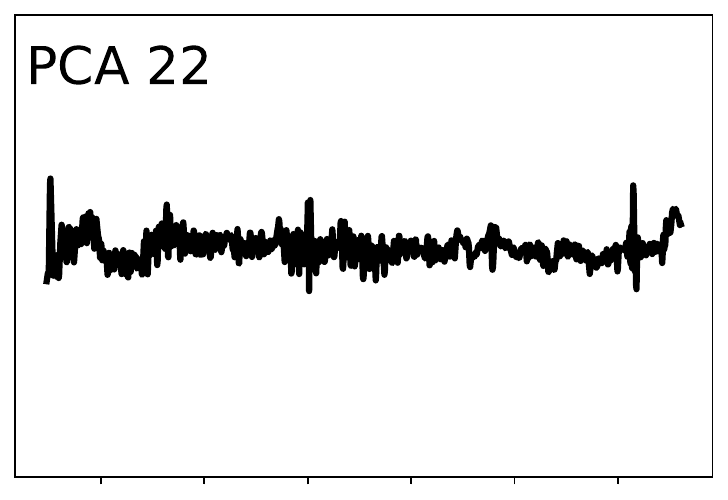} &
        \includegraphics[width=5.8cm,height=4.cm]{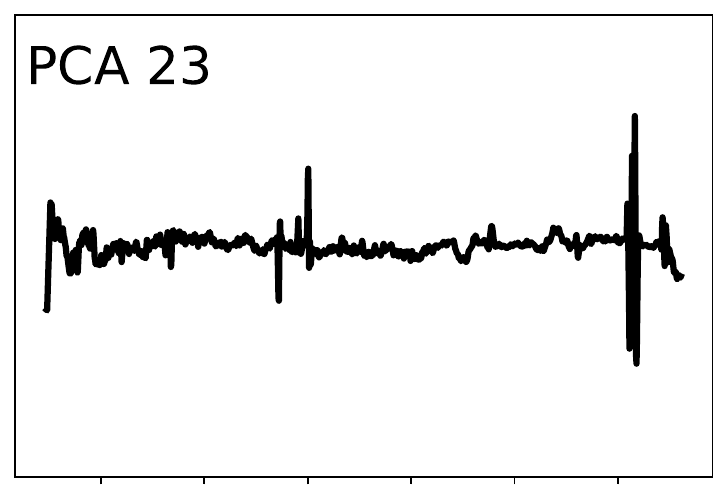} \\
        \\[-0.69cm]

        \raisebox{1.1cm}{\includegraphics[width=6.6cm,height=4.cm]{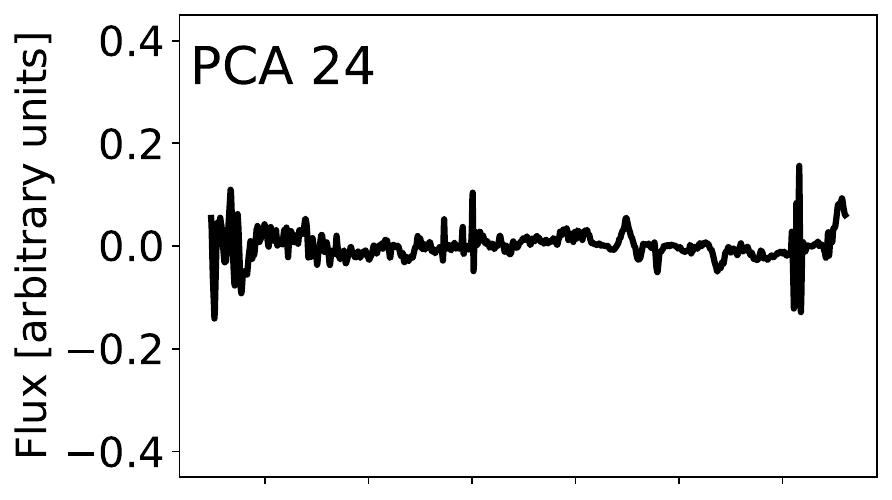}} & 
        \raisebox{1.1cm}{\includegraphics[width=5.8cm,height=4.cm]{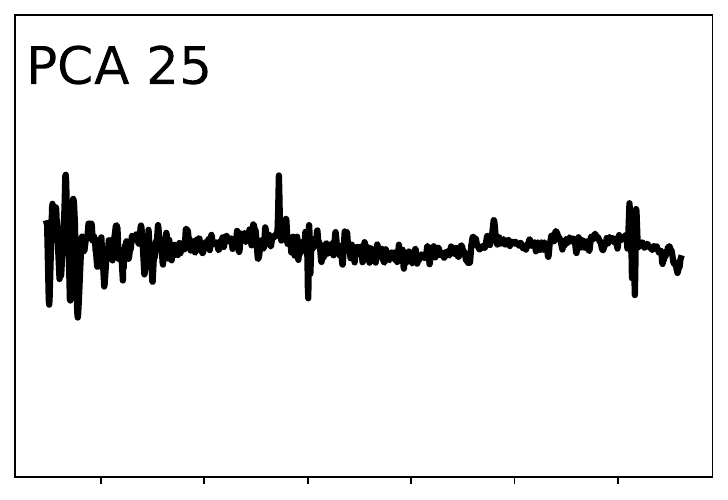}} &
        \raisebox{0.3cm}{\includegraphics[width=5.8cm,height=4.8cm]{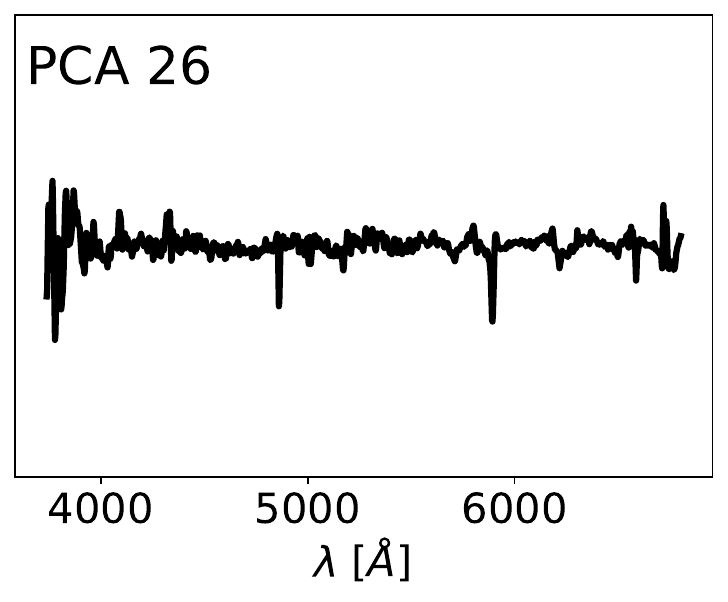}} \\
        \\[-1.75cm]       
        \includegraphics[width=6.6cm,height=4.8cm]{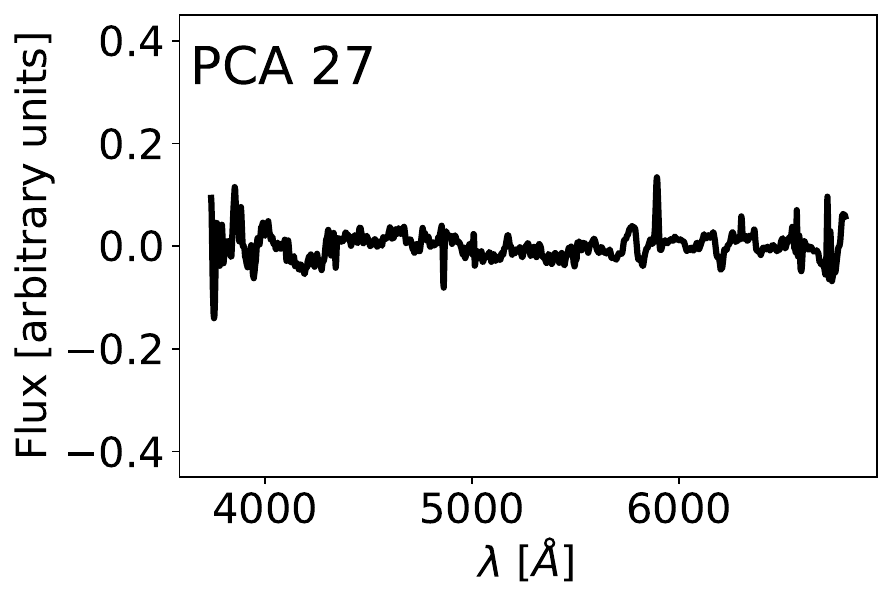} &
        \includegraphics[width=5.8cm,height=4.8cm]{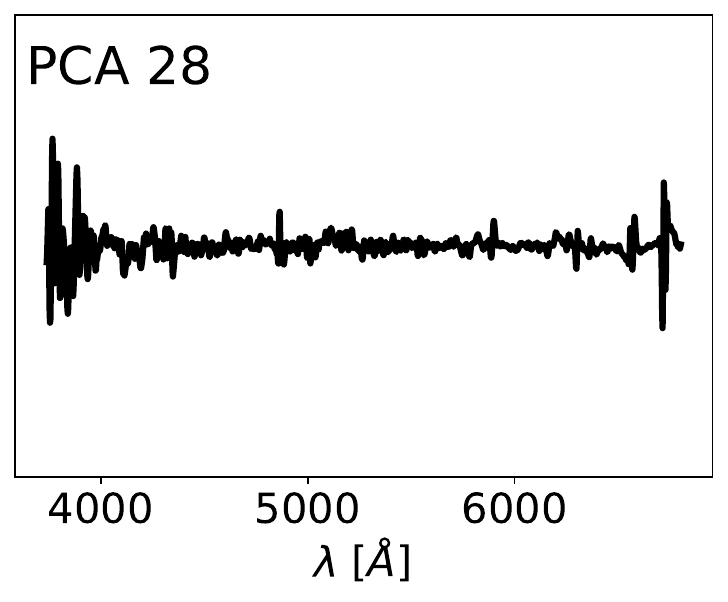} &
        \multicolumn{1}{c}{} 
        \end{tabular} 
    \caption{\small{Remaining PCA components up to the 29th component used reduce the dimensionality of eCALIFA spectra.}}
    \label{fig:PCA2}
\end{figure*}

\section{SNN input maps}\label{app:SNN_input}
\begin{figure*}[h!]
    \centering
    \begin{tabular}{ccc}
        \includegraphics[width=5cm,height=4.5cm]{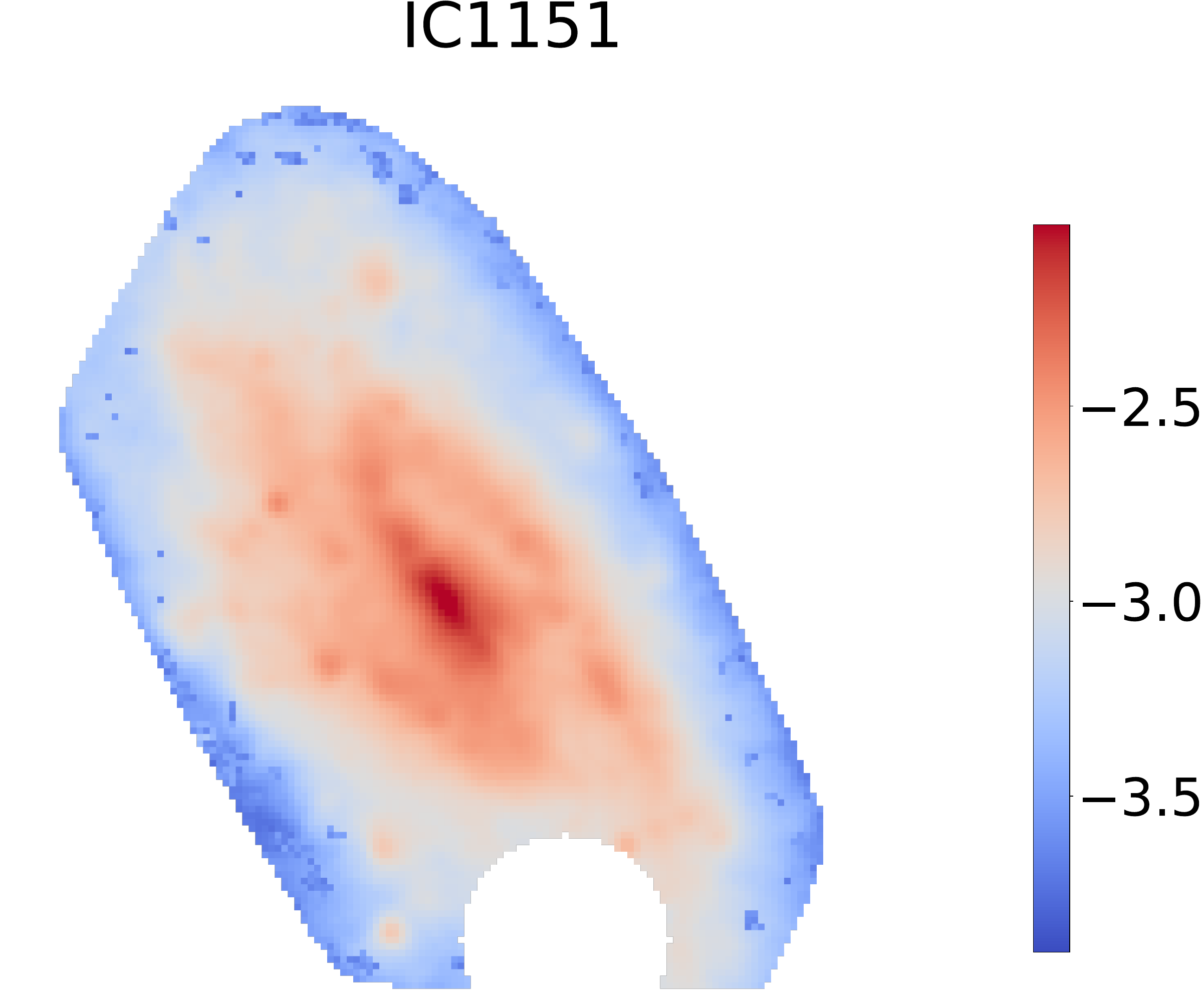} & 
    \includegraphics[width=5cm,height=4.5cm]{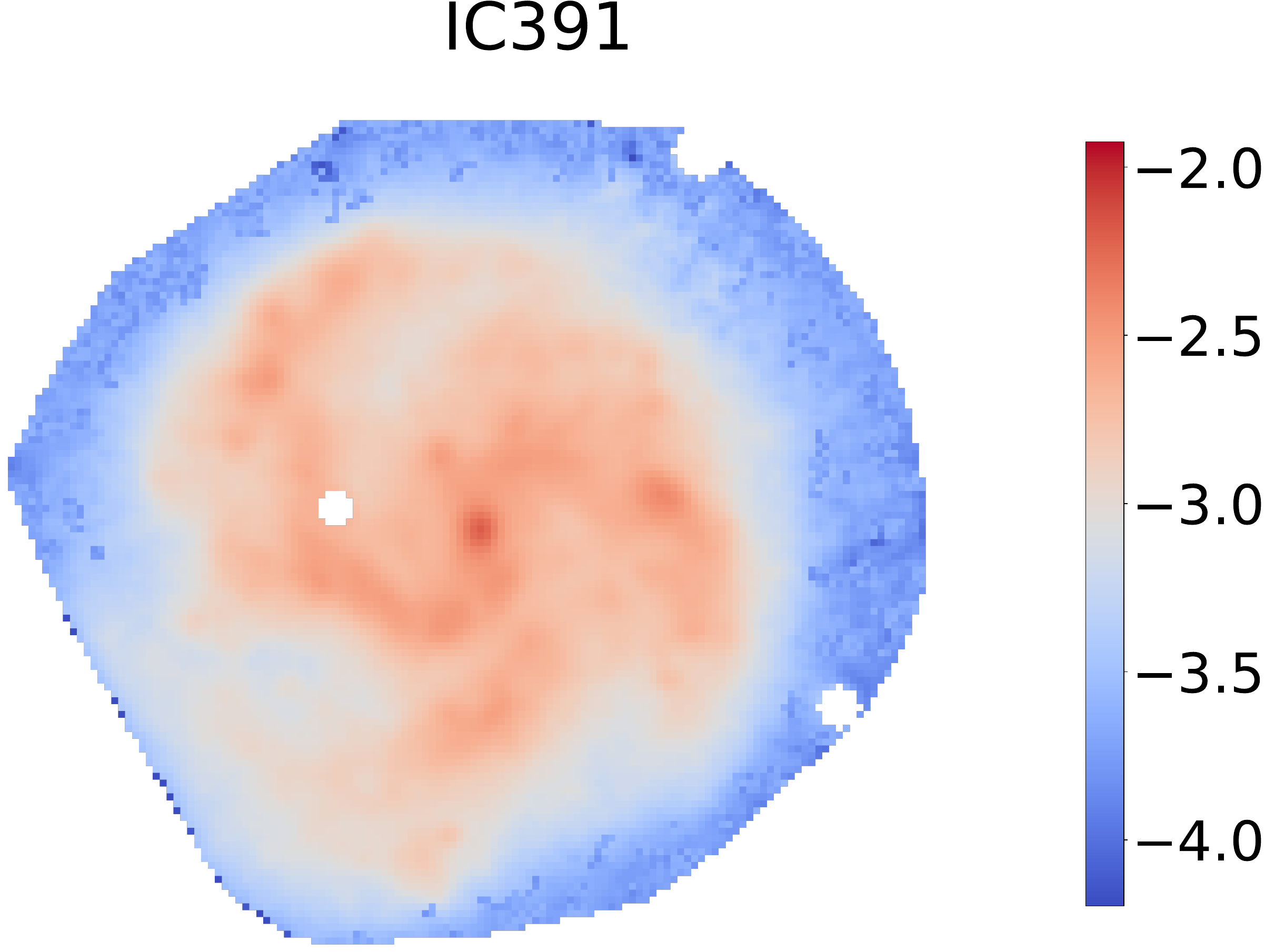} &
        \includegraphics[width=5cm,height=4.5cm]{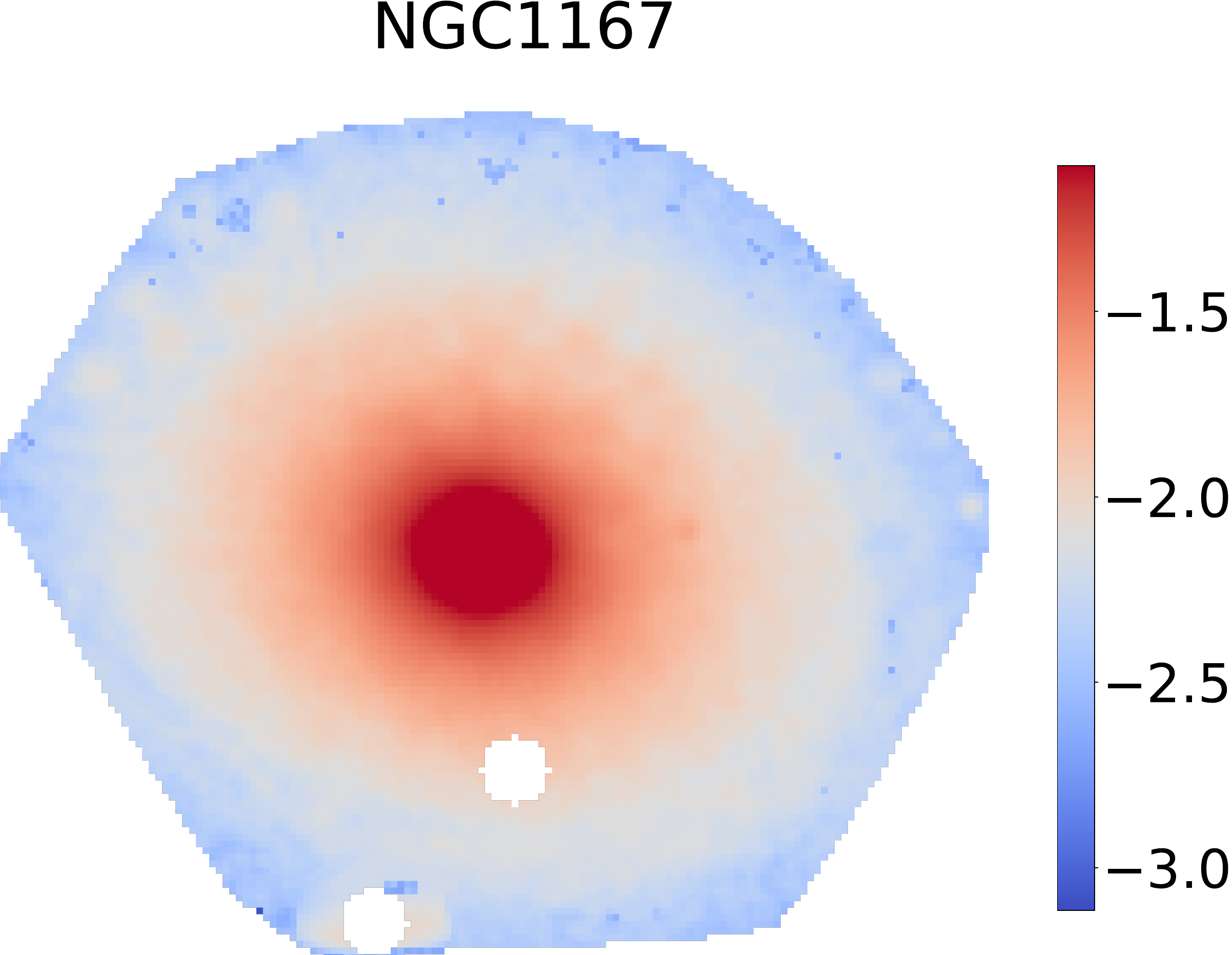}   

        \\
        \includegraphics[width=5cm,height=4.5cm]{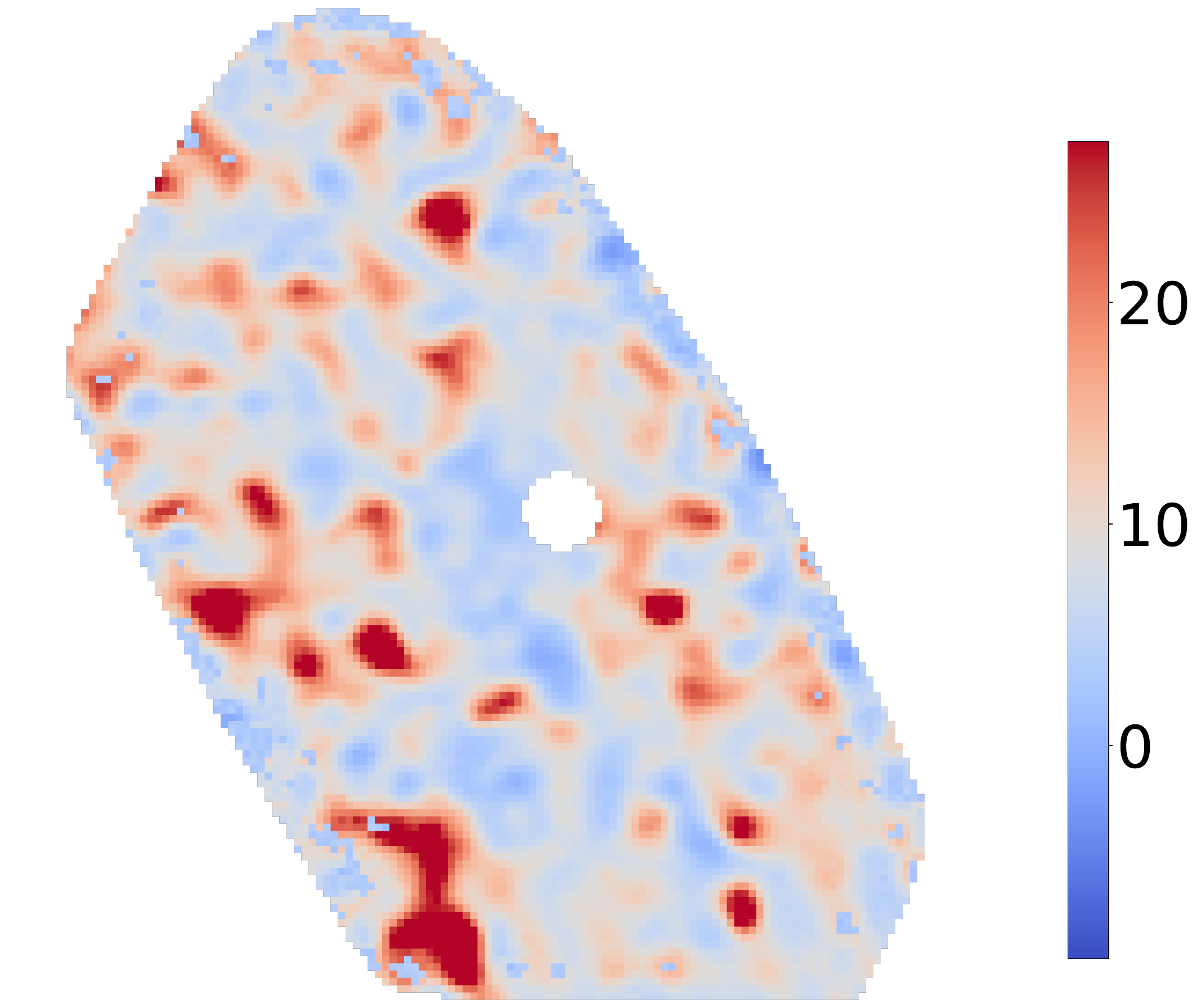} & 
        \includegraphics[width=5cm,height=4.5cm]{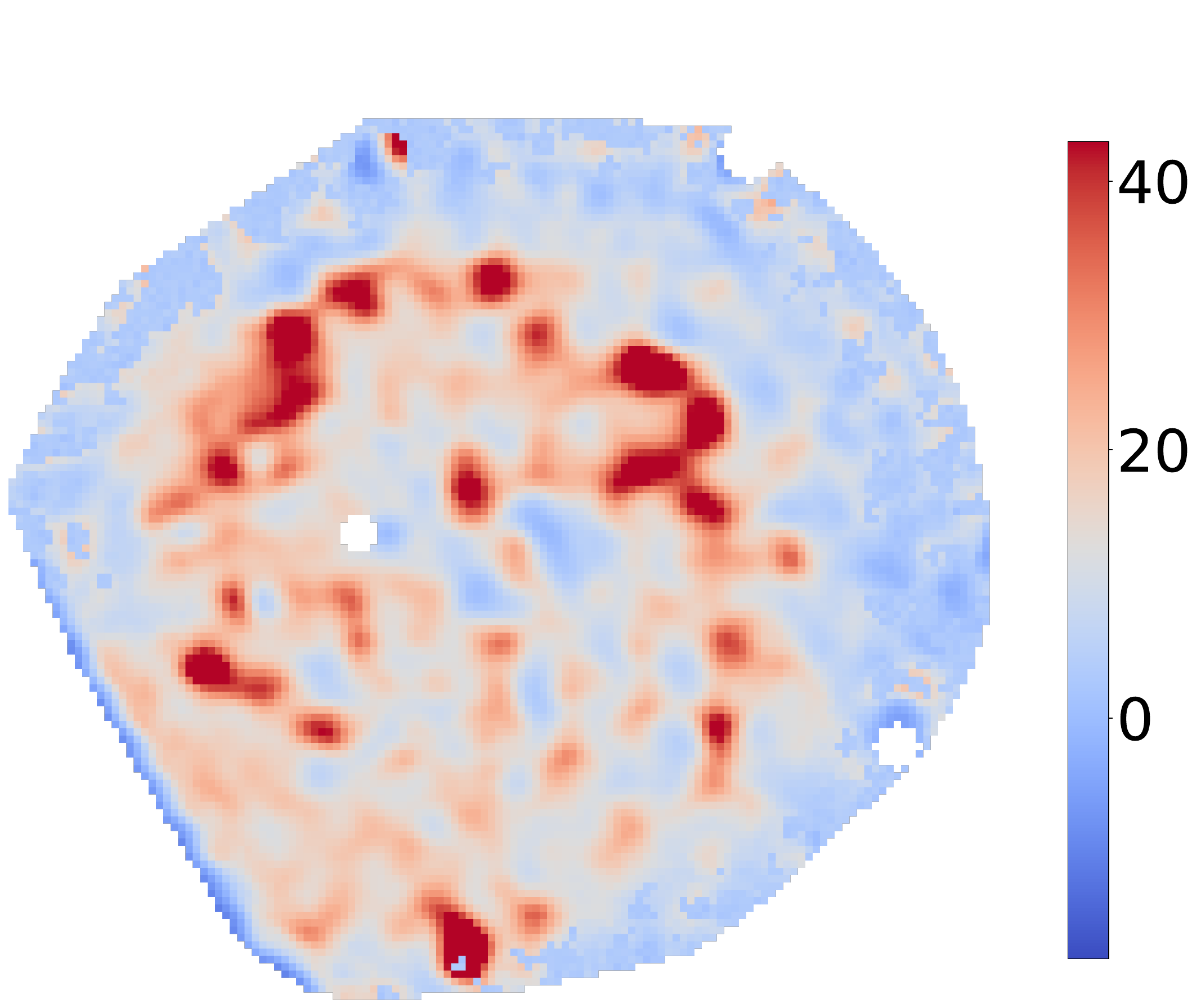} &
        \includegraphics[width=5cm,height=4.5cm]{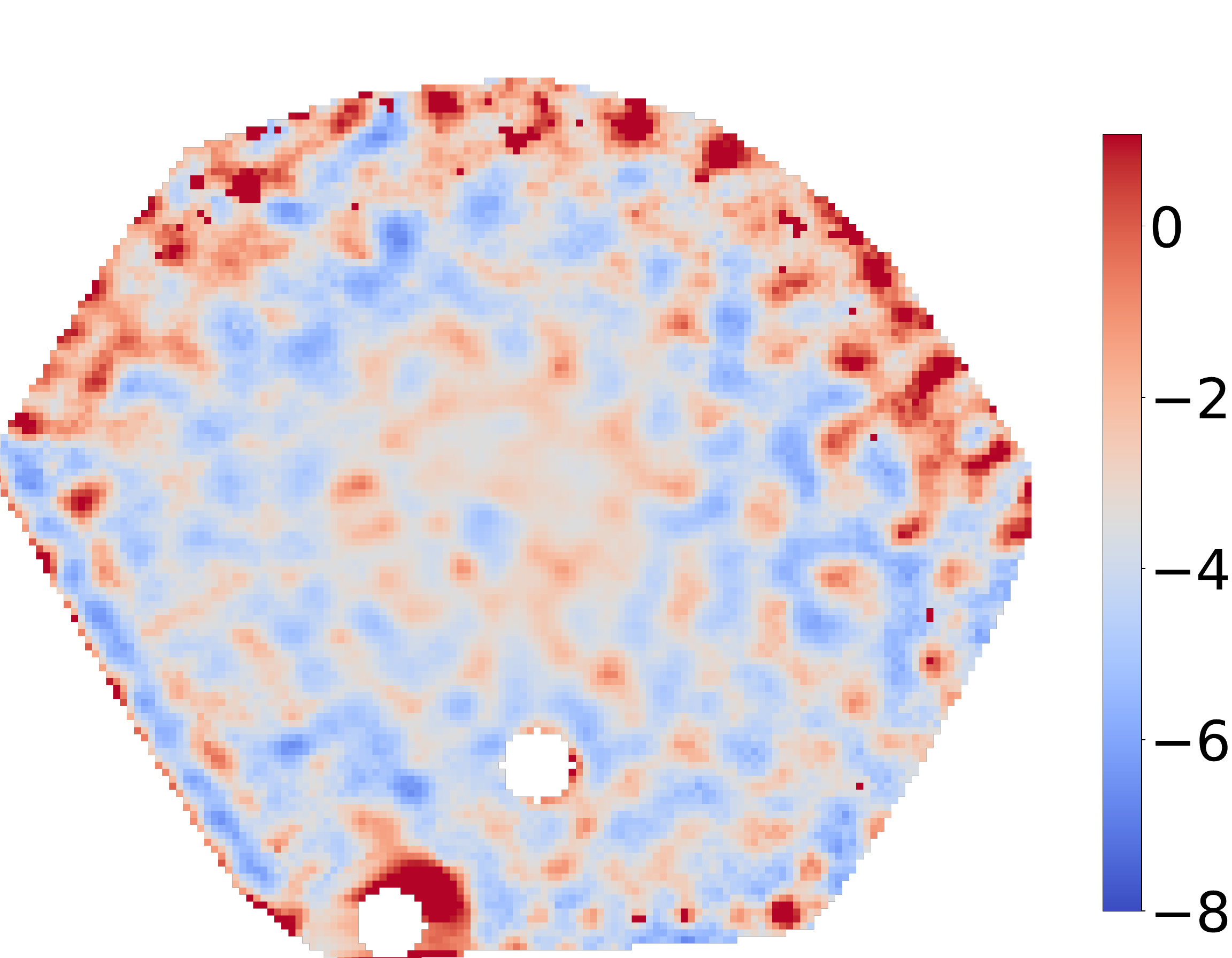}  

        \\
        \includegraphics[width=5cm,height=4.5cm]{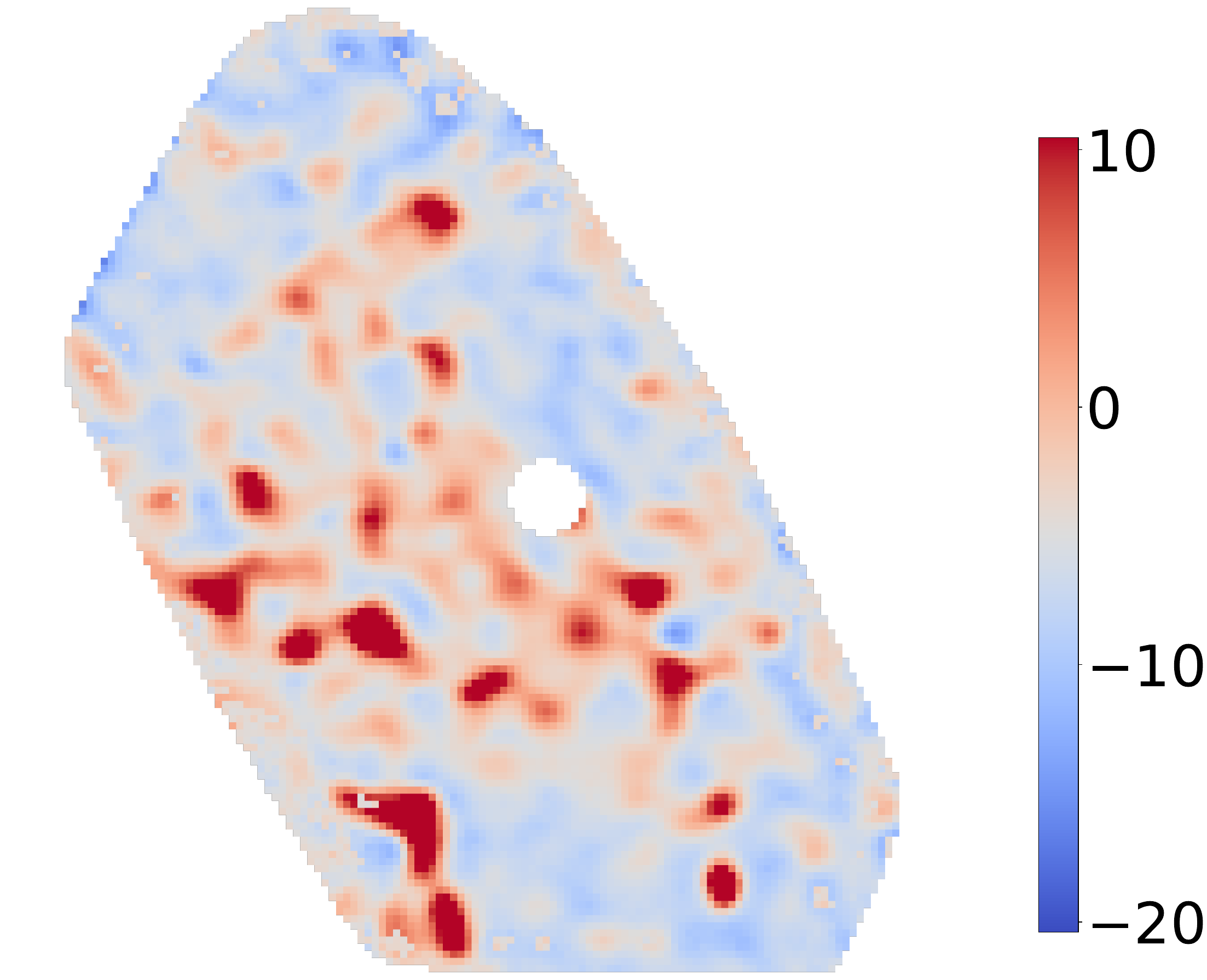} & 
        \includegraphics[width=5cm,height=4.5cm]{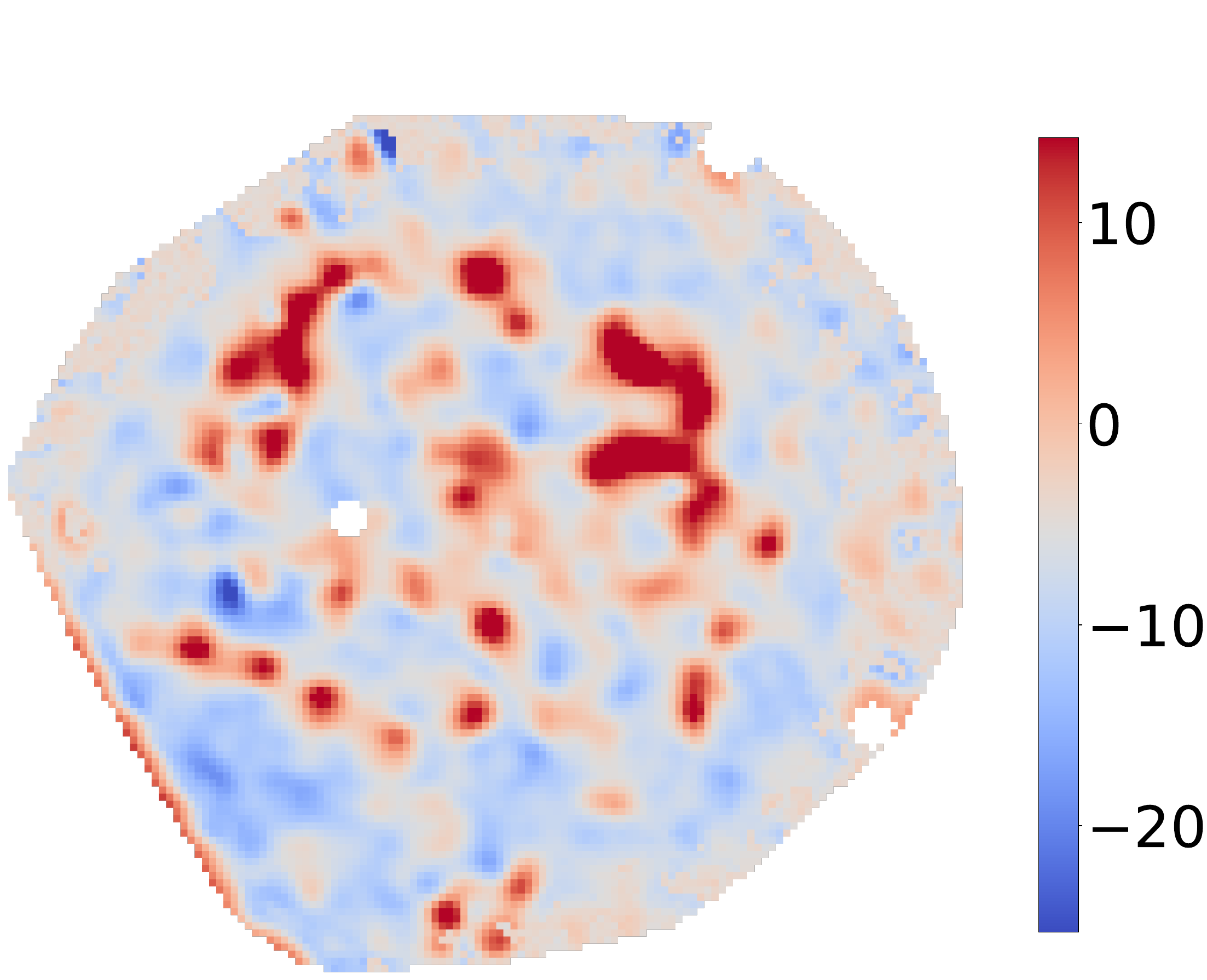} &
        \includegraphics[width=5cm,height=4.5cm]{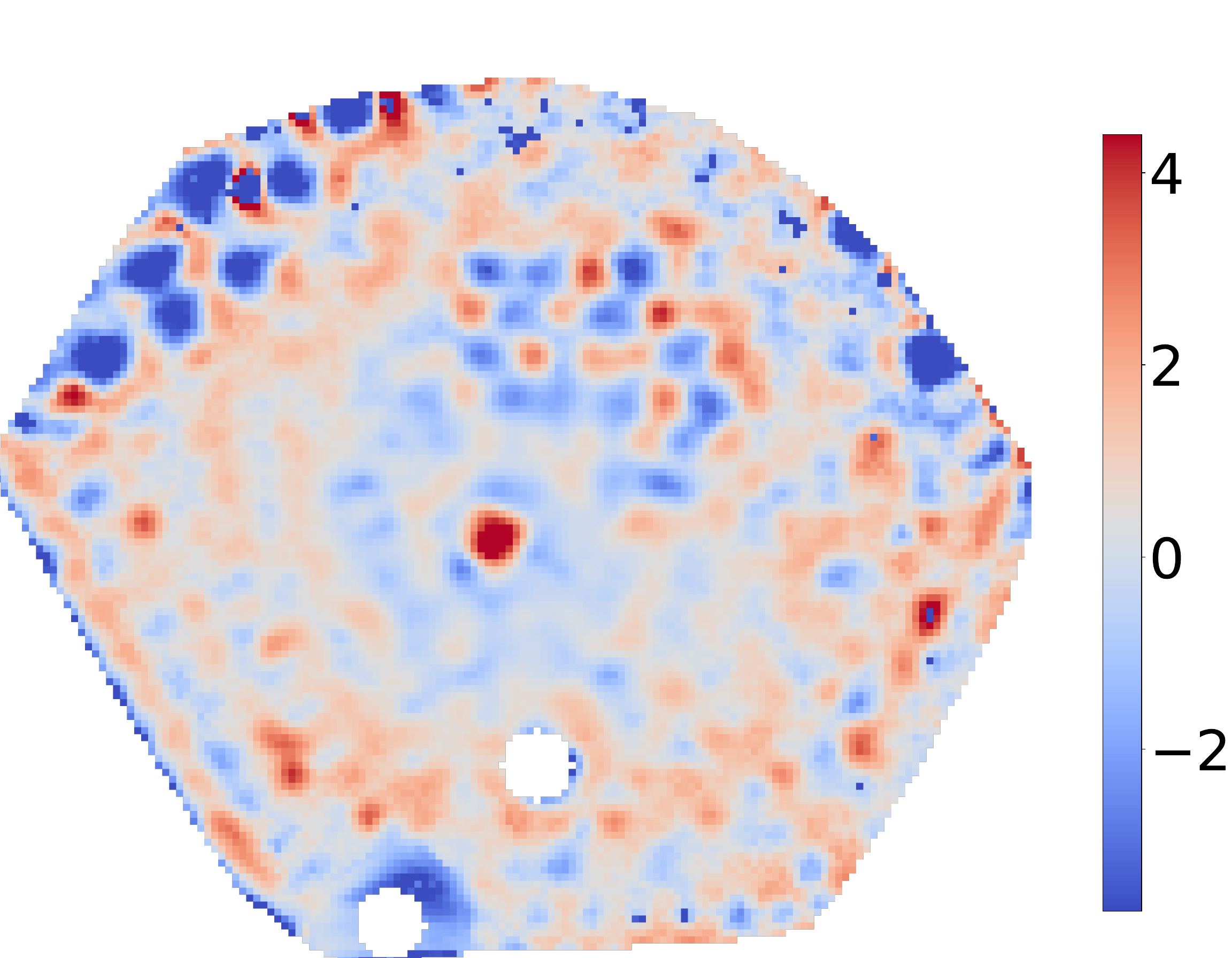}          
        \\
        \includegraphics[width=5cm,height=4.5cm]{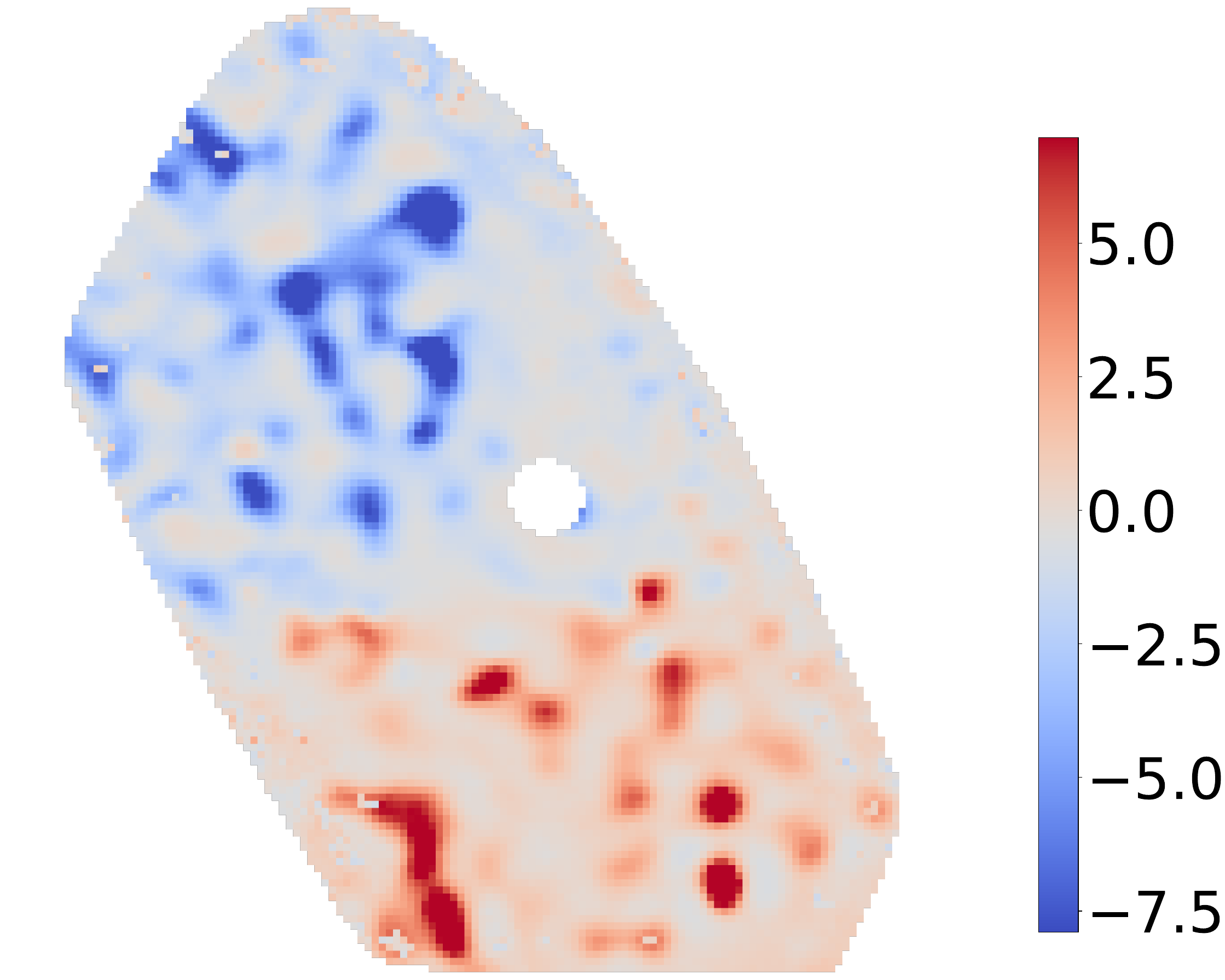} & 
        \includegraphics[width=5cm,height=4.5cm]{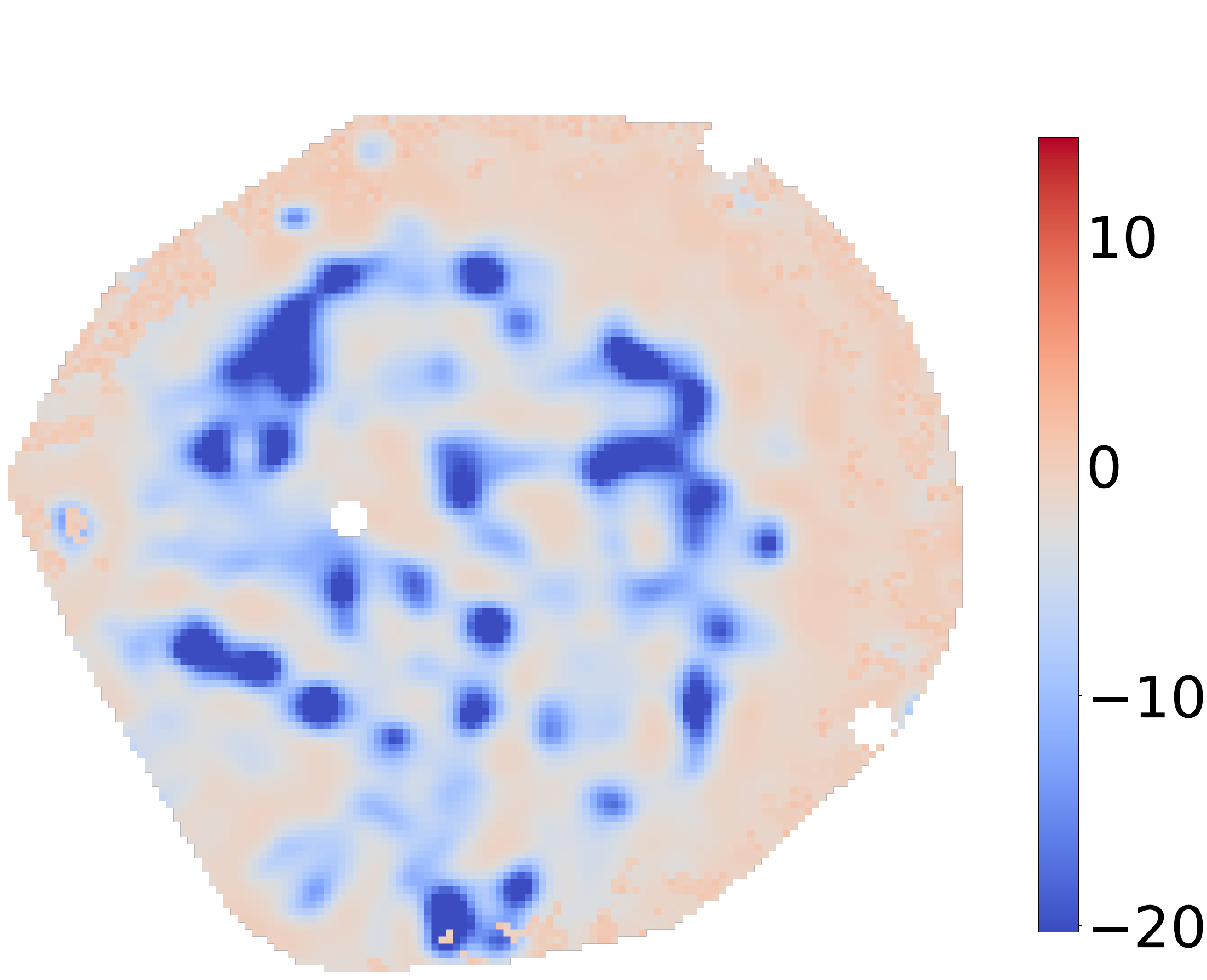} &
        \includegraphics[width=5cm,height=4.5cm]{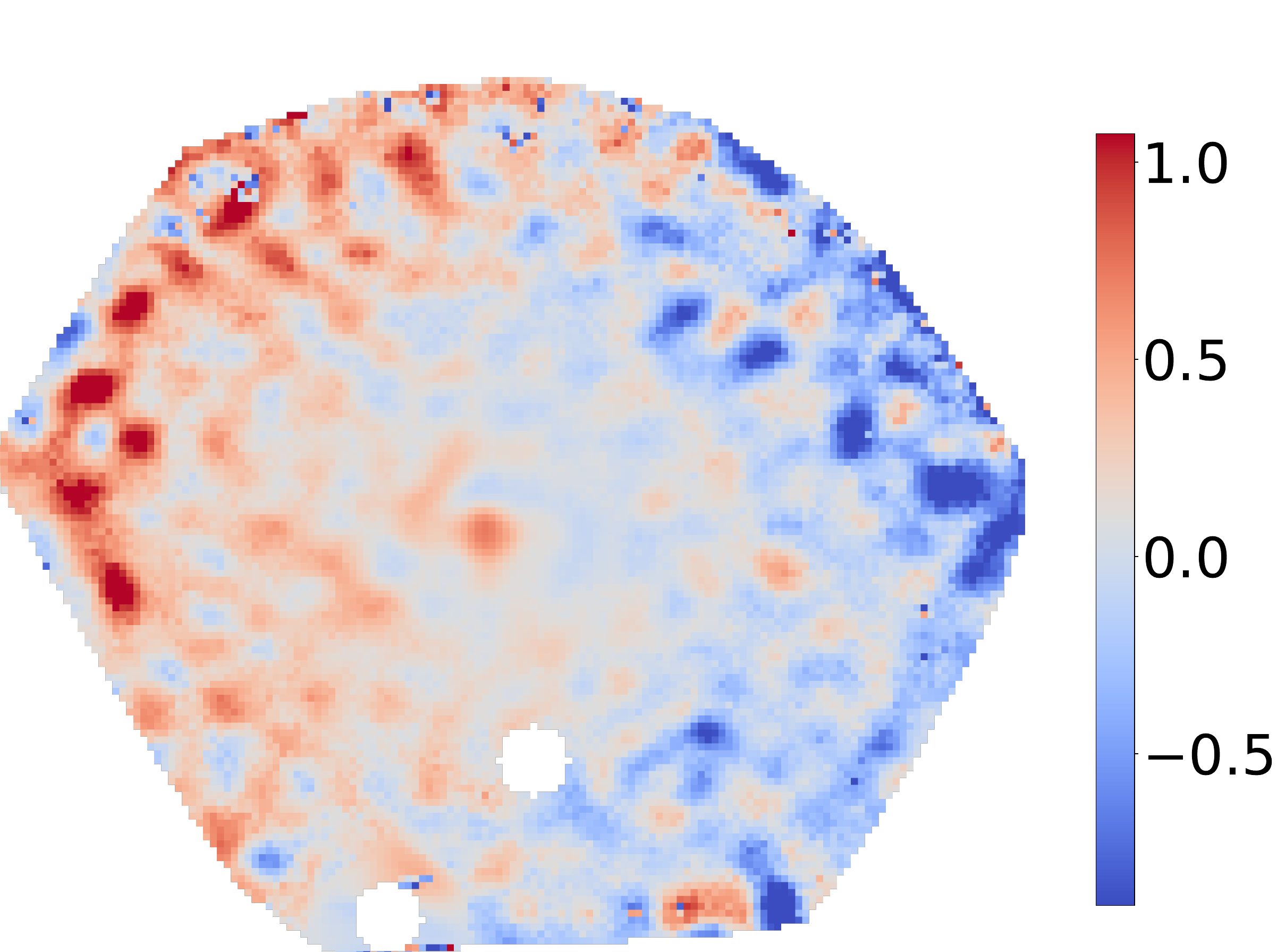}          
        \\
        \includegraphics[width=5cm,height=4.5cm]{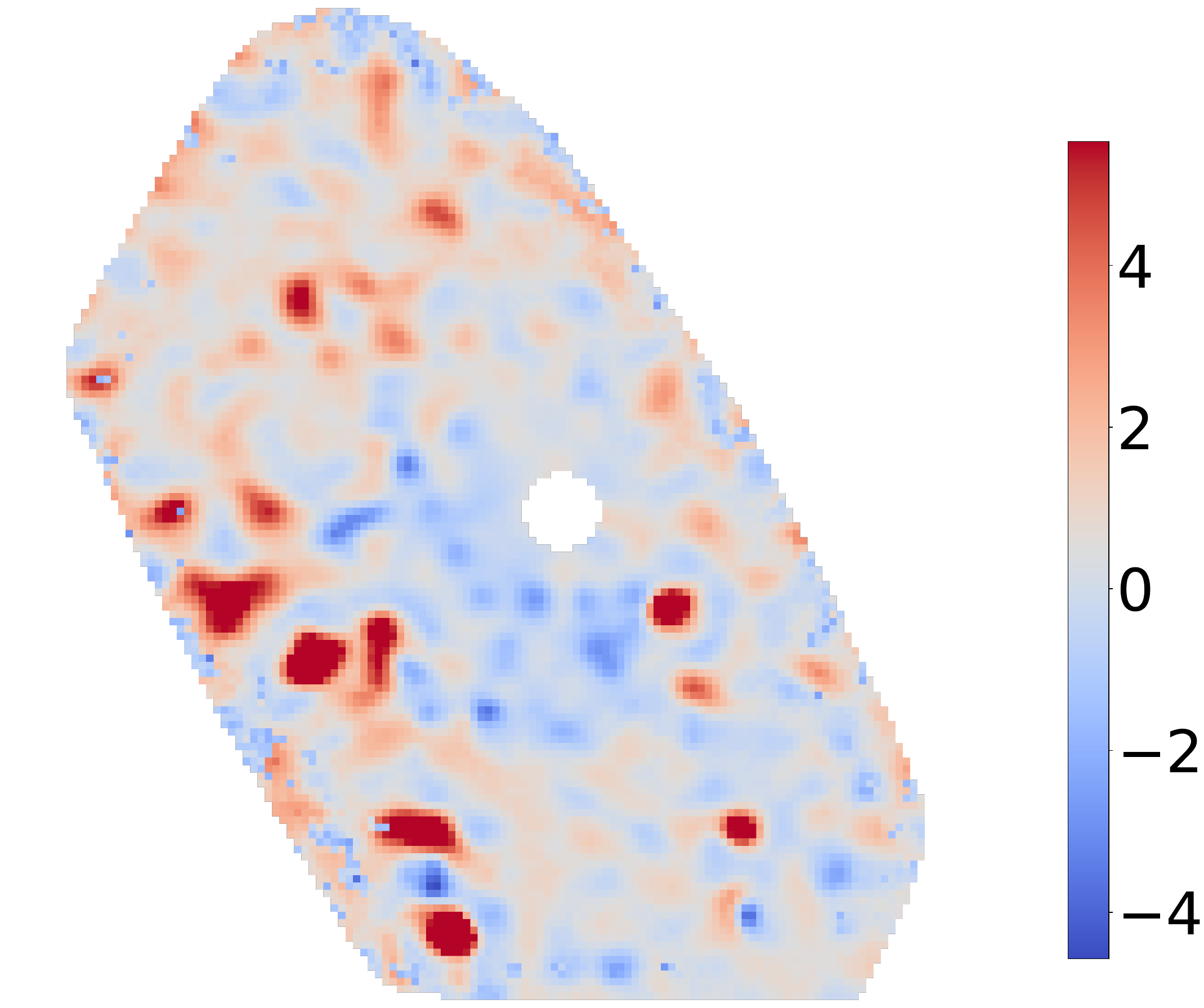} & 
        \includegraphics[width=5cm,height=4.5cm]{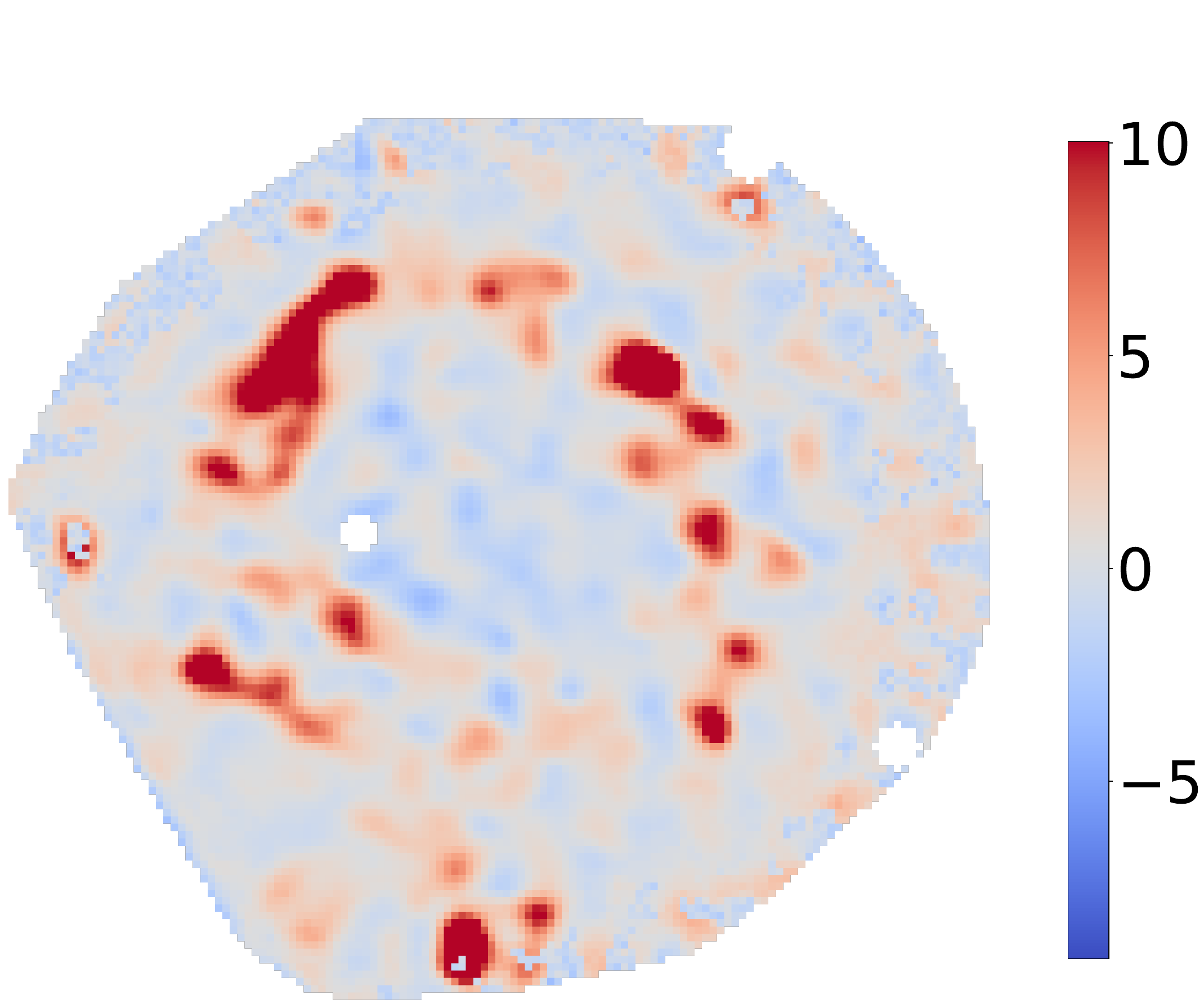} &
        \includegraphics[width=5cm,height=4.5cm]{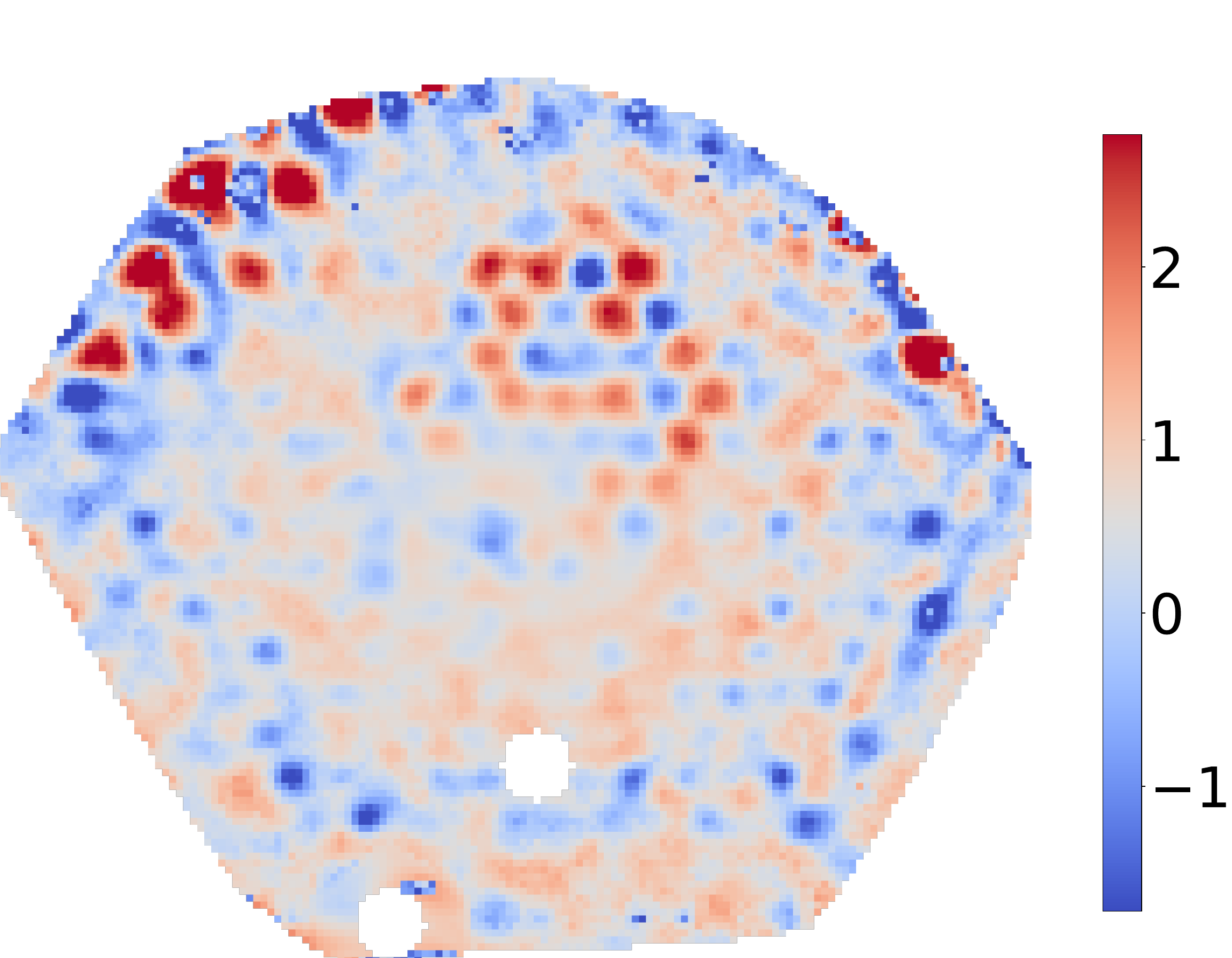}        
        \end{tabular} 
    \caption{\small{Visual representation of various data maps for three galaxies of eCALIFA survey, IC391, IC1151, and NGC1167. Each column corresponds to a galaxy, ordered from left to right. The first row displays the logarithmic luminosity maps at the normalise window (5700~$\AA$) highlighting the distribution of light across each galaxy. The subsequent rows represent PCA maps of differing components: the second row shows PCA 0 maps; the third row contains PCA 1 maps; the fourth row illustrates PCA 2 maps; and the fifth row shows PCA 3 maps.}}
    \label{fig:PCA_maps}
\end{figure*}

To better understand the data processed by the SNN, we present in Figure~\ref{fig:PCA_maps} the first four PCA maps alongside the luminosity map for the galaxies shown in Figure~\ref{fig:transformation}. IC391 and IC1151 are both spiral galaxies exhibiting SF, while NGC1167 is classified as an S0 galaxy with an AGN. In the PCA 0 maps, SF clumps are prominent features for IC391 and IC1151. Notably, IC391 exhibits more intense SF events compared to IC1151. In contrast, NGC1167 displays minimal SF, primarily at the periphery, and is predominantly quenched. Examining the PCA 1 map, both IC391 and IC1151 show similar intensities, reflecting the structural similarities observed in the PCA 0 map, which modulate the presentation of SF events. However, an intensity peak at the centre of NGC1167 in the PCA 1 map signals the presence of the AGN. Turning to PCA 3, which represents gas kinematics, we observe the rotational motion of the galaxies in IC391 and NGC1167. However, other PCA components also influence the kinematics, as we cannot recognise the rotation pattern in IC1151 directly. It is important to acknowledge that PCA components cannot be directly correlated with physical entities due to the mathematical basis of PCA; therefore, physical interpretations must be approached with caution. Nevertheless, PCA maps encapsulate the most relevant information from the eCALIFA data cubes, and it is the task of the SNN to interpret these data effectively.

\end{document}